\documentclass[prd,aps,eqsecnum,12pt]{revtex4}

\usepackage{amssymb,amsmath}
\usepackage{graphicx}
\begin{document}

\def\d{{\mathrm{d}}}
\def\nbox#1#2{\vcenter{\hrule \hbox{\vrule height#2cm \kern#1cm \vrule}
\hrule}}
\def\sq{\,\raise.5pt\hbox{$\nbox{.22}{.22}$}\,}
\def\sqb{\,\raise.5pt\hbox{$\overline{\nbox{.22}{.22}}$}\,}
\newcommand{\tab}{$\langle T_{ab} \rangle$}
\preprint{LA-UR-04-6142}

\title{Short distance and initial state effects in inflation: \\ stress
tensor and decoherence \\   \ }
\author{Paul R. Anderson}
\affiliation{Department of Physics,
Wake Forest University, Winston-Salem,
North Carolina, 27109}
\email{anderson@wfu.edu}
\author{Carmen Molina-Par\'{\i}s}
\affiliation{Department of Applied Mathematics,
University of Leeds,
Leeds LS2 9JT, UK and
Departamento de Matem\'aticas, F\'{\i}sica Aplicada y
F\'{\i}sico-qu\'{\i}mica,
Facultad de Farmacia,
Universidad San Pablo CEU, E-28660 Madrid,
Spain}
\email{carmen@maths.leeds.ac.uk}
\author{Emil Mottola}
\affiliation{Theoretical Division, T-8, Los Alamos
National Laboratory,\\ Los Alamos, New Mexico, 87545}
\email{emil@lanl.gov}

\maketitle
\centerline{LA-UR-04-6142}
\vspace{2cm}
\pagestyle{empty}
\pagebreak

\centerline{\bf Abstract}
\vspace{1cm}

We present a consistent low energy effective field theory
framework for parameterizing the effects of novel short distance
physics in inflation, and their possible observational signatures
in the Cosmic Microwave Background. We consider the class of
general homogeneous, isotropic initial states for quantum scalar
fields in Robertson-Walker (RW) spacetimes, subject to the
requirement that their ultraviolet behavior be consistent with
renormalizability of the covariantly conserved stress
tensor which couples to gravity. In the functional Schr\"odinger
picture such states are coherent, squeezed, mixed states
characterized by a Gaussian density matrix.  This Gaussian has
parameters which approach those of the adiabatic vacuum at large
wave number, and evolve in time according to an effective
classical Hamiltonian. The one complex parameter family of
$\alpha$ squeezed states in de Sitter spacetime does not fall into
this UV allowed class, except for the special value of the
parameter corresponding to the Bunch-Davies state. We determine
the finite contributions to the inflationary power spectrum and
stress tensor expectation value of general UV allowed adiabatic
states, and obtain quantitative limits on the observability and
backreaction effects of some recently proposed models of short
distance modifications of the initial state of inflation.  For all
UV allowed states, the second order adiabatic basis provides a good
description of particles created in the expanding RW universe. Due
to the absence of particle creation for the massless, minimally
coupled scalar field in de Sitter space, there is no phase
decoherence in the simplest free field inflationary models. We
apply adiabatic regularization to the renormalization of the
decoherence functional in cosmology to corroborate this result.

\pagebreak

\pagenumbering{arabic}
\pagestyle{plain}

\section{Introduction and Overview}
\label{sec:intro}

Inflationary models were introduced principally to account for the
observed large scale homogeneity, isotropy and flatness of the
universe in a causal way, independently of detailed initial
conditions~\cite{cosmology}. Because of the exponential expansion
of an initially small causal patch, the inflationary de Sitter
epoch dominated by the vacuum equation of state $p=-\varepsilon$
suppresses any classical inhomogeneities in the initial conditions
by many orders of magnitude, and leads to a primordial power
spectrum that is both scale invariant and featureless. Most
inflationary models assume that the quantum fluctuations which
lead to this scale invariant spectrum originate in the maximally
$O(4,1)$ symmetric Bunch-Davies state of scalar fields in de
Sitter spacetime~\cite{b-d}, although the possibility that other
states may play a role was considered by some
authors~\cite{VilFoLinSta,Mot,All}. In the last few years there
has been a renewed interest in the possible effects of different
initial states in
inflation~\cite{BraMarEas,DanBer,Dan,Tana,KKLS,disp,nc,Sus,StaTka,BNB,Por,Shnew,Danew,h-h,SriPad},
fueled largely by the speculation that more precise observations
of the Cosmic Microwave Background (CMB) might make such effects
observable, thus opening up the possibility of using CMB
observations to probe novel short distance physics in the very
early universe.

The primary purpose of this paper is to present a consistent low
energy effective field theory (EFT) framework for parameterizing
such short distance and initial state effects in cosmological
spacetimes. Although the elements of quantum field theory in
curved space upon which this EFT framework relies have been known
for some time, a comprehensive treatment of general homogeneous,
isotropic initial states in Robertson-Walker spacetimes has not
been given previously to our knowledge. Such a treatment of
general initial states requires both covariant and canonical
methods, as well as a dictionary to translate between them.
Establishing the EFT framework, the relationship between the
covariant and canonical approaches, and the form of the
state-dependent terms in the covariant stress tensor occupies
Secs. \ref{sec:gen-init}-\ref{sec:tmunu-allowed} of the paper. The
paper is designed so that after becoming acquainted with the
definitions and conventions in Sec. \ref{sec:gen-init}, the reader
may skip the detailed development of Secs.
\ref{sec:density}-\ref{sec:tmunu-allowed} if desired, and go
directly to the applications in later sections, referring back to
the previous sections for the derivation of the formulae as
necessary. Readers interested only in particular modifications of
the initial state of inflation and their effects on the CMB may
wish to skip directly to Sec. \ref{sec:shortdist}.

The secondary purpose of this paper is to apply the EFT methods developed
in Secs. \ref{sec:gen-init}-\ref{sec:tmunu-allowed} to the processes of particle
creation and decoherence in semi-classical cosmology. Certain problems with
the definitions of particle number and the decoherence functional are resolved by the
same adiabatic methods used to define the general class of UV allowed initial
states in the EFT approach.  Readers interested primarily in these applications
may wish to go directly to Secs. \ref{sec:adiabatic} or \ref{sec:decoherence}
respectively, likewise referring to the earlier sections as needed.
The remainder of this Introduction contains a general overview of
the issues addressed in the paper and our approach to them.
As a further guide to the content of the paper, the concluding Sec.
\ref{sec:conclusions} contains a point-by-point summary of our main results.

As we consider initial state modifications of inflation it is perhaps worth
emphasizing from the outset that any sensitivity of present day observations
to initial state effects runs counter to some of the original motivations
for and attractiveness of inflation. A scale invariant spectrum is one of
inflation's most generic predictions, precisely because of the
presumed late time {\it in}sensitivity to perturbations of the
initial state. If there are features in the power spectrum of the
CMB today which are not erased by the exponential redshift of the
inflationary epoch and which bear the imprint of new physics at
short distance scales, then one might ask what prevents short
distance physics from affecting other large scale properties of
the universe, such as its homogeneity, isotropy or flatness. Since
it is not clear which inflationary model (if any) is correct, fine
tuning a specific model to make particular modifications
observable in the CMB power spectrum results in a diminishing of
the overall predictive power of inflation. As long as it is
possible to accommodate any observable features in the power
spectrum by appropriately fine tuning the inflationary model, the
physical origin of these features as true signatures of new high
energy physics must remain uncertain~\cite{SriPad}. Finally, the
remarkable detection of a non-zero cosmological dark energy in the
universe today~\cite{dark}, at a level very different from
estimates based on considerations of ``naturalness" from short
distance physics, should caution that present cosmological models
are as yet far from complete, and the connection between microphysics
and macroscopic structure in the universe is still to be
elucidated.

Despite these fine tuning and naturalness problems, it is nevertheless
true that in almost any given inflationary model there can be
surviving initial state effects in the primordial power spectrum at
{\it some} level, and the advent of more precise CMB data makes
quantifying the sensitivity of inflationary models to such initial
state effects potentially worthwhile.

A quantitative treatment of short distance and initial state effects
in inflation requires a {\it predictive, low energy} framework in
which such effects can be parameterized and studied with a minimum of
assumptions about the unknown physics at ultrashort
distances. Effective field theory (EFT) provides exactly that
framework in other contexts, and we assume in this paper that EFT
methods may be applied in gravity and cosmology as well. The EFT
approach to perturbative gravitational scattering amplitudes was
discussed in Ref.~\cite{Don}. Here we extend EFT methods to the
non-perturbative regime of semi-classical cosmology.

Since a fully predictive quantum theory of gravity is still
lacking, we are virtually compelled to adopt an EFT approach to
cosmology. Since all scales are presumed to be redshifted to lower
energy scales where an EFT description eventually becomes applicable,
the EFT appropriate for cosmology which respects general coordinate
invariance and the Equivalence Principle is the Einstein theory
together with its quantum corrections. In the EFT framework of
semi-classical gravity we can study general ({\it i.e.,} scale
non-invariant) initial states in Robertson-Walker (RW) spacetimes in a
well-defined way, without detailed knowledge of the short distance
physics which may have generated them.  Although more general initial
states and more general matter EFT interactions could be considered,
we focus in this paper on the specific gravitational effects of
quantum matter fields, and restrict ourselves for simplicity to free
scalar fields in spatially homogeneous and isotropic initial states
consistent with the symmetry of the RW geometry.

The red-shifting of short distance scales to larger ones as the
universe expands distinguishes semi-classical gravity from other
effective field theories, which possess a {\it fixed} physical
cutoff. In the case of a fixed cutoff, the shorter distance modes
of the effective theory can be excluded from consideration, and
their effects subsumed into a finite number of parameters of the
low energy description. In practice, absorption of the cutoff
dependence of quantum corrections into a finite set of parameters
of the effective action is no different in an EFT from that in a
renormalizable theory, except for the allowance of higher
dimensional interactions and the corresponding parameters which
are suppressed by the cutoff scale in the EFT description.

Implicit in the EFT framework is the assumption that the effects
of short distance degrees of freedom {\it decouple} from the long
distance ones. However, in an expanding universe decoupling is a
delicate matter.  New short distance modes are continually coming
within the purview of the low energy description, and some
additional information is required to handle these short distance
degrees of freedom as they newly appear. If these ultraviolet
(UV) modes carry energy-momentum, as they should when their
wavelength becomes larger than the short distance cutoff and EFT
methods apply, then simply excluding their contribution to
the energy-momentum tensor at earlier times will lead to
violations of energy conservation, a well-known point that has been
emphasized anew in Ref.~\cite{Sus}.  Energy non-conservation
occurs with a physical momentum cutoff because energy is being
supplied (by an unspecified external mechanism) to the new degrees
of freedom as they redshift below the cutoff, although they
carried no energy-momentum formerly.  This is an essential point:
arbitrary short distance modifications that violate energy
conservation are unacceptable in a low energy EFT respecting
general coordinate invariance, since the resulting energy-momentum
tensor $\langle T_{ab}\rangle$ cannot be a consistent source for
the semi-classical Einstein equations at {\it large} distances.
Upsetting the macroscopic energy conservation law by a short
distance physical cutoff affects the cosmological evolution at all
scales, hence violating decoupling as well.

These considerations show that the necessary existence of a conserved
source for the semi-classical Einstein equations provides an important
constraint on the class of possible short distance modifications of
the initial state of inflation, quite independent of the matter field
content and its EFT. Let us emphasize that there is no problem
modifying the initial state of a quantum field at a {\it fixed} time
$t_0$ for momenta below some physical scale $M$ at that time. However
the quantum state is not completely specified and a conserved
energy-momentum tensor cannot be computed unambiguously, until
information is given also for physical momenta initially much greater
than $M$, which will redshift below $M$ at later times $t >
t_0$. Without some prescription consistent with general coordinate
invariance for dealing with these arbitrarily high energy
``trans-Planckian" modes, which will become physical low energy modes
eventually, the low energy effective theory of semi-classical gravity
is not complete or predictive.

General covariance of the low energy effective theory of gravity
coupled to quantum matter is the key technical assumption which we
make in this paper. General coordinate invariance determines the form
of the effective action, and therefore the counterterms which are
available to absorb the ultraviolet divergences of the energy-momentum
tensor of the quantum matter fluctuations. The renormalization of
$\langle T_{ab}\rangle$ with the standard local covariant counterterms
up to dimension four is possible if and only if the short distance
properties of the vacuum fluctuations are the standard ones, as
expressed for example in the Hadamard conditions on the two-point
function $\langle \Phi (x) \Phi(x')\rangle$ as $x\rightarrow
x'$~\cite{aln,wald}. These UV conditions on the structure of the
vacuum should be viewed as an extension of the Equivalence Principle
to semi-classical gravity, since they amount to assuming that the local
behavior of quantum matter fluctuations are determined in a curved
spacetime by those of flat spacetime and small (calculable) deviations
therefrom. With this physical assumption about the local properties of
the vacuum, the ultrashort distance modes are necessarily adiabatic
vacuum modes and are fully specified as they redshift below the UV
cutoff scale. Then covariant energy conservation is ensured, there are
no state dependent divergences in $\langle T_{ab}\rangle$, and the
effective field theory of semi-classical gravity applied to cosmology
becomes well-defined and predictive within its domain of validity.

Several authors have considered specific short distance modifications,
such as modified dispersion relations for the modes~\cite{disp} or
spacetime non-commutativity~\cite{nc}. We do not consider in this paper
these or any other possible specific short distance modifications that
would take us out of the framework of the covariant low energy EFT of
gravity, without providing a completely consistent quantum
alternative. Once the low energy EFT of semi-classical gravity applies,
any imprint of UV physics can be encoded only in the parameters of the
initial state up to some large but finite energy scale $M$. While many
papers have discussed possible imprints of new short distance physics
on the CMB power spectrum, a few authors have considered also the
constraints that may arise from the energy-momentum tensor of the
fields in a state other than the BD state, making use of various order
of magnitude estimates~\cite{Tana,StaTka,Por,Shnew,Danew,SriPad}. If the
initial state modifications are parameterized by adding irrelevant
higher dimensional operators of the scalar EFT at the boundary, there
is apparent disagreement between several authors about the order of
the corrections these modifications induce in the energy-momentum
tensor~\cite{Por,Shnew}. Quantitative control of the finite state
dependent terms in the energy-momentum tensor is potentially important
for determining whether the short distance modifications can be
observable in the power spectrum without upsetting other features of
inflation. If the scalar field energy density is too large, it could
prevent an inflationary phase from occurring at
all~\cite{Tana,Sus,SriPad}. In this paper we present the framework necessary
for the unambiguous evaluation of initial state effects in both the
power spectrum and the renormalized $\langle T_{ab}\rangle$ for any
homogeneous and isotropic state. We then use the general framework to
find specific constraints on initial states, such as those proposed in
the boundary action formalism of Refs.~\cite{Bound}. We find by
explicit computation that the terms in the matter EFT boundary action
linear in the first higher dimensional operators that one can add do
enter the energy-momentum tensor and may place more restrictive bounds
on the parameters than reported in~\cite{Shnew}.

A systematic study of more general initial states in RW spacetimes
and their stress tensor expectation values was initiated in two
recent papers~\cite{hm-pm,desitter}. Although it had generally
been assumed that initial state inhomogeneities in $\langle
T_{ab}\rangle$, different from the BD expectation value, would
redshift on an expansion time scale $H^{-1}$, the proper treatment
of the initially ultrahigh frequency (trans-Planckian) modes is
critical to proving this result, and also for demonstrating how it
may break down in certain special cases. The conditions on
homogeneous and isotropic initial states of a free scalar field in
de Sitter space, necessary for its two-point function and
energy-momentum tensor to be both finite in the IR and
renormalizable in the UV were defined in Ref.~\cite{desitter}.
When these conditions are satisfied it was shown that $\langle
T_{ab}\rangle$ for a free scalar field of mass $m$ and curvature
coupling $\xi$ does approach the BD value with corrections that
decay as $a_{\rm dS}^{-3 + 2\,{\rm Re}\,\nu}$ for ${\rm Re} \, \nu
< 3/2$, where $a_{\rm dS}$ is the RW scale factor in de Sitter
space, given by Eq.~(\ref{eq:RWds}) and $\nu$ is defined by
Eq.~(\ref{eq:nu}) below. For sufficiently massive fields $\nu$ is
pure imaginary and these fields have energy-momentum tensor
expectation values which decay to the BD value as $a_{\rm
dS}^{-3}$, just as would be expected for classical
non-relativistic dust with negligible pressure. However, the light
or massless cases in which $\nu^2 >0$, show a quite different late
time behavior. The massless conformally coupled scalar field, for
which $\nu = \pm 1/2$, has in addition to the expected subdominant
$a_{\rm dS}^{-4}$ behavior of classical massless radiation with $p
= \varepsilon/3$, a much more slowly falling $a_{\rm dS}^{-2}$
component, with $p = -\varepsilon/3$, arising from quantum
squeezed state effects. In the massless, minimally coupled case,
relevant for the slowly rolling inflaton field, as well as the
graviton itself, $ \nu = 3/2$, and there is an additional $a_{\rm
dS}^0$ constant coherent component in the late time behavior, with
$p=-\varepsilon$, signaling the breakdown of the $O(4,1)$
invariant BD state in the IR, and the change in the stress tensor
from the BD to the Allen-Folacci (AF) value~\cite{a-f}. These
features could not be so readily anticipated by purely classical
considerations, but are quite straightforward to obtain with the
general properly renormalized $\langle T_{ab}\rangle$.

We present here a comprehensive treatment of general initial states in
arbitrary RW spacetimes begun in Refs.~\cite{hm-pm,desitter}, from
both a canonical and covariant viewpoint. We emphasize throughout the
paper that in a general RW spacetime all such states are on an equal
footing {\it a priori}. There is no need to resolve the vacuum
``ambiguity" often said to exist in curved space field
theory. Different physical initial data will simply select different
physical states. Only the local behavior of these states at very short
distances needs to be restricted by the Equivalence Principle. Because
the structure of the vacuum is most explicit in its wave functional
representation, we review the Schr\"odinger description of arbitrary RW
states first.  The Schr\"odinger description in RW spacetimes has previously
been investigated for pure states in~\cite{GLH,EboPiSam} and for mixed
states in~\cite{density-RW}. The general state of a free scalar field
is that of a mixed, squeezed state Gaussian density matrix
$\hat{\rho}$, given by Eq.~(\ref{eq:gaussd}), evolving according to
the quantum Liouville equation~(\ref{eq:Lio}) in the Schr\"odinger
representation. The parameters that specify this general Gaussian
density matrix are in one-to-one correspondence with the amplitudes
that define the two-point Wightman function and power spectrum of the
field. We show that the quantum Liouville equation or the scalar wave
equation in the covariant description imply that these Gaussian
parameters evolve with time according to an effective classical
Hamiltonian~(\ref{eq:effHam}), in which $\hbar$ appears as a
parameter. This demonstrates that the evolution of an arbitrary
initial state is completely unitary and time reversible in any RW
spacetime. Any apparent discrepancy between the Hamiltonian and
covariant approaches is resolved by including the RW scale factor on
an equal footing with the matter field(s) in the Hamiltonian
description. In any case the correct source for Einstein's equations
and backreaction considerations is not the canonical or effective
Hamiltonian of the Schr\"odinger representation but the expectation
value of the covariant energy-momentum tensor, $\langle
T_{ab}\rangle$.

The fourth order adiabaticity condition~\cite{birrell-davies} on
the short distance components of the wave functional defines the
class of UV allowed RW states consistent with the low energy
effective field theory satisfying general covariance.  We exhibit
the finite state-dependent contributions to $\langle
T_{ab}\rangle$ for a general homogeneous, isotropic RW state
in~(\ref{eq:etrw}). The $O(4,1)$ invariant Bunch-Davies (BD) state
is a UV allowed fourth order adiabatic vacuum state in de Sitter
space, but the one complex parameter family of squeezed state
generalizations of the BD state~\cite{Mot,All} (sometimes called
$\alpha$ vacuua) are {\it not} UV allowed RW states~\cite{BerFol}.
All such states except the BD state are therefore excluded as
possible modified initial states in the low energy description,
unless they are cut off at some physical momentum scale
$M$~\cite{Dan,Sus}, and are thence no longer de Sitter invariant.
Various possible modifications of the inflaton initial state up to
some physical scale $M$ at the initial time $t_0$ are considered
in Sec.~\ref{sec:shortdist}, and their power spectrum and
backreaction effects are computed in a consistent way. We compare
our treatment of short distance and initial state effects with
previous work involving $\alpha$ vacuua~\cite{Dan}, adiabatic
states~\cite{DanBer}, and a boundary action approach
~\cite{Shnew,Bound}. The precise connection between the boundary
action approach and the initial state specification is
established. Readers interested in only these initial state
effects in inflation may wish to skip directly to
Section~\ref{sec:shortdist}, where these short distance effects
are considered, and bounds on the short distance modifications are
obtained from backreaction considerations.

In addition to the power spectrum and energy-momentum source for
the gravitational field, the adiabatic method provides a
consistent framework to discuss particle creation and decoherence
in semi-classical cosmology. Although the definition of particle
number in an expanding universe is inherently non-unique, we show
that the adiabatic number basis matched to the second adiabatic
terms in $\langle T_{ab}\rangle$ is the minimum one that allows
for a finite total particle number with conserved energy-momentum.
Matching the particle basis to fourth or higher adiabatic order is
possible but of decreasing physical and practical importance. The
EFT description also suggests that one should limit the particle
number definition to the minimal one that requires the fewest
number of derivatives of the scale factor, {\it i.e., two}, which
are sufficient to eliminate all power law cutoff dependences in
the stress tensor. Hence the second order adiabatic basis is
selected by the short distance covariance properties of the
vacuum, together with the local derivative expansion
characteristic of a low energy EFT description of Einstein's
equations, which are themselves second order in derivatives of the
metric. The second order definition of adiabatic particle number,
matched to the form of the conserved $\langle T_{ab}\rangle$ is
also the minimal one needed to render the total number of created
particles in Eq.~(\ref{eq:finN}) finite. As is well known, in the
case of the massless, conformally coupled scalar field in an {\it
arbitrary} RW spacetime the zeroth order adiabatic vacuum modes
become {\it exact} solutions of the wave equation, and hence no
mixing between positive and negative frequency modes occurs. We
show that no particle creation occurs also for the massless,
minimally coupled scalar field (which sometimes serves as the
inflaton field) in the special case of de Sitter spacetime.

Particle creation may be described as a squeezing of the density
matrix parameters and corresponds to a basis in which the off-diagonal
elements of $\hat{ \rho}$ are rapidly oscillating in phase, and may be
replaced by zero with a high degree of accuracy.  To the extent that
this approximation is valid and the information contained in these
rapidly oscillating phases cannot be recovered, the evolution is {\it
effectively} dissipative at a macroscopic level, despite being
microscopically time reversible. The macroscopic irreversibility is
measured by the von Neumann entropy~(\ref{eq:vN}) of the phase
averaged density matrix in the adiabatic particle basis. In the two
special massless cases of cosmological interest mentioned above,
namely the conformally coupled, scalar field in arbitrary RW spacetime
and the minimally coupled scalar field in de Sitter spacetime, this
phase averaging effect is absent, since no particle creation occurs.

The decoherence functional is defined in the Schr\"odinger
representation as the wave function overlap between two states with
similar initial conditions but different macroscopic RW scale factors,
{\it i.e.,} it measures the quantum (de)coherence between different
semi-classical realizations of the universe~\cite{KieLaf}. Physical
expectations of a very nearly classical universe suggest that this
quantum overlap between different macroscopic states in cosmology
should be finite in principle but extremely small. However, a naive
computation of the decoherence functional in RW cosmology is plagued
by divergences, qualitatively similar to those encountered in $\langle
T_{ab}\rangle$. Moreover, previous authors have found that the exact
form and degree of these divergences depend upon the parameterization
used for the matter field variables~\cite{KieLaf,PazSin,BarKam}. These
divergences and ambiguities have prevented up until now the
straightforward application and physical interpretation of the
decoherence functional in semi-classical cosmology. By analyzing the
general form of the divergences in the decoherence functional and
relating them to the divergences in the effective closed time path
(CTP) action of semi-classical gravity~\cite{CTP-ref}, we show that
these divergences can be regulated and removed by a slightly modified
form of the adiabatic subtraction procedure~\cite{parker} used to
define both the renormalized $\langle T_{ab}\rangle$ and the finite
particle number basis. This gives an unambiguous definition of a
physical, UV {\it finite} decoherence functional for RW spacetimes
which is displayed in Eq.~(\ref{eq:decren}) and which is free of field
parameterization dependence and other ambiguities previously noted in
the literature. The finite decoherence functional does fall rapidly to
zero with time in the general case in which particle creation takes
place, in accordance with physical intuition. In the special massless
cases in which no particle creation occurs, the renormalized
decoherence functional {\it vanishes}, showing that no decoherence of
quantum fluctuations between different semi-classical RW universes
occurs in these cases. This corroborates the close connection between
the particle creation and decoherence effects which has been found in
other contexts~\cite{CKHP}, and shows that the emergence of a
classical universe from initial conditions on a massless field must be
due to other effects, such as interactions, which are neglected in the
free field treatment presented in this paper.

The outline of the paper is as follows. In the next section we
establish notation and define the general class of homogeneous,
isotropic RW states in a RW spacetime. In section~\ref{sec:density},
we review the Hamiltonian description of the evolution of these states
and give the form of the mixed state Gaussian density matrix of the
Schr\"odinger representation, as well as the Wigner function and
effective classical Hamiltonian which describes the evolution. In
section~\ref{sec:tmunu-allowed} we evaluate the expectation value of
the energy-momentum tensor, and the low energy effective action for
gravity of which it is part. We specify the conditions on the short
distance components of a general RW state in order for $\langle
T_{ab}\rangle$ to be UV renormalizable with geometric counterterms of
the same form as the effective action, and obtain an expression for
the finite contributions of arbitrary UV allowed RW states. In
section~\ref{sec:shortdist} we consider three types of modified
initial states in inflation, evaluating the power spectrum and
energy-momentum tensor for each in turn. In section~\ref{sec:adiabatic}
we define the adiabatic particle number basis and show how particle creation
leads to an effective dissipation in the density matrix description.
In section~\ref{sec:decoherence} we define a finite renormalized decoherence
functional for semi-classical cosmology, and corroborate the non-decoherence
of massless inflaton fluctuations in de Sitter space. We conclude with a
detailed summary and discussion of our results. Technical details of the
Gaussian parameterization of the density matrix and its properties, the
evaluation of integrals needed in Sec.~\ref{sec:shortdist}
and the comparison of adiabatic bases used in squeezing and decoherence
calculations by previous authors are relegated to Appendices A, B
and C respectively. Throughout we set $c=1$ and use the metric and curvature
conventions of MTW~\cite{mtw}.

\section{General RW initial states}
\label{sec:gen-init}

Homogeneous and isotropic RW spacetimes can be described by the line
element,
\begin{equation}
\d s^2 =  -\d t^2 + a^2(t)\, \d\Sigma^2
= - \d t^2 + a^2(t)\, \gamma_{ij} \;  \d x^i \d x^j\,,
\label{eq:FRWs}
\end{equation}
with $t$ the comoving (or cosmic) time, and $\gamma_{ij}$ the metric
of the three-dimensional spacelike sections $\Sigma$ of constant
spatial curvature, which may be open, flat, or closed.  It is also
useful to introduce the conformal time coordinate,
\begin{equation}
\eta = \int^t \frac{\d t} {a(t)}\,,
\end{equation}
so that the line element~(\ref{eq:FRWs}) may be expressed in the
alternative form,
\begin{equation}
\d s^2 =  a^2 (\eta)(- \d \eta^2 + \d\Sigma^2),
\label{eq:FRWc}
\end{equation}
where $a$ is now viewed as a function of conformal time $\eta$.  We
take $a$ to have dimensions of length with $\eta$ dimensionless.
The scalar curvature is
\begin{equation}
R= 6\,\left( \dot H + 2H^2 + {\frac{\epsilon}{a^2}}\right)
\,,\qquad H \equiv \frac{\dot a}{a}\,.
\label{eq:Riccis}
\end{equation}
The overdot denotes differentiation with respect to $t$ and
$\epsilon=-1,0,+1$ depending on whether the spatial sections are
open, flat, or closed, respectively.

A free scalar field with mass $m$ obeys the scalar wave equation,
\begin{equation}
\left(- \sq + m^2 + \xi R\right)\, \Phi =0\,,
\label{eq:waveq}
\end{equation}
where $\sq = g^{ab} \nabla_a \nabla_b$ and $\xi$ is the arbitrary
dimensionless coupling to the scalar curvature. Since the RW
three-geometry $\Sigma$ is spatially homogeneous and isotropic, the
wave equation~(\ref{eq:waveq}) may be solved by decomposing $\Phi(t,
{\bf x})$ into Fourier modes in the general form,
\begin{eqnarray}
\Phi (t, {\bf x}) &=& \int [\d{\bf k}]
\left(a_{\rule{0mm}{2.3mm}\bf k} \phi_{\rule{0mm}{2.3mm}k}(t)
   Y_{\rule{0mm}{2.3mm}\bf k} ({\bf x})
+ a_{\bf k}^\dag \phi_{\rule{0mm}{2mm}k}^*(t)
Y_{\rule{0mm}{2.3mm}\bf k}^* ({\bf x})\right) \; .
\label{eq:phiquant}
\end{eqnarray}
The $Y_{\bf k}$ are the eigenfunctions of the three-dimensional
Laplace-Beltrami operator $\Delta_3$ on $\Sigma$, satisfying
\begin{equation}
- \Delta_3\,Y_{\bf k}({\bf x}) \equiv
   -\frac{1}{\sqrt \gamma}\frac{\partial}{\partial x^i}\,
\gamma^{ij} \,\sqrt \gamma\, \frac{\partial}{\partial x^j}\,Y_{\bf
k}({\bf x}) = (k^2 - \epsilon) Y_{\bf k}({\bf x}) \; ,
\label{eq:lapbel}
\end{equation}
and the $\phi_k (t)$ are functions only of time and the magnitude of
the wave vector, $k\equiv \vert {\bf k}\vert$.

For flat spatial sections, $\epsilon = 0$, $\gamma_{ij} = \delta_{ij}$,
$\gamma \equiv {\rm det}\, \gamma_{ij} = 1$, and the $Y_{\bf k}({\bf x})$
are simply plane waves $e^{i {\bf k \cdot x}}$. The integration measure
in Eq.~(\ref{eq:phiquant}) for this case is
$\int [\d {\bf k}] = \int \d^3{\bf k}/(2\pi)^3$.

In the case of compact spatial sections, $\epsilon =+1$, the wave
number $k$ takes on discrete values which we label by the positive
integers $k\ge 1$, and the harmonic functions in Eq.~(\ref{eq:lapbel})
are the spherical harmonics of the sphere $S^3$. These $S^3$ harmonics
denoted by $Y_{klm}$ depend on three integers ${\bf k} \leftrightarrow
(k,\ell, m)$, the first of which may be identified with $\vert {\bf
k}\vert$, while $(\ell, m)$ refer to the usual spherical harmonics on
$S^2$ with $\ell \le k-1$. Since $\sum_{\ell=0}^{k-1} (2\ell + 1) =
k^2$, a given eigenvalue of the Laplacian~(\ref{eq:lapbel}) labelled
by $k$ is $k^2$-fold degenerate. The scalar spherical harmonics
$Y_{klm}$ may be chosen to satisfy $Y_{\bf k}^*({\bf x}) = Y^{\
}_{-\bf k}({\bf x}) \equiv Y_{k\ell\,-m}({\bf x})$, and normalized on
$S^3$ so that~\cite{bunch}
\begin{equation}
\int_{S^3} \d^3\Sigma \ Y^*_{k'\ell'm'}({\bf x})\, Y^{\ }_{k\ell
m}({\bf x}) = \delta_{\rule{0mm}{2mm}{\bf k} {\bf k'}} =
\delta_{kk'} \delta_{\ell\ell'}\delta_{mm'}\,,
\end{equation}
and
\begin{equation}
\sum_{\ell = 0}^{k-1}\sum_{m=-\ell}^{\ell} \vert
Y_{\rule{0mm}{2.3mm}k\ell m}({\bf x})\vert^2 =
\frac{k^2}{2\pi^2}\,,
\label{eq:lmsum}
\end{equation}
which is independent of ${\bf x}$.

In the open case $\epsilon = -1$, the sums over $(\ell, m)$ remain,
but $k$ becomes a continuous variable with range $[0, \infty)$.  After
integration over the direction of ${\bf k}$ in the $\epsilon =0$ case,
one is also left with the integration over the magnitude $k$ with the
scalar measure $\int \d k\, k^2/(2\pi^2)$. Because of
Eq.~(\ref{eq:lmsum}), the compact $\epsilon = +1 $ case is simply
related to the non-compact cases of $\epsilon = 0, -1$ by the
replacement of the integral $\int \d k\, k^2/(2\pi^2)$ by the discrete
sum, $\sum_{k=1} k^2/(2\pi^2)$.  Beginning the sum from $k=1$ (so that
the spatially homogeneous mode on $S^3$ has eigenvalue $k=1$ instead
of $k=0$) makes this correspondence between the discrete and
continuous cases most immediate. We define the scalar measure,
\begin{equation}
\int [\d k] \equiv  \left\{ \begin{array}{ll}
\int_0^\infty \d k \quad & {\rm if}\quad \epsilon =
0, -1  \\
\sum_{k=1}^\infty  \quad & {\rm if}\quad \epsilon = 1
\end{array}\right.
\label{eq:mutilde}
\end{equation}
in order to cover all three cases with a single notation.

The time dependent mode functions $\phi_k (t)$ satisfy the ordinary
differential equation,
\begin{equation}
\frac {\d^2\phi_k}{\d t^2} + 3H\, \frac{\d\phi_k}{\d t}
+ \frac{(k^2 - \epsilon)}{a^2}
\,\phi_k + (m^2 + \xi R)\phi_k = 0\,.
\label{eq:phikeq}
\end{equation}
If one defines $f_k (t) \equiv a^{\frac{3}{2}} \phi_k$, then this
equation is equivalent to
\begin{equation}
\ddot f_k + \left[\omega_k^2 + \left(\xi - \frac{1}{6}\right) R - \frac{1}{2}
\left( \dot H + \frac{H^2}{2}  \right) \right]f_k=0\,,
\label{eq:modefns}
\end{equation}
where
\begin{equation}
\omega_k^2(t) \equiv \frac{k^2}{a^2} + m^2 \,.
\label{eq:Omega}
\end{equation}
Eq.~(\ref{eq:modefns}) is the equation for a harmonic oscillator with
time dependent frequency. Note that the comoving momentum index $k$ of
the mode is constant, while the physical momentum $p = k/a$ redshifts
as the universe expands.

An analogous time dependent harmonic oscillator equation may be
derived also in conformal time under the substitution,
$\chi_{\rule{0mm}{2.1mm}k}(\eta) \equiv a \phi_{\rule{0mm}{2.1mm}k}$,
{\it viz.,}
\begin{equation}
\chi_k'' + \left[ k^2 + m^2 a^2 + (6\xi -1) \left( \frac{a''}{a\,}
+ \epsilon\right)
\right] \chi_{\rule{0mm}{2.1mm}k} = 0\,,
\label{eq:chiosc}
\end{equation}
where the primes denote differentiation with respect to conformal time
$\eta$.

In view of the completeness and orthonormality of the spatial harmonic
functions $Y_{\bf k} ({\bf x})$, it is easily verified that the equal
time commutation relation
\begin{equation}
\left[\Phi (t,{\bf x}), \frac{\partial \Phi}{\partial t}
(t,{\bf x'})\right] =
\frac{i\hbar} {a^3}\,\delta_{\Sigma}({\bf x}, {\bf x'}) \equiv
\frac{i\hbar} {a^3}\,\sqrt{\gamma}\ \delta^3({\bf x}-{\bf x'})\,,
\label{eq:etc}
\end{equation}
is satisfied, provided the creation and annihilation operators
obey
\begin{eqnarray}
[a_{\raisebox{-0.6mm}{\scriptsize\bf k}}, a^{\dagger}_{\bf k'}] =
\delta_{\rule{0mm}{-2.3mm}{\bf k} {\bf k'}}\,,
\end{eqnarray}
in the discrete notation, and the complex mode functions satisfy
the Wronskian condition,
\begin{eqnarray}
a^3(\dot \phi_{\raisebox{-0.4mm}{\scriptsize \it k}} \phi_k^* -
\phi_{\raisebox{-0.4mm}{\scriptsize \it k}}\dot \phi_k^*)
= \dot f_{\raisebox{-0.4mm}{\scriptsize \it k}} f_k^* -
f_{\raisebox{-0.4mm}{\scriptsize \it k}} \dot f_k^*
= \chi^{\prime}_k\chi_{\rule{0mm}{-2.3mm}k}^* -
\chi_{\raisebox{-0.4mm}{\scriptsize \it k}} \chi_k^{*\prime} = -i\hbar\,.
\label{eq:wron}
\end{eqnarray}
{From} the equation of motion~(\ref{eq:phikeq}),~(\ref{eq:modefns})
or~(\ref{eq:chiosc}) this Wronskian condition is preserved under
time evolution.  Hence any initial condition for the second order
equation of motion satisfying~(\ref{eq:wron}) is {\it a priori}
allowed by the commutation relations. Given any two solutions
of~(\ref{eq:phikeq}), we define their Klein-Gordon inner product
as
\begin{equation}
(\psi_k, \phi_k) \equiv  \frac{ia^3}{\hbar}
(\psi_k^* \dot \phi_{\raisebox{-0.4mm}{\scriptsize \it k}}
   - \dot\psi_k^* \phi_{\raisebox{-0.4mm}{\scriptsize \it k}}) \, ,
\label{eq:KG}
\end{equation}
which is independent of time.

Let $v_{\rule{0mm}{2.1mm}k}(t)$ be some particular set of time
dependent mode functions satisfying Eq.~(\ref{eq:phikeq}) and the
Wronskian condition~(\ref{eq:wron}). These can be used to define
a vacuum state. Any other set of solutions $\phi_{\rule{0mm}{2.1mm}k}$
satisfying the same Wronskian condition can be expressed
as a linear superposition of $v_k$ and its complex conjugate,
\begin{equation}
\phi_{\raisebox{-0.4mm}{\scriptsize \it k}}=
A_{\raisebox{-0.4mm}{\scriptsize \it k}}\,v_{\raisebox{-0.4mm}{\scriptsize \it k}} +
B_{\raisebox{-0.4mm}{\scriptsize \it k}} v_k^*\,,
\label{eq:bog}
\end{equation}
which is the form of a Bogoliubov transformation. Because of Eq.~(\ref{eq:wron})
the time independent complex Bogoliubov coefficients must satisfy
\begin{equation}
\vert A_k\vert^2 - \vert B_k\vert^2 = 1\,,
\label{eq:bognorm}
\end{equation}
for each $k$. This is one real condition on the two complex numbers
$A_k$ and $B_k$. Since multiplication of both $A_k$ and $B_k$
by an overall constant phase has no physical consequences, there
are only two real parameters needed to specify the mode function
for each $k$.

The inner product~(\ref{eq:KG}) is preserved under the Bogoliubov
transformation~(\ref{eq:bog}), {\it i.e.,}
\begin{subequations}
\begin{eqnarray}
(\phi_k, \phi_k) = (v_k, v_k)  &=& 1\,,\\
(\phi_k^*, \phi_{\rule{0mm}{2.1mm}k}) = (v_k^*, v_{\rule{0mm}{2.1mm}k})&=& 0\,,
\end{eqnarray}
\end{subequations}
Thus, we can invert~(\ref{eq:bog}) and solve for the Bogoliubov
coefficients at an arbitrary initial time, $t=t_0$ or $\eta =
\eta_0$, with the result,
\begin{subequations}
\begin{eqnarray}
A_k &=& (v_k, \phi_k) = \frac{ia_0^3}{\hbar}
(v_k^*\,\dot \phi_{\raisebox{-0.4mm}{\scriptsize \it k}} -
\dot v_k^* \,\phi_{\raisebox{-0.4mm}{\scriptsize \it k}})_0\, ,\\
B_k &=& (v_k^*, \phi_{\rule{0mm}{2.1mm}k}) = -\frac{ia_0^3}{\hbar}
(v_k\,\dot \phi_k - \dot v_k \,\phi_k)_0\,.
\end{eqnarray}
\label{eq:abk}
\end{subequations}
\hspace{-0.25cm} Interactions may be incorporated in this treatment as
well~\cite{coop-mott,CKHM}, within the semi-classical large $N$
approximation, but in order to keep the discussion as simple as
possible we shall not consider scalar self-interactions in this paper.

We restrict our attention to initial states of a free scalar field,
which like the RW geometry~(\ref{eq:FRWs}) itself, are also spatially
homogeneous and isotropic, and call such states {\it RW
states}. Spatial homogeneity of the RW states, {\it i.e.,} invariance
under spatial translations in $\Sigma$ implies that the bilinear
expectation values, $\langle a^{\dagger}_{\bf
k}a_{\rule{0mm}{2.3mm}\bf k'}\rangle$ and $\langle
a_{\raisebox{-0.4mm}{\scriptsize \bf k}} a^{\dagger}_{\bf k'}\rangle$,
can be non-vanishing if and only if ${\bf k} = {\bf k'}$, while the
expectation values, $\langle a^{\dagger}_{\bf k}a^{\dagger}_{\bf
k'}\rangle$ and $\langle a_{\rule{0mm}{2.3mm}\bf
k}a_{\rule{0mm}{2.3mm}\bf k'}\rangle$, can be non-vanishing if and
only if ${\bf k} = - {\bf k'}$. In addition, isotropy of the RW states
under spatial rotations implies that the expectation value of the
number operator for ${\bf k} = {\bf k'}$,
\begin{eqnarray}
\langle a^{\dagger}_{\bf k}a_{\raisebox{-0.4mm}{\scriptsize \bf k}}\rangle =
n_k \equiv \langle a_{\raisebox{-0.4mm}{\scriptsize \bf k}}a^{\dagger}_{\bf k}\rangle - 1\,,
\label{eq:part}
\end{eqnarray}
can be a function only of the magnitude $k$. This constant number in
each Fourier mode is the consequence of the unmeasurable $U(1)$ phase
of the mode functions, and becomes the third real parameter needed for
each $k$ to specify the initial quantum state of the scalar field. We
show in the next section that if the free field density matrix for a
RW state is described by the general Gaussian ansatz in the
Hamiltonian description, then the state is necessarily a {\it mixed}
state if $n_k \neq 0$.

Because of the two parameter freedom to redefine $\phi_k$ according to
the Bogoliubov transformation,~(\ref{eq:bog}) and~(\ref{eq:bognorm}),
it is always possible to fix the parameters so that the remaining
bilinears are equal to zero~\cite{CKHM}, {\it i.e.,}
\begin{eqnarray}
\langle a_{\rule{0mm}{2.3mm}\bf k}a_{\rule{0mm}{2.3mm}-\bf
k}\rangle = \langle a^{\dagger}_{\bf k}a^{\dagger}_{-\bf k}\rangle
= 0\,.
\end{eqnarray}
If $\Phi$ is expanded in terms of the vacuum modes $v_k$ instead of
$\phi_k$, then the corresponding annihilation and creation operators
are
\begin{subequations}
\begin{eqnarray}
c_{\rule{0mm}{2.3mm}\bf k} &=& A_{\rule{0mm}{2.3mm}k} a_{\rule{0mm}{2.3mm}\bf k}
+ B_{\rule{0mm}{2mm}k}^* a_{-\bf k}^\dag \,,\\
c_{\bf k}^\dag &=& A_{\rule{0mm}{2.3mm}k}^* a_{\bf k}^\dag
+ B_{\rule{0mm}{2.3mm}k} a_{-\rule{0mm}{2.3mm}\bf k} \,.
\end{eqnarray}
\end{subequations}
\hspace{-0.2cm}
This is the characteristic form of a Bogoliubov transformation to a
squeezed state. If the arbitrary overall phase is fixed by requiring $A_k$
to be real, then the general squeezed state parameters $r_k$ and $\theta_k$
are defined by
\begin{subequations}
\begin{eqnarray}
A_k &=& \cosh r_k\,,\\
B_k &=& e^{i\theta_k}\,\sinh r_k\,.
\end{eqnarray}
\label{eq:squeeze}
\end{subequations}
\hspace{-0.25cm}
The bilinear expectation values,
\begin{subequations}
\begin{eqnarray}
&& \langle c_{\rule{0mm}{2.3mm}\bf k}c_{\rule{0mm}{2.3mm}-\bf k}\rangle
= (2n_k +1)\, A_{\rule{0mm}{2.3mm}k} B_{\rule{0mm}{2mm}k}^* =
\sigma_k A_{\rule{0mm}{2.3mm}k} B_{\rule{0mm}{2mm}k}^*\,,
\\
&& \langle c^{\dagger}_{\bf k}c^{\dagger}_{-\bf k}\rangle = (2n_k + 1)\,
A_{\rule{0mm}{2mm}k}^* B_{\rule{0mm}{2.3mm}k}
= \sigma_k A_{\rule{0mm}{2mm}k}^* B_{\rule{0mm}{2.3mm}k}\,,
\end{eqnarray}
\label{eq:ccbi}
\end{subequations}
\hspace{-0.25cm} are non-zero, and
\begin{equation}
N_k \equiv \langle c^{\dagger}_{\bf k} c_{\raisebox{-0.4mm}{\scriptsize \bf k}}\rangle
= \vert B_k\vert^2 + (2 \vert B_k\vert^2 + 1)\,n_k =
n_k + \sigma_k\, \vert B_k\vert^2\,,
\label{eq:numvac}
\end{equation}
is the average occupation number of the general mixed, squeezed state
with respect to the vacuum modes $v_k$. We have introduced the
shorthand notation $\sigma_k \equiv 2n_k +1$ for the Bose-Einstein
factor in Eqs.~(\ref{eq:ccbi}) and~(\ref{eq:numvac}).

The two-point (Wightman) function of the scalar field may be expressed
in terms of the mode functions $\phi_k$ and $n_k$ in the form,
\begin{equation}
\langle \Phi(t,{\bf x}) \Phi(t', {\bf x'})\rangle = \int [\d {\bf k}] \,
\biggl(n_{\rule{0mm}{2.1mm}k}\,\phi_{\rule{0mm}{-2.3mm}k}^*(t)
\phi_{\raisebox{-0.4mm}{\scriptsize \it k}}(t') +
(n_{\rule{0mm}{2.1mm}k} + 1)\, \phi_{\raisebox{-0.4mm}{\scriptsize
\it k}}(t) \phi_{\rule{0mm}{-2.3mm}k}^*(t') \biggr)
   Y_{\raisebox{-0.4mm}{\scriptsize \bf k}}({\bf x}) Y^*_{\bf k}({\bf x'}) \,,
\label{eq:wight}
\end{equation}
in the case that the expectation value $\langle \Phi(t, {\bf x})
\rangle =0$. When ${\bf x} = {\bf x'}$, the sums or integrals over
the angular part of {\bf k} can be evaluated with the result that
\begin{equation}
\langle \Phi^2 (t, {\bf x}) \rangle = \int\, [\d k]\,
\frac{P_{\phi}(k;t)}{k}\,,
\end{equation}
is independent of $\bf x$.  Here
   \begin{equation} P_{\phi}(k;t) \equiv \frac{k^3}{2 \pi^2} \,
\sigma_k \, \vert \phi_k (t) \vert^2
\label{eq:powerphi}
\end{equation}
is the power spectrum of scalar fluctuations in the general RW
mixed, squeezed initial state. It may also be
expressed in terms of $v_k$ in the form,
\begin{subequations}
\begin{eqnarray}
&&P_{\phi}(k;t) = P_v (k;t) + \frac{k^3}{\pi^2}\Bigl( N_k\,\vert
v_k(t) \vert^2\, + \sigma_k\,{\rm Re}\, [A_{\rule{0mm}{2.1mm}k}
B_k^* v_k^2(t)]\Bigr)
\end{eqnarray}
where \begin{eqnarray}
P_v (k;t) \equiv
\frac{k^3}{2\pi^2}\, \vert v_k (t)\vert^2\,,
\end{eqnarray}
\label{eq:powerv}
\end{subequations}
\hspace{-0.25cm}
is the fluctuation power spectrum in the selected vacuum state,
and Eqs.~(\ref{eq:bog}) and~(\ref{eq:numvac}) have been used.

In inflationary models this scalar power spectrum becomes
the source for scalar metric fluctuations, and it is the power
spectrum of these metric fluctuations that is actually measured
in the CMB~\cite{padmanabhan}. Due to the linearity of the
metric fluctuations and the assumed spatial homogeneity of
the classical inflaton field, the resulting power spectrum
in the CMB is the same (up to an overall normalization)
as the quantum scalar field spectrum (\ref{eq:powerv}) that
generates it. For example, in slow
roll inflationary models, the
relation between the linearized
curvature perturbation ${\cal R}_{\bf k}$
and the quantum scalar
  field perturbation $\delta\phi_{\bf k}$ is~\cite{cosmology}
\begin{equation}
{\cal R}_{\bf k} = -\left[\frac{H}{\dot\phi}\,\delta\phi_{\bf k}\right]_{t=t_*}\,,
\label{eq:curvscal}
\end{equation}
where $\phi(t)$ is the classical inflaton field (assumed independent of
position) and $t_*$ is a time a few e-folds after the perturbation
has exited the horizon. From (\ref{eq:curvscal}) we find that the
power spectrum of Gaussian curvature fluctuations $P_{\cal R} (k;t_*)$
that is actually observed in the temperature fluctuations of
the CMB is related to the power spectrum of scalar field fluctuations
$P_{\phi}(k;t_*)$ by
\begin{eqnarray}
P_{\cal R} (k;t_*) &=& \left(\frac{H}{\dot\phi}\right)^2\,P_{\phi}(k;t_*)\nonumber\\
&=& \frac{1}{8\pi^2\epsilon}\ \left(\frac{H}{M_{Pl}}\right)^2\,
\frac{P_{\phi}(k;t_*)}{P_{\phi}^{\rm BD}(k;t_*)}\,.
\label{eq:slowroll}
\end{eqnarray}
In the latter relation we have introduced the standard definition of
the slow roll parameter $\epsilon$~\cite{cosmology}, not to be
confused with the $\epsilon = 0,\pm 1$ defined in (\ref{eq:Riccis})
denoting flat, closed or open spatial sections. We have also normalized the
spectrum to the scale invariant Bunch-Davies vacuum power spectrum $P_v(k;t_*)=
P_{\phi}^{\rm BD}(k;t_*)$, given explicitly by Eq.~(\ref{eq:PBD}) of
Sec.~\ref{sec:shortdist} below. Since our intention in this paper is
to address the short distance and initial state effects of the scalar
field in a general, model independent way, we focus on the scalar
field power spectrum $P_{\phi}(k;t)$ of Eqs.~(\ref{eq:powerv})
exclusively in the succeeding sections, leaving the model dependent
connection to the scalar metric power spectrum $P_{\cal R} (k;t)$
unspecified. We do not discuss tensor perturbations of the metric at
all in this paper.

\section{Density Matrix and Hamiltonian}
\label{sec:density}

The development of the previous section in terms of expectation values
of the Heisenberg field operator $\Phi(t,{\bf x})$, with initial data
specified by the time dependent complex mode functions $\phi_k$ is
well suited to a treatment of the covariant energy-momentum tensor and
its renormalization.  We consider this in detail for general RW states
in the next section. However, the evolution of a quantum system from
given initial conditions may be expressed just as well in the
Schr\"odinger representation, and it is the latter approach which
makes most explicit the specification of the initial state and its
unitary evolution in configuration space. The Schr\"odinger
representation is also the one best suited to discussions of
decoherence of cosmological perturbations and the quantum to classical
transition in inflationary models, which we discuss in
Sec.~\ref{sec:decoherence}. We present the Hamiltonian form of the
evolution and corresponding density matrix for a general RW initial
state in this section, demonstrating its full equivalence with the
covariant formulation.

For a free field theory (or an interacting one treated in the
semi-classical large $N$ or Hartree approximations) it is clear that
the two-point function~(\ref{eq:wight}) contains all the non-trivial
information about the dynamical evolution of the general RW initial
state. Even if non-zero, higher order connected correlators do not
evolve in time either in a free field theory or in the leading order
large $N$ approximation to an interacting theory. Hence at least as
far as the time evolution is concerned, a time dependent Gaussian
ansatz for the Schr\"odinger wave functional (or density matrix) of
the scalar field about its mean value is sufficient in both cases. A
proof of the equivalence between the large $N$ semi-classical equations
and the evolution of a Gaussian density matrix has been given for flat
Minkowski spacetime in Refs.~\cite{init,CKHM}. The general RW
case differs from the flat spacetime case mainly by the appearance of
the time dependent scale factor, so that with the appropriate
modifications a mixed state Gaussian density matrix also exists for
the scalar field evolution in cosmology.

In order to obtain this Gaussian density matrix, let us first derive
the Hamiltonian form of the evolution Eqs.~(\ref{eq:waveq})
and~(\ref{eq:phikeq}). In its Hamiltonian form, the variational
principle both for the matter and metric degrees of freedom, should
begin with a classical action that contains only {\it first} time
derivatives of both $\Phi$ and $a$. However, the standard classical
action for the scalar field plus gravity system, {\it viz.,}
\begin{equation}
S_{\Phi} + S_{EH} = -\frac{1}{2}\int \d^4 x \;  \sqrt{-g}\,
[ (\nabla_a \Phi) g^{ab} (\nabla_b \Phi) +m^2 \Phi^2+ \xi R\Phi^2]
+ \frac{1}{16\pi G_{N}}\int \, \d^4x\, \sqrt{-g}\, (R-2\Lambda)
\label{eq:Scl}
\end{equation}
contains the Ricci scalar $R$ in both the matter action $S_{\Phi}$
(when $\xi \neq 0$) and the Einstein-Hilbert action $S_{EH}$. For RW
spacetimes $R$ contains second order time derivatives of the scale
factor in the $\dot H$ term of Eq.~(\ref{eq:Riccis}). In order to
remove these second order time derivatives from the action one should
add a surface term to the action functional above, thereby replacing
$S_{\Phi} + S_{EH}$ in Eq.~(\ref{eq:Scl}) by
\begin{equation}
S_{\rm cl}[\Phi,\dot\Phi; a, \dot a] = S_{\Phi} + S_{EH} +
3 \int \,\d t\, \d^3\Sigma\ \frac{\d}{\d t}
\left[a^3 H \left(\xi\Phi^2 - \frac{1}{8\pi G_N}\right)\right] \,,
\label{eq:surf}
\end{equation}
which modifies both the matter and gravitational parts of the
classical action at the endpoints of the time integration, but
otherwise leaves the Lagrangian evolution equations away from the
endpoints unchanged. In fact, it is this classical action $S_{\rm cl}$
modified by the surface term in Eq.~(\ref{eq:surf}), and {\it not}
$S_{\Phi} + S_{EH}$ whose Euler-Lagrange variation (which by
definition has vanishing $\delta \Phi$ and $\delta a$ at the
endpoints) leads to the scalar field equation of
motion~(\ref{eq:waveq}), as well as the Friedman equation for the
scale factor. The surface term addition to the gravitational action
for a general spacetime has been given in Ref.~\cite{GibHaw}.

With this corrected classical action $S_{\rm cl}$, the conjugate
momentum for the scalar field is
\begin{eqnarray}
\Pi_\Phi &\equiv& \frac{\delta S_{\rm cl}} {\delta \dot
\Phi} = a^3 (\dot \Phi + 6 \xi H \Phi )\,. \label{eq:Piphi}
\end{eqnarray}
If we ignore the Friedman equation for the scale factor for the
moment, treating $a(t)$ as an externally specified function of time,
then the classical Hamiltonian density of the scalar field alone is
\begin{eqnarray}
&&{\cal H}_{\Phi} \equiv \dot \Phi \Pi_{\Phi} - {\cal L}_{\rm cl} =
\frac{\Pi_\Phi^2}{2a^3} - 3 \xi H (\Pi_\Phi \Phi + \Phi \Pi_\Phi)
\nonumber\\
&& \qquad + \frac{a}{2} \gamma^{ij} (\partial_i \Phi) (\partial_j  \Phi)
+  \frac{a^3}{2} m^2 \Phi^2
+ 3 \xi a^3 \left[(6 \xi -1) H^2 + \frac{\epsilon}{a^2} \right]\Phi^2 \,,
\label{eq:Hphi}
\end{eqnarray}
where we have symmetrized the second term in this expression involving
$\Pi_\Phi \Phi$, in anticipation of the replacement of $\Phi$ and
$\Pi_\Phi$ by non-commuting quantum operators.

In the Hamiltonian framework the three independent symmetric quadratic
variances, $\langle \Phi^2 \rangle$, $\langle \Phi \Pi_\Phi + \Pi_\Phi
\Phi\rangle$ and $\langle \Pi_\Phi^2 \rangle$ at coincident times
determine the Gaussian density matrix $\hat\rho$. The one
antisymmetric variance, $\langle \Phi \Pi_\Phi - \Pi_\Phi \Phi\rangle$
is fixed by the canonical commutation relation,
\begin{equation}
[\Phi(t, {\bf x}), \Pi_\Phi (t, {\bf x'})]
= i \, \hbar \, \delta_{\Sigma}({\bf x}, {\bf x'})\,,
\end{equation}
which using Eq.~(\ref{eq:Piphi}) and $[\Phi,\Phi] =
[\Pi_\Phi,\Pi_\Phi] =0$ is equivalent to Eq.~(\ref{eq:etc}). Let us
introduce the definitions,
\begin{subequations}
\begin{eqnarray}
\sigma_k &\equiv& 2n_k + 1 \, ,
\label{eq:sigdef}\\
\zeta_k(t) &\equiv &\sqrt{\sigma_k}\ \vert \phi_k\vert \, ,
\label{eq:zkdefonly}\\
\pi_k (t) &\equiv& a^3 (\dot \zeta_k + 6 \xi H \zeta_k)\; ,
\end{eqnarray}
\label{eq:zetakdef}
\end{subequations}
\hspace{-0.25cm}
for the time independent Bose-Einstein factor, $\sigma_k$, and the two
real functions of time, $\zeta_k(t)$ and $\pi_k(t)$. We show in
Appendix A that these definitions allow us to express the three
bilinear Fourier field mode amplitudes in the form,
\begin{subequations}
\begin{eqnarray}
&& \sigma_k |\phi_k|^2 = \zeta_k^2 \; , \\ &&\sigma_k \,{\rm Re}\,
(\phi_k^* \dot \phi_{\raisebox{-0.3mm}{\scriptsize \it k}})=
\zeta_k \,\dot\zeta_k \; ,\\
&& \sigma_k |\dot \phi_k|^2 = \dot\zeta_k^2 + \frac{\hbar^2\sigma_k^2}
{4a^6 \zeta_k^2} \, ,
\end{eqnarray}
\label{eq:phibil}
\end{subequations}
\hspace{-0.25cm}
and this allows in turn for the three independent Gaussian variances
at coincident spacetime points to be written as
\begin{subequations}
\begin{eqnarray}
&& \langle \Phi^2\rangle = \frac{1}{2 \pi^2} \int\, [\d k]\, k^2\,
\zeta_k^2\,,
\label{eq:Phisq}\\
&& \langle \Phi \Pi_\Phi + \Pi_\Phi \Phi \rangle =
\frac{1}{\pi^2} \int\,[\d k]\,  k^2\, \zeta_k \pi_k \, ,\\
&& \langle  \Pi_\Phi^2\rangle = \frac{1}{2\pi^2} \int\, [\d k]\,
k^2\, \left(\pi^2_k + \frac{\hbar^2\sigma^2_k}{4 \zeta^2_k}\right)\, .
\end{eqnarray}
\label{eq:piphibi}
\end{subequations}
\hspace{-0.2cm}
Thus, the three independent bilinears depend on a set of three real
functions of $k$, $(\zeta_k, \pi_k; \sigma_k)$, as expected from our
discussion of the initial data in the Heisenberg representation of
the previous section. The usefulness of this particular set is that
$(\zeta_k, \pi_k)$ will turn out to be canonically conjugate variables
of the effective Hamiltonian that describes the semi-classical time
evolution of the general Gaussian density matrix, while $\sigma_k$ is
strictly a constant of the motion. Notice also that the power spectrum
defined in~(\ref{eq:powerphi}) can be written in terms of the Gaussian
width parameter $\zeta_k (t)$ directly as
\begin{equation}
P_{\phi}(k;t) = \frac{k^3}{2\pi^2}\,\zeta_k^2(t)\,,
\end{equation}
for both the pure and more general mixed state cases. It is
independent of ${\bf x}$ and the direction of ${\bf k}$ by the spatial
homogeneity and isotropy of the RW state.

The Hamiltonian and corresponding pure state Schr\"odinger wave
functional for scalar field evolution in cosmology has
been previously given in Ref.~\cite{GLH,EboPiSam}. However, a pure
state Gaussian ansatz for the wave functional imposes a constraint
on the three parameters $(\zeta_k, \pi_k; \sigma_k)$, in fact
implying $\sigma_k = 1$ for all $k$~\cite{init,CKHM}. To remove
this restriction one must allow for the Gaussian ansatz also to
contain mixed terms, so that the density matrix $\hat\rho \neq
\vert\Psi\rangle \langle \Psi \vert$ in general.  By simply
keeping track of the powers of $a(t)$ and its derivatives, it is
straightforward to generalize the Minkowski spacetime density
matrix to the RW case with the result,
\begin{eqnarray}
&&\langle q|\hat{\rho}(t)|q'\rangle =\langle q_{\rule{0mm}{2.1mm}0}
\vert\hat \rho_0 (\bar\phi, \bar p; \zeta_0, \pi_0; \sigma_0)
\vert q'_0\rangle \times \prod_{\bf k \neq 0}
\langle q_{\rule{0mm}{2.1mm}\bf k}\vert \hat{\rho}(\zeta_k, \pi_k; \sigma_k)
\vert q_{\bf k}^{\prime}\rangle \nonumber\\
&&=\rho_0 \prod_{\bf k \neq 0}(2\pi \zeta_k^2)^{-\frac{1}{2}} \exp
\left\{ - \frac{\sigma_k^2 + 1}{8 \zeta_k^2} \left(\vert
q_{\rule{0mm}{2.1mm}\bf k}\vert^2 + \vert q_{\bf
k}^{\prime}\vert^2 \right) \right. \nonumber \\
&& \;\;\;\;  \left. + \frac{i\pi_k}{2 \hbar\zeta_k} \left( \vert
q_{\bf k}\vert^2 -\vert q_{\bf k}^{\prime}\vert^2 \right) +
\frac{\sigma_k^2 - 1}{8 \zeta_k^2} \left(  q_{\bf k}^{\ } q_{\bf
k}^{\prime *} +  q_{\bf k}' q_{\bf k}^*\right)\right\} \,,
\label{eq:gaussd}
\end{eqnarray}
where the $\{ q_{\bf k}\}$ are the set of complex valued Fourier
coordinates of the scalar field which are time independent in the
Schr\"odinger representation, {\it i.e.,} the matrix elements of
the Heisenberg field operator $\Phi (t_0, {\bf x})$
at an arbitrary initial time $t_0$ are defined by
\begin{equation}
\langle  q\vert \Phi(t_0, {\bf x})\vert q'\rangle = \left(\int [\d
{\bf k}] Y_{\bf k}({\bf x})\,  q_{\bf k}\right) \langle q\vert
q'\rangle \,.
\label{eq:coorphi}
\end{equation}
The latter matrix element is non-vanishing only for $q_{\bf k} =
q'_{\bf k}$ and is defined precisely by Eq.~(\ref{eq:qdel}) below.  In
this Schr\"odinger coordinate representation the action of the
conjugate momentum operator $\Pi_\Phi$ is given by
\begin{equation}
\langle  q\vert \Pi_\Phi(t_0, {\bf x})\vert q'\rangle = -i\hbar
\left(\int [\d {\bf k}] Y_{\bf k}({\bf x}) \frac{\partial}{\partial
q_{\bf k}^*}\right) \,\langle q\vert q'\rangle \, .
\label{eq:coorpi}
\end{equation}
Since $\Phi$ is a real field, the complex coordinates are related by
$ q_{\bf k}^* =  q_{\raisebox{-.5mm}{\scriptsize -\bf k}}$ and occur
in conjugate pairs. Hence we have the rule,
\begin{equation}
\frac{\partial  q_{\rule{0mm}{2.1mm} \bf k}} {\partial  q_{\bf
k'}^*} = \delta_{\rule{0mm}{2mm}{\bf k}, -{\bf k'}}
\end{equation}
and the $\pm {\bf k}$ terms in the density matrix~(\ref{eq:gaussd})
are identical, and may be combined. The $ q_{{\bf k} ={\bf 0}}$ field
coordinate is real, and we have separated off the ${\bf k} = {\bf 0}$
component of the density matrix $\hat\rho$ in Eq.~(\ref{eq:gaussd}),
denoting it by $\rho_0$. In this spatially homogeneous and isotropic
Fourier component we may also allow for the possibility of a
non-vanishing real mean value of the scalar field,
\begin{equation}
\bar\phi(t) = \langle \Phi(t, {\bf x})\rangle \equiv {\rm Tr}
(\Phi(t, {\bf x}) \hat \rho) = \int_{-\infty}^{\infty}\, \d q_0 \
q_0 \  \langle q_{\rule{0mm}{2.1mm}0}\vert\hat
\rho_{\rule{0mm}{2.1mm} 0}
   (\bar\phi, \bar p; \zeta_0, \pi_0; \sigma_0)
\vert  q_0\rangle \,,
\end{equation}
which because of the RW symmetry is a function of time only. This
spatially homogeneous expectation value is the classical inflaton
field in inflationary models. Separating this mode explicitly from the
rest is possible strictly only in a discrete basis, such as that
corresponding to closed spatial sections, $\epsilon = +1$. The density
matrix in the spatially homogeneous sector is
\begin{eqnarray}
\rho_0(t) &\equiv& \langle  q_{\rule{0mm}{2.1mm} 0} \vert
\hat\rho_{\rule{0mm}{2.1mm} 0} ( \bar \phi, \bar p ; \zeta_0,
\pi_0; \sigma_0) \vert  q_0^{\prime}\rangle \nonumber \\
   &=& (2\pi \zeta_0^2)^{-\frac{1}{2}} \exp \biggl\{i \,\frac{\bar p}{\hbar}\,(
q_{\rule{0mm}{2.1mm}0} - q_0^{\prime})- \frac{\sigma_0^2 + {1}}{8
\zeta_0^2} \left[ ( q_{\rule{0mm}{2.1mm} 0}-\bar\phi)^2
+ ( q_0^{\prime}-\bar\phi)^2\right]\nonumber\\
&& \qquad + \frac{i\pi_0}{2 \hbar\zeta_0} \left[(
q_{\rule{0mm}{2.1mm} 0}-\bar\phi)^2
   -  ( q_0^{\prime }-\bar\phi)^2\right]
+ \frac{\sigma_0^2 - 1}{4\zeta_0^2}\,
( q_{\rule{0mm}{2.1mm} 0}-\bar\phi)( q_0^{\prime}-\bar\phi)\biggr\}\,,
\label{gauss0}
\end{eqnarray}
where
\begin{equation}
\bar p (t) \equiv a^3 \left(\dot{\bar\phi} (t) +6\xi H\bar\phi (t)\right)
\label{eq:pbardef}
\end{equation}
is the momentum conjugate to the spatially homogeneous mean field
$\bar\phi (t)$.

Real field coordinates $(q_{\bf k}^{\rm R}, q_{\bf k}^{\rm I})$ for
the ${\bf k} \neq {\bf 0}$ modes may be introduced by
\begin{equation}
   q_{\bf k} = \frac{1}{\sqrt 2} ( q_{\bf k}^{\rm R} - i q_{\bf k}^{\rm I})
\,,\qquad {\bf k} \neq {\bf 0}
\end{equation}
and the functional integration measure over the field coordinate space
defined by
\begin{equation}
[{\cal D} q] \equiv \d  q_0
\prod_{{\bf k} > {\bf 0}} \d q_{\bf k}^{\rm R} \,\d q_{\bf k}^{\rm I} \,.
\label{eq:meas}
\end{equation}
The inner product appearing in~(\ref{eq:coorphi}) is defined by
\begin{equation}
\langle q \vert q' \rangle \equiv \delta(q_0-q'_0)\,\prod_{{\bf k} > {\bf 0}}
\delta (q_{\bf k}^{\rm R} - q_{\bf k}^{\prime \rm R})
\delta (q_{\bf k}^{\rm I} - q_{\bf k}^{\prime \rm I})
\label{eq:qdel}
\end{equation}
and the general mixed state Gaussian density matrix~(\ref{eq:gaussd})
is properly normalized,
\begin{equation}
{\rm Tr}\, \hat \rho
= \int [{\cal D} q] \langle  q|\hat \rho |q\rangle = 1\,,
\label{eq:rhonorm}
\end{equation}
with respect to this measure.

It is clear from~(\ref{eq:gaussd}) that if $\sigma_k =1$ for all $k$,
then the mixed terms vanish and the density matrix reduces to the
product,
\begin{equation}
\langle  q|\hat{\rho}|q'\rangle\Big\vert_{\sigma_k =1} =
\langle q\vert\Psi\rangle\langle \Psi\vert q'\rangle
   \equiv\prod_{\bf k}\Psi( q_{\rule{0mm}{2.1mm}\bf k})\Psi^*( q'_{\bf k})\,,
\label{eq:prodpur}
\end{equation}
characteristic of a pure state, with
\begin{subequations}
\begin{eqnarray}
&&\Psi(\rule{0mm}{2.3mm} q_{\bf k})
= \Psi(\rule{0mm}{2.3mm} q_{-\bf k})
= (2\pi \zeta_k^2)^{-\frac{1}{4}} \exp \left\{
-\frac {\vert q_{\bf k}\vert^2} {4 \zeta_k^2}
+ i\pi_k\,\frac{\vert q_{\bf k}\vert^2}{2 \hbar\zeta_k} \right\}\,,
\qquad {\bf k} \neq {\bf 0}\,,\\
&&\Psi(\rule{0mm}{2.3mm} q_0) =(2\pi \zeta_0^2)^{-\frac{1}{4}}
\exp \left\{i\, \frac{\bar p}{\hbar}\,( q_0 -\bar\phi) -\frac {(
q_0 -\bar\phi)^2} {4 \zeta_0^2} + i\pi_0\,\frac{( q_0^{\
}-\bar\phi)^2} {2 \hbar\zeta_0} \right\}\,, \; {\bf k} ={\bf 0}\,,
\end{eqnarray}
\label{eq:pure}
\end{subequations}
\hspace{-0.25cm}
which is the Gaussian pure state Schr\"odinger wave functional in the
Fourier representation of the complex field coordinate
basis~(\ref{eq:coorphi}). The pure state case corresponds to $n_k
=0$, $\sigma_k = 1$, and requires only the two real functions of
$k$ and $t$, $(\zeta_k, \pi_k)$ for its full specification.

The Wigner function(al) representation of the Gaussian density matrix
is obtained by shifting $ q_{\bf k} \rightarrow q_{\bf k} + x_{\bf
k}/2$ and $ q_{\bf k}' \rightarrow q_{\bf k} - x_{\bf k}/2$
in~(\ref{eq:gaussd}), and Fourier transforming $\hat \rho$ with
respect to the difference variables $x_{\bf k}$, {\it viz.,}
\begin{eqnarray}
&& F_W[q, p] \equiv \int\,[{\cal D}x]
\prod_{\bf k}\,(2\pi\hbar)^{-1}\,\exp\left(-\frac{i}{\hbar}\,
p_{\bf k}^*x_{\bf k}^{\,}\right)
\left\langle q_{\bf k} + \frac{x_{\bf k}}{2}\,\Big\vert\, \hat{\rho}
\,\Big\vert\,  q_{\bf k} - \frac{x_{\bf k}}{2}\right\rangle \nonumber\\
&&\quad = F_0( q_0,p_0)
\prod_{{\bf k}\neq {\bf 0}} (\pi \hbar \sigma_k)^{-1}
\exp \left\{ - \frac{\vert q_{\bf k}\vert^2}{2\zeta_k^2} -
\frac{2}{\hbar^2\sigma_k^2}\, \big\vert\zeta_k p_{\bf k} -
\pi_k q_{\bf k}\big\vert^2\right\}\,,
\label{eq:wigner}
\end{eqnarray}
where $p_{\bf k}^* = p_{\raisebox{-0.3mm}{\scriptsize -\bf k}}$, and
\begin{equation}
F_0( q_0,p_0) = (\pi \hbar \sigma_0)^{-1}
\exp \left\{ - \frac{(q_0 -\bar\phi)^2}{2\zeta_0^2} -
\frac{2}{\hbar^2\sigma_0^2}\, \big[\zeta_0 \,(p_0 -\bar p)
- \pi_0\, (q_0-\bar\phi)\big]^2\right\}
\end{equation}
is the Wigner function in the spatially homogeneous ${\bf k} = {\bf
0}$ sector.  Note that the normalization of the Gaussian Wigner
functional is constant in time, as required for a Hamiltonian
evolution in phase space.  For a given $ q_{\bf k}$ this Gaussian
function is peaked on the phase space trajectory,
\begin{subequations}
\begin{eqnarray}
p_{\bf k} &\approx& \frac{\pi_k}{\zeta_k} \, q_{\bf k}\,,
\qquad {\bf k} \neq {\bf 0}\, ,\\
p_0 &\approx& \bar p + \frac{\pi_0}{\zeta_0}\,  (q_0-\bar\phi)\,,\qquad
{\bf k} = {\bf 0}\,,
\end{eqnarray}
\end{subequations}
becoming very sharply peaked on this trajectory in the formal
classical limit $\hbar \rightarrow 0$, $\sigma_k$ fixed, although the
width of the peak becomes larger for mixed states with larger
$\sigma_k$ (with $\hbar$ fixed)~\cite{PolSta}. The Wigner
functional~(\ref{eq:wigner}) is positive definite for Gaussian states
and may be interpreted as a normalized probability distribution for
any $\hbar\sigma_k$~\cite{SriPad}.

The functional integration measure~(\ref{eq:meas}) implies an
inner product,
\begin{equation}
\langle \Psi_2 \vert \Psi_1\rangle = \int\,
[{\cal D}q]\,\langle \Psi_2 \vert q\rangle\langle
q\vert\Psi_1\rangle \equiv \exp\,(i\Gamma_{12})
\label{eq:inner}
\end{equation}
between pure states, and a coherence probability functional,
\begin{equation}
{\rm Tr} (\hat\rho_1\,\hat\rho_2) = \int\,[{\cal D}q]\,\int\, [{\cal D}q']\,
\langle q\vert \hat \rho_1 \vert q'\rangle\,
\langle q'\vert\hat \rho_2\vert q \rangle
\equiv \exp\,(-2\,\tilde\Gamma_{12})
\label{eq:prob}
\end{equation}
for general mixed states in the Schr\"odinger picture. In the case of
pure states the real functional $\tilde\Gamma_{12}$ becomes Im
$\Gamma_{12}$ of~(\ref{eq:inner}). In the case $\hat \rho_1 = \hat
\rho_2$, performing the Gaussian integrations in the coordinate
representation gives
\begin{eqnarray}
{\rm Tr} (\hat\rho ^2) &=& \int\,[{\cal D}q]\,\int\, [{\cal D}q']\,
\langle q\vert \hat \rho \vert q'\rangle\,
\langle q'\vert\hat \rho\vert q \rangle
\nonumber\\
&=& \left(\prod_{\bf k} \sigma_{\bf k}\right)^{-1} \nonumber\\
&=& \exp\left( - \frac{1}{\pi^2} \int\,[\d k]\, k^2 \ln
\sigma_k\right) \le 1\,.
\label{eq:mixed}
\end{eqnarray}
The inequality is saturated if and only if $\sigma_k = 1$ for all
$k$, in which case $\hat \rho = \vert\Psi \rangle \langle
\Psi\vert$, and the equality is simply a consequence of the
normalization condition, $\langle \Psi \vert \Psi \rangle = 1$ on
the pure state wave functional. If for any $k$, $\sigma_k > 1$,
Tr $\hat\rho^2 < 1$, which is characteristic of a mixed state
density matrix.

For either a pure or mixed state the Gaussian density matrix satisfies
the quantum Liouville equation,
\begin{equation}
i \hbar\frac{\partial \hat \rho}{\partial t}  =
[{\cal H}_{\Phi},\hat \rho]\,,
\label{eq:Lio}
\end{equation}
provided the time dependent parameters $(\bar \phi, \bar p; \zeta_k,
\pi_k)$ appearing in $\hat \rho$ satisfy the first order equations,
\begin{subequations}
\begin{eqnarray}
\dot{\bar\phi}&=& \frac{\bar p}{a^3} - 6 \xi H \bar\phi \, ,
\label{eq:hamevo} \\
\dot{\bar p}&=&  6 \xi H  \bar p - a^3 \left[ m^2 + 6 \xi
\frac{\epsilon}{a^2} + 6 \xi (6\xi -1){H^2} \right]\bar \phi \; ,
\label{eq:hamevo2} \\
\dot\zeta_k &=& = \frac{\pi_k}{a^3} - 6 \xi H \zeta_k \; ,
\label{eq:hamevo3} \\
\dot\pi_k &=& 6 \xi H  \pi_k - a^3 \left[\frac{k^2}{a^2} + m^2 +
(6 \xi -1)\left(\frac{\epsilon}{a^2} +6 \xi {H^2}\right) \right]\zeta_k
+ \frac{\hbar^2\sigma_k^2}{4 a^3 \zeta_k^3}\,.
\label{eq:hamevo4}
\end{eqnarray}
\label{eq:hamev}
\end{subequations}
\hspace{-0.25cm}
The first two of these equations are equivalent to the
second order equation for the spatially homogeneous mean field,
\begin{equation}
\ddot{\bar\phi} + 3H \dot{\bar\phi} + (m^2 + \xi R)\bar\phi = 0\,,
\end{equation}
while the latter two may be combined to yield the second order
equation for the Gaussian width parameter,
\begin{equation}
\ddot \zeta_k + 3 H \dot \zeta_k
+ \left(\frac{k^2 -\epsilon}{a^2} + m^2 + \xi R\right)\zeta_k
= \frac{\hbar^2\sigma_k^2}{4 a^6 \zeta_k^3}\,.
\label{eq:zdd}
\end{equation}
The last equation is derived in Appendix A from Eq.~(\ref{eq:phikeq})
and the defining relation~(\ref{eq:zkdefonly}).

This establishes the equivalence between the Hamiltonian evolution of
the Gaussian density matrix~(\ref{eq:gaussd}), according to~(\ref{eq:Lio}),
and the Lagrangian evolution of general RW initial states described in
the previous section. Since the Hamiltonian~(\ref{eq:Hphi}) is Hermitian
with respect to the field coordinate measure~(\ref{eq:meas}), the time
evolution of $\hat \rho$ is unitary and the normalization~(\ref{eq:rhonorm})
is preserved. Hence there is no dissipation in the system and the evolution
is fully time reversible in principle. The time evolution of the density
matrix parameters $(\bar \phi, \bar p; \zeta_k, \pi_k)$ may also be derived
from an effective {\it classical} Hamiltonian, $H_{\rm eff} =
{\rm Tr}\, ({\cal H}_{\Phi} \hat \rho)$, given explicitly by Eq. (\ref{eq:effHam})
of Appendix A. This effective Hamiltonian is just the expectation value of
the quantum Hamiltonian ${\cal H}_\Phi$ in the general Gaussian state,
in which $\hbar$ appears as a parameter. Notice that the role of the $\hbar$
term in Eq.~(\ref{eq:zdd}) is to act as a ``centrifugal barrier'' for
the coordinate $\zeta_k$, preventing the Gaussian width parameter from
ever shrinking to zero. This width depends on the product $\hbar \sigma_k$,
so that the classical high temperature limit $\hbar \sigma_k
\rightarrow k_BT/\omega_k$ is treated on the same footing as the
quantum zero temperature limit $\hbar \sigma_k \rightarrow
\hbar$. Both classical thermal and quantum uncertainty principle
effects contribute to the width of the Gaussian in general.

The Hamiltonian evolution and the density matrix description of RW
states is not manifestly covariant under general
coordinate transformations, depending as it does on a particular
slicing of the four dimensional geometry into three dimensional
surfaces $\Sigma$. Since initial data must be specified on such a
spacelike Cauchy surface, this is the natural $3+1$ splitting for
initial value problems in RW cosmology. The equation of motion
(\ref{eq:waveq}) is certainly invariant under general
coordinate transformations, whereas the initial data must be specified
on some three surface $\Sigma$ for any particular physical initial state.

We note that the canonical effective Hamiltonian generating the time
evolution of the wave functional in the Schr\"odinger representation
is not simply related to the expectation value $\langle T_{tt}\rangle$
of the covariant energy-momentum tensor.  The key point in reconciling
the canonical and covariant energy densities is that the {\it full}
system of matter plus metric fields must be taken into account. After
the addition of the surface terms to the standard classical action in
Eq.~(\ref{eq:Scl}) to remove the second derivatives of the metric, the
momentum conjugate to the scale factor $a$ is
\begin{equation}
\Pi_a \equiv \frac{\delta S_{\rm cl}}{\delta \dot a}
= - \frac{3}{4\pi G_N}\, \dot a a +
6 \xi \dot a a \Phi^2 + 6 \xi a^2 \dot \Phi \Phi\,.
\label{eq:pia}
\end{equation}
Then the total Hamiltonian density, constructed in the canonical
prescription is
\begin{eqnarray}
{\cal H}_{\rm tot} &=& \Pi_\Phi \dot \Phi + \Pi_a \dot a - {\cal L}_{\rm cl}
\nonumber\\
&=& - \frac{1}{8\pi G_N} a^3 G_{tt} + a^3\, T_{tt}\,,
\label{eq:Htot}
\end{eqnarray}
where
\begin{equation}
G_{tt} = 3 \left(H^2 + \frac{\epsilon}{a^2}\right)\,,
\end{equation}
is the $tt$ component of the Einstein tensor, $G_{ab} = R_{ab} -
g_{ab} R/2$ and
\begin{equation}
T_{tt} = \frac{1}{2} \dot\Phi^2 + 6\xi H \dot \Phi\,\Phi +
\frac{1}{2a^2} \gamma^{ij} (\partial_i\Phi)(\partial_j\Phi)
+ \frac{m^2}{2}\Phi^2 +
3 \xi \left(H^2 + \frac{\epsilon}{a^2}\right)\Phi^2\,,
\end{equation}
is the $tt$ component of the covariant energy-momentum tensor
\begin{eqnarray}
T_{ab} &\equiv& -\frac{2}{\sqrt{-g}} \frac{\delta}{\delta g^{ab}} S_{\Phi}
= (\nabla_a \Phi) (\nabla_b \Phi) - \frac{g_{ab}}{2}
g^{cd} (\nabla_c \Phi) (\nabla_d \Phi)\nonumber\\
&&- 2\xi \nabla_a(\Phi\nabla_b \Phi)
+ 2 \xi g_{ab} g^{cd}\,\nabla_d(\Phi\nabla_c \Phi)
+ \xi G_{ab} \Phi^2  - \frac{m^2}{2} g_{ab} \Phi^2\,.
\label{eq:Tcl}
\end{eqnarray}
Invariance under transformations of the time coordinate leads
to the classical equation of constraint,
\begin{equation}
{\cal H}_{\rm tot} = 0\,.
\end{equation}
Because of Eq.~(\ref{eq:Htot}) this coincides with the $tt$ component
of the classical Einstein equations, {\it viz.,} the Friedman equation
for RW cosmologies. This constraint equation is equivalent to the
requirement that the classical theory be invariant under arbitrary time
reparameterizations, $t \rightarrow t'(t)$, a condition which
Hamiltonian evolution in a fixed background $a(t)$ does not
require. The three momentum constraints of spatial coordinate
transformations, equivalent to the $ti$ components of Einstein's
equations are automatically satisfied in any homogeneous,
isotropic RW state.

Hence the Hamiltonian and covariant approaches agree, only for the
{\it full} system of gravity plus matter, {\it i.e.,} provided that
the RW scale factor is treated as a dynamical degree of freedom, on
the same footing as the matter field $\Phi$. In contrast, the
Hamiltonian ${\cal H}_{\Phi}$ of Eq.~(\ref{eq:Hphi}) generates the
correct evolution of the $\Phi$ field in a fixed RW background,
whether or not the scale factor $a(t)$ satisfies Einstein's equations,
and with no requirement of invariance under reparameterizations of
time. It is the covariant energy-momentum tensor $T_{ab}$
that is conserved and should be used whenever the full cosmological
theory of matter and gravitational degrees of freedom are under
consideration. With this important proviso the canonical and
covariant formulations of the initial value problem are
completely equivalent.

\section{Energy-Momentum Tensor of UV Allowed RW States}
\label{sec:tmunu-allowed}

In the previous sections we have defined and described general
homogeneous and isotropic RW initial states, with no restriction on
the set of three density matrix parameters $(\zeta_k, \pi_k;\sigma_k)$ which
describes the state and its evolution.  However, because of the
Wronskian condition~(\ref{eq:wron}), that enforces the canonical
commutation relations of the quantum field, the state parameters do
not approach zero rapidly enough at large $k$ for the integrals
in~(\ref{eq:piphibi}) or the expectation value of
the covariant energy-momentum tensor $\langle T_{ab}\rangle$, to converge.
Hence these expressions are purely formal, and a definite renormalization
prescription is necessary to extract the finite state dependent
terms. This is a necessary prerequisite for any discussion of short
distance, initial state, or backreaction effects in inflation, at
least within a conventional effective field theory framework.

Because the energy-momentum tensor is an operator of mass dimension
four, it contains divergences up to fourth order in the comoving
momentum cutoff $k_M$. Requiring that the forms
of the integrands at large $k$ match those expected for the vacuum
up to fourth order in derivatives of the metric, allows for all
the divergences in $\langle T_{ab}\rangle$ to be absorbed into
counterterms of the relevant and marginally irrelevant terms of the
local gravitational effective action~\cite{birrell-davies}. This
adiabatic order four condition on the initial state imposes
restrictions on the set of parameters $(\zeta_k, \pi_k;\sigma_k)$ at
large $k$, and guarantees that the renormalized expectation value
$\langle T_{ab}\rangle_{_R}$ will remain finite and well-defined at
all subsequent times~\cite{parker}. Conversely, failure to impose
these short distance restrictions on the initial state leads to cutoff
dependence which cannot be identified with covariant local
counterterms up to dimension four in the gravitational action, and
which violate the assumptions of a low energy EFT for gravity
consistent with the symmetries of general covariance implied by the
Equivalence Principle.

The available counterterms up to dimension four in the coordinate
invariant effective action for gravity are the four local geometric
terms $\Lambda, R, R^2$ and $C_{abcd}C^{abcd}$ (the square of the Weyl
conformal tensor), which can be added to the one-loop action of the
scalar field, $S^{(1)}[g]$. Hence the low energy gravitational effective
action is formally
\begin{equation}
S_{\rm eff}[g] = S^{(1)}[g] + \frac{1}{16\pi G_N}\int \,
\d^4x\, \sqrt{-g}\, (R-2\Lambda)
- \frac{1}{2} \int\, \d^4x\, \sqrt{-g} \,
\left(\alpha\, C_{abcd} C^{abcd} + \beta\, R^2 \right)
\; ,
\label{eq:effS}
\end{equation}
where
\begin{equation}
S^{(1)}[g] = \frac{i\hbar}{2}\,{\rm Tr} \ln (-\sq + \xi R + m^2)
\; ,
\end{equation}
and $\Lambda$, $G_N$, $\alpha$, and $\beta$ are bare parameters which
are chosen to cancel the corresponding divergences in $S^{(1)}[g]$.
A fully covariant renormalization procedure is one that removes all
divergences in $S^{(1)}[g]$ by adjustment of the scalar parameters
$\Lambda$, $G_N$, $\alpha$, and $\beta$ of~(\ref{eq:effS}), and {\it
only} those parameters, in such a way that the total effective action
$S_{\rm eff}[g]$ and the renormalized energy-momentum tensor derived
from it,
\begin{equation}
\langle T_{ab}\rangle_{_R}
= -\frac{2}{\sqrt{-g}} \frac{\delta}{\delta g^{ab}} S_{\rm R}^{(1)}[g]\,,
\label{eq:Tren}
\end{equation}
is finite ({\it i.e.,} independent of the cutoff $k_M$) and covariantly
conserved:
\begin{equation}
\nabla^b \langle T_{ab}\rangle_{_R} = 0\,.
\label{eq:cocons}
\end{equation}
Thus the renormalized expectation value $\langle
T_{ab}\rangle_{_R}$ is strictly well defined only by reference to
the full low energy effective action $S_{\rm eff}[g]$ and the
equations of motion of the gravitational field following from it,
\begin{equation}
\frac{1}{8\pi G_N}\left(G_{ab} +
\Lambda g_{ab}\right) =  \langle T_{ab}\rangle_{_R}
+ \alpha_{_R}\ ^{(C)}H_{ab} + \beta_{_R}\ ^{(1)}H_{ab}\, ,
\label{eq:scF}
\end{equation}
of which it is a part.

The local conserved tensors,
\begin{subequations}
\begin{eqnarray}
^{(1)}H_{ab}&&\equiv \frac{1}{\sqrt{-g}}\frac{\delta}{\delta g^{ab}}
\int\, \d^4\,x\,\sqrt{-g}\, R^2 =
2 g_{ab}\sq R  -2\nabla_a\nabla_b R + 2 R R_{ab} - \frac{g_{ab}}{2}R^2\,,
\label{eq:Hone}\\
^{(C)}H_{ab}&&\equiv \frac{1}{\sqrt{-g}}\frac{\delta}{\delta g^{ab}}
\int\, \d^4\,x\,\sqrt{-g}\, C_{abcd}C^{abcd} = 4\nabla^c \nabla^d C_{acbd}
+ 2R^{cd}C_{acbd}\,,
\end{eqnarray}
\label{eq:Hdef}
\end{subequations}
\hspace{-0.2cm}
derived from the fourth order terms in the effective action
are similar to those which appear in any EFT, whose
equations of motion involve a local expansion in the number of
derivatives. Provided that we restrict our attention to the low
momentum region of validity of the EFT, these terms in Eq.~(\ref{eq:Tren})
should be negligible compared to those involving fewer derivatives
of the metric. Their only role is to provide the
covariant UV counterterms necessary to remove the subleading
logarithmic divergences in $\langle T_{ab}\rangle$.  Conversely, if
when multiplied by finite renormalized parameters $\alpha_{_R}$ and
$\beta_{_R}$ of order unity, they are not negligibly small, then the
applicability of the EFT framework for low energy gravity is in
question. In cosmology certainly a necessary condition for this framework
to be applicable is that the Riemann curvature tensor components and
their contractions are negligibly small in Planck length units, {\it i.e.,}
\begin{equation}
\ell_{\rm Pl}^2\,|R_{ab}^{\ \ cd}| =
\hbar G_N\,|R_{ab}^{\ \ cd}|\ll 1\,,
\label{eq:subPl}
\end{equation}
and we restrict ourselves to this regime.

The independent (unrenormalized) components of the energy-momentum
tensor with non-vanishing expectation values in a general
RW initial state are the energy density,
\begin{subequations}
\begin{eqnarray}
&&\varepsilon_u\equiv \langle T_{tt}\rangle_u =
\frac{1}{4 \pi^2} \int\,
   [\d k]\, k^2 \, \sigma_k \left[ |\dot \phi_k|^2
+ \left(\frac{k^2 -\epsilon}{a^2} + m^2 \right) |\phi_k|^2 \right]
\nonumber\\
&& \qquad + \frac{3\xi }{2 \pi^2} \int\,[\d k]\, k^2 \, \sigma_k
\left[2H \,{\rm Re} (\phi_{k}^* \dot \phi_{\rule{0mm}{-2.3mm}k}) +
\left(\frac{\epsilon}{a^2} + H^2 \right) |\phi_k|^2\right]\,,
\label{eq:energy}
\end{eqnarray}
and the trace,
\begin{eqnarray}
T_u &=& \frac{(6 \xi-1) }{2 \pi^2} \int  [\d k]\, k^2 \, \sigma_k
\left[ -|\dot \phi_k|^2 +\left(\frac{k^2 -\epsilon}{a^2} + m^2 +
\xi R\right) |\phi_k|^2\right] \nonumber \\
&& \quad - \frac{m^2}{2 \pi^2} \int \, [\d k]\, k^2
\,\sigma_k |\phi_k|^2\,.
\label{eq:trace}
\end{eqnarray}
\label{eq:unren}
\end{subequations}
\hspace{-0.25cm}
The other non-vanishing components of $\langle T_{ab} \rangle$ in
a general RW state are the diagonal spatial components,
$\langle T_{ij}\rangle = p g_{ij}$.  The isotropic pressure $p$
may be obtained from the energy density $\varepsilon$ and trace $T$,
by $p=(\varepsilon + T)/3$. The conservation equation~(\ref{eq:cocons})
in the case of RW symmetry has only a time component which is non-trivial,
namely,
\begin{equation}
\dot\varepsilon + 3H\, (\varepsilon + p) =
\dot\varepsilon + H\, (4 \varepsilon + T) = 0\,.
\label{eq:cons}
\end{equation}
The unrenormalized expressions~(\ref{eq:unren}) satisfy this
relation by use of the equation of motion~(\ref{eq:phikeq}),
provided that a comoving momentum cutoff $k_M$, introduced to
render the integrals finite, is itself independent of time. An
important criterion for any renormalization procedure is that it
preserve this property so that~(\ref{eq:cons}) remains valid for
the fully renormalized quantities as well. Notice that a fixed
cutoff in the physical momentum $p_M = k_M/a$ will {\it not}
preserve the form of the covariant conservation
Eq.~(\ref{eq:cons}), because of the non-vanishing time derivative
operating on the upper limit of the integrals at $k_M = p_M a$, if
$p_M=M$ is assumed to be independent of time.

In the case of spatially homogeneous and isotropic RW spacetimes the
adiabatic method has been shown to be equivalent to a fully covariant
treatment of the divergences of the energy-momentum tensor which
preserves its conservation~\cite{BirAndPar}. The starting point
of this method is the WKB-like form of the exact mode functions,
\begin{equation}
\phi_k(t) \equiv {\sqrt\frac{\hbar}{2a^3 \Omega_k}}\,\exp\left(-i\int^t\,\d t'
\Omega_k(t')\right)\,,
\label{eq:adbmode}
\end{equation}
which when substituted into~(\ref{eq:phikeq}), yields the second order
equation for $\Omega_k$,
\begin{equation}
\Omega_k^2 = \omega_k^2 + \left(\xi - \frac{1}{6}\right) R - \frac{1}{2}
\left( \dot H + \frac{H^2}{2}  \right) + \frac{3}{4}
\frac{\dot\Omega_k^2}{\Omega_k^2}
- \frac{\ddot \Omega_k}{2\,\Omega_k}\,.
\label{eq:freq}
\end{equation}
{From} this expression a systematic asymptotic expansion of the
frequency $\Omega_k$ in time derivatives of the metric scale
factor $a(t)$ can be developed. At leading order, neglecting all
time derivatives, $\Omega_k \simeq \omega_k$. Substituting this
into the right hand side of~(\ref{eq:freq}), one finds to second
order,
\begin{equation}
\Omega_k \simeq \omega_k + \frac{\left(\xi - \frac{1}{6}\right)} {2\,\omega_k}\,R
- \frac{m^2}{4\,\omega_k^3}\,(\dot H + 3H^2) +
\frac{5}{8}\,\frac{m^4}{\omega_k^5} H^2
+ \dots \,,
\label{eq:adbtwo}
\end{equation}
where the ellipsis consists of terms third and higher order in
derivatives of the metric. It is clear that this asymptotic expansion
is valid at large $k$, {\it i.e.,} at distance scales much shorter
than the characteristic scale of the variation of the geometry
$H^{-1}$. Hence requiring the exact solutions of the mode
equation~(\ref{eq:phikeq}) to match this asymptotic expansion to some
order implies that the quantum state density matrix of the scalar
field~(\ref{eq:gaussd}) should match that of the local vacuum to that
order. It is a statement of the Equivalence Principle in the low
energy EFT that the {\it local}, short distance
properties of the quantum vacuum at a point $x$ should approximate
that of the nearly flat space vacuum constructed in a local
neighborhood of $x$. Hence the wave functional~(\ref{eq:pure}) must
have large $k$ components characterized by $\{\zeta_k, \pi_k \}$ which
are {\it universal}, corresponding to local geometric invariants at $x$ in
the effective action, and the same for all physically realizable
states, independent of the geometry of the spacetime at larger scales.

When~(\ref{eq:adbtwo}) is substituted into~(\ref{eq:adbmode}),
and the resulting mode function is substituted into~(\ref{eq:unren})
with $\sigma_k$ set equal to one, one obtains integrands which match
the quartic and quadratic divergent behavior of the unrenormalized
stress tensor components~\cite{parker}. Up to adiabatic
order two these are explicitly given by
\begin{subequations}
\begin{eqnarray}
\varepsilon^{(2)} &=&\frac{\hbar}{4 \pi^2 a^3} \int\,[\d k]\, k^2 \,
\varepsilon^{(2)}_k \,,\\
T^{(2)}&=&\frac{\hbar}{4 \pi^2 a^3} \int \,[\d k]\, k^2 \,
T^{(2)}_k\,,
\end{eqnarray}
\end{subequations}
with
\begin{subequations}
\begin{eqnarray}
&&\varepsilon^{(2)}_k = \omega_k + \frac{m^4}{8\, \omega_k^5}\,H^2
+\frac{(6 \xi -1)}{2 \,\omega_k}
\left[\frac{\epsilon}{a^2} - H^2 - \frac{m^2}{\omega_k^2}\,H^2\right] \,,
\label{eq:epstwo}
\\
&& T^{(2)}_k = -\frac{m^2}{\omega_k} - \frac{m^4}{4\,\omega_k^5}
\,(\dot H + 3 H^2) + \frac{5\,m^6}{8\,\omega_k^7}\,H^2 \nonumber \\
&& \qquad + \frac{(6 \xi -1)}{\omega_k} \left[ \dot H + H^2 +
\frac{m^2}{2\,\omega_k^2}\, \left(2\dot H + 3 H^2 +
\frac{\epsilon}{a^2} \right) - \frac{3\,m^4}{2\,\omega_k^4}\,H^2
\right]\,.
\label{eq:Ttwo}
\end{eqnarray}
\label{eq:epsTtwo}
\end{subequations}
\hspace{-0.3cm} Notice that these expressions are state independent
and universal, depending only upon the RW geometry and the parameters
$m, \xi$ of the matter Lagrangian. Although not manifestly covariant
in form, Refs.~\cite{BirAndPar} show that subtracting these second
order asymptotic terms from the unrenormalized energy
density~(\ref{eq:energy}) and trace~(\ref{eq:trace}) corresponds to
adjustment of the generally covariant counterterms up to two
derivatives in the low energy effective
action~(\ref{eq:effS}). Consistent with this, it may be checked that
the second order energy density $\varepsilon^{(2)}$ of
Eq.~(\ref{eq:epstwo}) and the second order pressure $p^{(2)}$, satisfy
the covariant conservation equation~(\ref{eq:cons}), provided any
cutoff of the $k$ integrals is again independent of time. Hence the
(partially) renormalized energy density $\varepsilon_u -
\varepsilon^{(2)}$, and trace $T_u - T^{(2)}$, which are free of
quartic and quadratic divergences, also obey the conservation
equation~(\ref{eq:cons}).

In order to remove the remaining logarithmic divergences in the
energy density and trace in a general RW spacetime, the terms
containing up to four derivatives of the metric must be subtracted
as well in four spacetime dimensions. The expressions for the adiabatic
order four terms in the mode expansion~(\ref{eq:adbtwo}), or
$\varepsilon^{(4)}$ and $T^{(4)}$ can be found in~\cite{bunch,hm-pm}.
We shall not need their explicit form here, and simply assume that one
can identify a particular solution $v_k$ to~(\ref{eq:phikeq}), whose
frequency function $\Omega_k$ possesses an asymptotic expansion for
large $k$ which agrees with~(\ref{eq:adbtwo}), up to fourth adiabatic
order, and stress tensor components~(\ref{eq:unren}), which agree
with $\varepsilon^{(4)}$ and $T^{(4)}$ up to fourth adiabatic order.

In the general case, this is the necessary and sufficient condition for
the renormalized stress tensor to be finite and conserved in the RW
state corresponding to this particular mode function $v_k$.  For example,
in de Sitter spacetime, these modes, $v_k$, could be taken to be the
Bunch-Davies (BD) modes~\cite{b-d}, since these are adiabatic
order four modes and the BD state is a candidate vacuum state. A
general set of modes $\phi_k$ can be written then as a Bogoliubov
transformation~(\ref{eq:bog}) of these vacuum modes. The difference
of the renormalized stress tensor in this general state with that
given by the particular choice of $\phi_k=v_k$ and $\sigma_k=1$
then define the finite state dependent terms in the stress tensor
in the general RW initial state. In order for the initial state
defined by this general set of modes to remain a UV
allowed RW initial state, the state dependent terms in the
renormalized stress tensor should not spoil the fourth order
approach to the local vacuum which we required of the vacuum modes
$v_k$. Hence we impose the condition that the integrals with state
dependent integrands must be convergent as well. Pure or mixed states
satisfying this condition will be called {\it UV allowed RW states}.

These UV allowed states are described by mode functions, $\phi_k$ and
corresponding
density matrix parameters, $(\zeta_k, \pi_k; \sigma_k)$ for
which the integrands in the stress tensor components (\ref{eq:unren}) agree
with the fourth order adiabatic integrands $\varepsilon^{(4)}_k$ and $T^{(4)}_k$
at large $k$. We may choose any particular fourth order adiabatic
$v_k$ with respect to which to define the renormalized vacuum
energy-momentum tensor components,
\begin{subequations}
\begin{eqnarray}
\varepsilon_v &\equiv \ \varepsilon_u\raisebox{-1.1ex}{\Big\vert}
_{\stackrel{\scriptstyle \phi_k = v_k}
{\scriptstyle \sigma_k =1\ }}
- \ \varepsilon^{(4)}\,,\\ [2ex]
T_v &\equiv \ T_u\raisebox{-1ex}{\Big\vert}_{\stackrel{\scriptstyle \phi_k = v_k}
{\scriptstyle \sigma_k =1\ }} - \ T^{(4)}\,.
\end{eqnarray}
\label{eq:vrenorm}
\end{subequations}
\hspace{-0.2cm}
The definition of the class of UV allowed states then guarantees that
the difference of stress tensors for any UV allowed RW state with respect
to this choice of vacuum are
  well-defined and finite. To identify these
terms we have only to
  introduce the form of the Bogoliubov transformation~(\ref{eq:bog})
for the general mode function $\phi_k$ into~(\ref{eq:unren}), and
using~(\ref{eq:bognorm}), separate off the vacuum terms evaluated at
$A_k =1$, $B_k =0$, and $\sigma_k =1$, which are renormalized
by~(\ref{eq:vrenorm}).  The remaining terms are the
finite terms for arbitrary UV allowed RW states with respect to the
given vacuum choice. Collecting these remaining state dependent
terms gives the fully renormalized result,
\begin{subequations}
\begin{eqnarray}
&&\varepsilon \equiv \langle T_{tt} \rangle_{_R} =
\varepsilon_v +
\frac{1}{2 \pi^2} \int\,
   [\d k]\, k^2 \, \Big(
N_k\, \vert \dot v_k  \vert^2
+ \sigma_k\,{\rm Re}[A_k B_k^* \dot v_k^2] \Big)
\nonumber\\
&& \qquad + \frac{1}{2 \pi^2} \int\,
   [\d k]\, k^2 \,
\Big(\frac{k^2 -\epsilon}{a^2} + m^2 \Big)
   \Big( N_k\, \vert  v_k  \vert^2
+ \sigma_k\,{\rm Re}[A_k B_k^*  v_k^2] \Big)
\nonumber\\
&& \qquad \qquad + \frac{6\xi H}{\pi^2} \int\,[\d k]\, k^2 \,
   \Big(N_k\, {\rm Re} [v_k^* \dot v_k]
+ \sigma_k\,{\rm Re}[A_k B_k^*  v_k  \dot v_k] \Big)
\nonumber\\
&& \qquad \qquad + \frac{3\xi }{ \pi^2} \int\,[\d k]\, k^2 \,
\Big(\frac{\epsilon}{a^2} + H^2 \Big)
   \Big( N_k\, \vert  v_k  \vert^2
+ \sigma_k\,{\rm Re} [A_k B_k^*  v_k^2] \Big)\,,
\label{eq:enerrw}
\end{eqnarray}
and
\begin{eqnarray}
&& T \equiv \langle T \rangle_{_R} = T_v + \frac{(1-6\xi)}{\pi^2}
\int  [\d k]\, k^2\, \Big( N_k\,\vert \dot v_k \vert^2
+\sigma_k \,{\rm Re}[A_k B_k^* \dot v_k^2] \Big)
\nonumber\\
&& \qquad + \frac{(6 \xi - 1) }{\pi^2} \int  [\d k]\, k^2 \,
\Big(\frac{k^2 -\epsilon}{a^2} + m^2 + \xi R\Big)
   \Big( N_k\,\vert  v_k  \vert^2
+ \sigma_k \,{\rm Re}[A_k B_k^*  v_k^2]
\Big)
\nonumber\\
&& \qquad \qquad - \frac{m^2}{\pi^2} \int \, [\d k]\, k^2\,
   \Big( N_k\,\vert  v_k  \vert^2
+ \sigma_k \,{\rm Re}[A_k B_k^*  v_k^2]
\Big)\,,
\label{eq:tracrw}
\end{eqnarray}
\label{eq:etrw}
\end{subequations}
\hspace{-0.3cm} where $N_k$ is defined by Eq.~(\ref{eq:numvac}).
The vacuum terms denoted by the subscript $v$ defined by
Eqs.~(\ref{eq:vrenorm}) and the additional state dependent terms
in Eqs.~(\ref{eq:etrw}) are separately conserved. Because the
state is assumed to be UV allowed, $N_k$ also approaches zero faster
than $k^{-4}$ as $k\rightarrow
\infty$, and all terms in Eq.~(\ref{eq:etrw})
are finite, {\it
i.e.,} there is no cutoff dependence and the integrals
may be
extended to infinity. Note also that a pure state with $n_k=0,
\sigma_k=1$ remains a pure (coherent) state under the Bogoliubov
transformation~(\ref{eq:bog}), notwithstanding the non-zero value
of $N_k = |B_k|^2$ for this state in the $v_k$ basis. The quantum
coherence effects of the Bogoliubov transformation appear also in
the rapidly oscillating interference terms involving $A_k B_k^*$
in Eq.~(\ref{eq:etrw}), which must be retained in the general UV
allowed RW coherent state in order to retain the strict time
reversibility of the evolution, as we shall see in
Sec.~\ref{sec:adiabatic}.

\section{Short Distance Effects in Inflation}
\label{sec:shortdist}

The development of the previous sections applies to general RW initial
states of the scalar field of any mass and $\xi$ in an arbitrary RW
spacetime. In this section we apply this general framework to the
special case relevant for slow roll inflationary models, namely de Sitter
space with a massless minimally coupled inflaton field. If spatially flat
sections are used, then the scale factor for de Sitter space takes the
form,
\begin{equation}
a_{\rm dS} = \frac{1}{H}\,e^{Ht} = - \frac{1}{H\eta}\,, \qquad
-\infty < \eta < 0\,,
\label{eq:RWds}
\end{equation}
with $H$ a spacetime constant related to the scalar curvature by
$R=12H^2$. The entire de Sitter manifold may be represented as a
hyperboloid of revolution embedded in a five dimensional flat
Minkowski spacetime~\cite{birrell-davies}. The hyperboloid
possesses an $O(4,1)$ invariance group of isometries, which can
be made manifest if spatially closed coordinates ($\epsilon =1$)
are used. The flat coordinates ($\epsilon =0$) with the scale
factor given by~(\ref{eq:RWds}) cover only one half of the full
de Sitter hyperboloid. None of the results presented in this
section will depend on the choice of flat, open, or geodesically
complete closed spatial sections, so we treat only the
flat sections ($\epsilon = 0$) in detail.

In the flat sections under the transformation of variables
$y= -k\eta = k \exp(-Ht)$, the mode equation~(\ref{eq:modefns}) becomes
Bessel's equation with index,
\begin{equation}
\nu^2 = \frac{9}{4} - \frac{m^2}{H^2} - 12 \xi\,.
\label{eq:nu}
\end{equation}
The Bunch-Davies (BD) state~\cite{c-t,d-c,b-d} is the unique
RW allowed state which is completely invariant under the full $O(4,1)$
isometry group of de Sitter space. In the coordinates where the scale
factor is given by~(\ref{eq:RWds}) the BD state is specified by the
particular solution of~(\ref{eq:phikeq}) given by
\begin{eqnarray}
\phi_k^{BD} &\equiv& \sqrt{ \frac{\pi\hbar}{4Ha^3}}\,
e^{i\pi\nu/2}\,e^{i\pi/4} \,H_{\nu}^{(1)}(y)\  \\
   &\rightarrow& \frac{1}{a} {\sqrt \frac{\hbar}{2k}}\
   e^{-ik\eta}\,,\qquad {\rm as}\quad  y \rightarrow \infty \,. \nonumber
\label{eq:BDmode}
\end{eqnarray}
For $\nu^2 <0$, $\nu$ is pure imaginary and Eq.~(\ref{eq:BDmode}) is
independent of the choice of sign of ${{\rm Im}}(\nu)$. Note that
the asymptotic form for $y=\vert k\eta\vert \rightarrow \infty$,
holds independently of the value of $\nu$. From the subleading terms
in this asymptotic expansion of the Hankel function
$H_{\nu}^{(1)}$ for large values of its argument, it is
straightforward to show that the BD state (with $n_k = 0$) is an
adiabatic order four UV allowed RW state for any $\nu$.
Hence taking $v_k = \phi_k^{BD}$ is allowed and the adiabatic
order four subtractions of Eqs.~(\ref{eq:vrenorm}) yield a UV
finite vacuum energy-momentum tensor expectation value in the BD
state, which satisfies $T_v = - 4 \varepsilon_v$ or $p_v = -
\varepsilon_v$, as a consequence of the de Sitter invariance of
this state. The calculation of the renormalized $\varepsilon_v$ as
a function of the parameters $m, H, \xi$ is given in
Refs.~\cite{d-c,b-d,BerFol}.

The massless minimally coupled field is of particular interest both
because slow roll inflationary models rely on such a field, and
because it obeys the same mode equation in a RW spacetime as gravitons
in a certain gauge~\cite{grischuk}. For this field $m=\xi = 0$, $\nu =
3/2$ and Eq.~(\ref{eq:BDmode}) becomes
\begin{equation}
\phi_k^{BD}\raisebox{-1ex}{\Big\vert}_{\stackrel{\scriptstyle m = 0}
{\scriptstyle \xi=0\ }} = H {\sqrt\frac{\hbar}{2k^3}}\,
e^{-ik\eta}\,(i - k\eta)\,.
\label{eq:mmc}
\end{equation}
Although this state is perfectly UV finite, an {\it infrared}
divergence occurs in the two-point function, Eq.~(\ref{eq:wight}).
The BD state must therefore be modified at very small values of
$k$~\cite{ford1} which means that, strictly speaking, it is
not possible to take $v_k = \phi_k^{BD}$ for all $k$. An IR finite
vacuum state is the Allen-Folacci (AF) state~\cite{a-f}, which is
actually a family of IR finite states. Because these states are not de
Sitter invariant, the energy-momentum tensor for the massless
minimally coupled scalar field in any of these states is also not
de Sitter invariant~\cite{a-f,k-g}. Nevertheless it has been
proven~\cite{desitter} that the energy-momentum tensor of the
$m=0$, $\xi =0$ scalar field for any of the AF states and
indeed for any UV finite, homogeneous, isotropic
state, asymptotically approaches the de Sitter invariant
energy-momentum tensor found by Allen and Folacci~\cite{a-f},
namely, $p_v=-\varepsilon_v = 119\, \hbar H^4/960 \pi^2$.

Since an AF state is just the BD state modified at very small values
of $k$, the short distance or UV properties of the AF and BD states
are identical.  In this paper we are only concerned with these short
distance effects, so we take as our preferred vacuum state $v_k =
\phi_k^{BD}$, even in the $m=\xi=0$ case, ignoring the infrared
divergences in the two point function which this generates. This is
not a problem for the power spectrum provided that the $k\approx 0$
modes are not the ones that dominate today.  There are no infrared
divergences in the energy-momentum tensor of the BD state and the
finite difference in $\langle T_{ab}\rangle_R$ between the BD state
and any realistic AF state are of order $H^4$ and small if
$H \ll M_{\rm Pl}$. Hence this distinction will play no role
in our analysis of short distance modifications of the initial state.

With $v_k = \phi_k^{BD}$ the power spectrum for a general
UV allowed pure state is given by~(\ref{eq:powerv})
becomes
\begin{equation}
P_{\phi}(k;t)\raisebox{-1ex}{\Big\vert}_{n_k = 0} = P_{\phi}^{\rm
BD} + \frac{k^3}{\pi^2} \Big(\vert B_k \vert^2 \vert
\phi_k^{BD}\vert^2 + {\rm Re} [A_k B_k^* \, (\phi_k^{BD})^2]\Big)
\label{eq:powpert}
\end{equation}
where
\begin{equation}
P_{\phi}^{\rm BD} = \frac{k^3}{2\pi^2} \vert \phi_k^{BD}\vert^2 =
\hbar \left( \frac{H}{2 \pi}\right)^2 \left(1 + k^2 \eta^2\right)
\,,
\label{eq:PBD}
\end{equation}
is the spectrum of the Bunch-Davies state for the massless, minimally
coupled field.  In the late time limit, $\eta \rightarrow 0^-$,
\begin{equation}
P_{\phi}^{\rm BD} \rightarrow \frac{\hbar H^2} {4 \pi^2} \,,
\label{eq:PBDlate}
\end{equation}
the BD power spectrum becomes completely independent of $k$, {\it
i.e.,} scale invariant. If one evaluates the power spectrum at the
time of horizon crossing instead, $k |\eta| = k/aH \sim 1$, one
also obtains a scale invariant spectrum with a normalization
differing slightly from~(\ref{eq:PBDlate}) by a constant factor of
order unity~\cite{SriPad,padmanabhan}.

The terms in~(\ref{eq:powpert}) dependent on $|B_k|^2$ and on $A_k
B_k^*$ are the contributions to the value of the power spectrum
for states different from the BD vacuum state. The first important
point to notice is that if the state is UV allowed then $|B_k|$
must approach zero and $P_{\phi}$ must approach
$P_{\phi}^{\rm BD}$ at large $k$. From this fact we can draw an
immediate conclusion, namely, if $P_{\phi}$ is evaluated at
horizon crossing, $k=Ha= e^{Ht}$, then {\it the power present in
any UV allowed initial state always reverts to its scale invariant
BD value for fluctuations with large enough $k$ which cross the
horizon at sufficiently late times}.  To make this statement more
quantitative suppose that the initial state is non-adiabatic up to
some physical scale $M$ at the initial time $t_0$ with $a(t_0)
\equiv a_0$, while above that scale is the same as the BD state.
The comoving wave number corresponding to this scale is $k_M =Ma_0$.
The horizon crossing time for a mode with this wave number is
\begin{equation}
t_M = H^{-1} \ln k_M = t_0 + H^{-1}\, \ln \left(\frac{M}{H}\right)
\,. \label{eq:tM}
\end{equation}
Fluctuations which leave the horizon at times $t> t_M$ will have
$k > k_M$ and the standard BD power spectrum, $P_{\phi}^{\rm BD}$.
Thus, if inflation goes on for longer than $\ln (M/H)$ e-foldings,
the initial state effects at the physical scale $M$ inflate to scales
far outside the horizon.  If the scales we observe in the CMB now
correspond to $k > k_M$, {\it i.e.,} to modes which left the
horizon of the de Sitter epoch at times $t>t_{M}$, then there will
be {\it no} imprint of the short distance initial state effects at
scale $M$ in the present day CMB observations. Conversely, if $k
\le k_M$ for the currently observable modes then the initial state
modifications of the spectrum at scale $M$ may be observable.
Taking the present horizon crossing scale to be of the order of
the present Hubble parameter $H_{now}$, this would imply that the
condition
\begin{equation}
M a_0/a_{now} \simeq  H_{now}
\label{eq:tuning}
\end{equation}
is satisfied.  This condition on $M a_0/a_{now}$ is a general
constraint on the present observability of any initial state
effects in inflation in the low energy EFT framework, regardless
of their short distance origin.
Additional constraints and
additional parameters may arise in any given inflationary model.
For example in the slow roll scenario the measured CMB power
spectrum depends on the slow roll parameter $\epsilon$ in Eq.
(\ref{eq:slowroll}), so that observational constraints on the
CMB power spectrum generally depend on more parameters than simply
those of the initial state of the scalar field.

A constraint which does not depend on other parameters of the
inflationary model is that arising from the energy-momentum tensor
of initial states different from the BD state. If these contributions
to the stress tensor are too large the model will deviate significantly
from de Sitter space and may not inflate at all. Specializing our
general results~(\ref{eq:etrw}) to the case of de Sitter space
with flat spatial sections and a scalar field that is massless and
minimally coupled, the relevant energy-momentum tensor components
are:
\begin{subequations}
\begin{eqnarray}
&&\varepsilon = \varepsilon_{\rm BD} + I_1 + I_2\,,\\
&&T = T_{\rm BD} + 2I_1 - 2I_2\,,\\
&&p = \frac{\varepsilon + T}{3} = p_{\rm BD} + I_1 -
\frac{I_2}{3}\,,
\end{eqnarray}
\label{eq:finT}
\end{subequations}
\hspace{-0.2cm}
where the finite state dependent integrals $I_1$ and $I_2$ for the case $n_k
=0,\,\sigma_k=1$ are
\begin{subequations}
\begin{eqnarray}
&& I_1 \equiv \frac{1}{2 \pi^2} \int_0^\infty \d k \; k^2
\Big(\vert B_k \vert^2 \vert \dot \phi_k^{\rm BD}  \vert^2
+ {\rm Re}[A_k B_k^* (\dot \phi_k^{\rm BD})^2]\Big)\,, \\
&& I_2 \equiv \frac{1}{2 \pi^2a^2} \int_0^\infty \d k \; k^4
\Big(\vert B_k \vert^2 \vert \phi_k^{\rm BD}  \vert^2 + {\rm Re}
[A_k B_k^* (\phi_k^{\rm BD})^2]\Big)\, .
\end{eqnarray}
\label{eq:I1I2}
\end{subequations}
\hspace{-0.25cm}
In the following subsections we consider various examples of
modifications of
the initial data for inflation and make use
of these general
expressions to compute the modified power
spectrum and
energy-momentum tensor components they generate.

\subsection{de Sitter Invariant $\alpha$ States}
\label{sec:alpha}

The BD state for the massive scalar field described by the mode
functions~(\ref{eq:BDmode}) is a special RW allowed state, since
it is invariant not only under spatial translations and rotations,
but also under the full $O(4,1)$ isometry group of globally
extended de Sitter spacetime~\cite{b-d}. For that reason it has
seemed the most natural analog of the Poincar\'e invariant
vacuum state of QFT in flat Minkowski spacetime, and is usually
assumed, explicitly or implicitly, to be the relevant ``vacuum"
state of the scalar field in inflationary models. However, as
maximally extended de Sitter spacetime is very different from
flat spacetime globally, global $O(4,1)$ invariance is a much
stronger condition than local flat space behavior.

Because it is a RW allowed state the BD state indeed has a
two-point correlation function $\langle \Phi(x) \Phi(x')\rangle$ with
short distance properties as $x \rightarrow x'$ that depend only on
the local geometry at $x$, and more specifically are of the Hadamard
form~\cite{BerFol}. Since $\langle \Phi(x) \Phi(x')\rangle$ is of mass
dimension two, this is equivalent to the statement that the BD mode
functions~(\ref{eq:BDmode}) possess an asymptotic expansion for large
$k$ which agrees with Eq.~(\ref{eq:adbmode}) and~(\ref{eq:adbtwo}) up
to second adiabatic order. However, {\it any UV allowed state} satisfies
this property. What is special about the
BD state is that its asymptotic
expansion for large $k$ agrees with
the adiabatic expansion~(\ref{eq:adbtwo})
to {\it all} orders. This
is a much stronger statement than that it goes over
to the local
Poincar\'e invariant vacuum state in the flat space limit, since
an
infinite order adiabatic state carries information about the geometry
of the background spacetime at {\it all} scales, including correlations
on causally disconnected scales much larger than that of the horizon
$H^{-1}$, a situation which has no analog in flat space.

The fourth order adiabatic condition on the state guarantees
that the stress tensor in that state possesses no new
divergences, and can be renormalized accordingly by the
standard local counterterms of the low energy EFT of gravity. None of
these RW allowed states are de Sitter invariant except the BD state.
Nevertheless, we showed in Ref.~\cite{desitter} that as a consequence
of the redshifting of short distance modes to large distances in de Sitter
space, all RW allowed initial states for Re $\nu < 3/2$ have a renormalized
$\langle T_{ab}\rangle$ which approaches the BD value at late times. All
such fourth order RW states are equivalent locally and are {\it a priori}
equally possible initial states for an inflationary model.

The only possible way to generalize the BD state while maintaining
de Sitter invariance, for Re $\nu < 3/2$, would be to require $|A_k|= |A|$
and $|B_k|=|B|$ be independent of $k$ and satisfy~(\ref{eq:bog})~\cite{c-t,Mot,All}.
Because of the unmeasurable overall phase of the Bogoliubov coefficients, we
can choose $A$ to be purely real and parameterize these squeezed states by a
single complex number, $z=re^{i\theta}$ as
\begin{subequations}
\begin{eqnarray}
&&A_k = \cosh r = \frac{1} {\sqrt{1 - |\lambda|^2}}
= \frac{1} {\sqrt{1 - e^{\alpha + \alpha^*}}}\,
\,,\\
&&B_k = e^{i\theta}\sinh r = \frac{\lambda} {\sqrt{1 - |\lambda|^2}}=
\frac{e^{\alpha}} {\sqrt{1 - e^{\alpha + \alpha^*}}}\,.
\end{eqnarray}
\label{eq:alpha}
\end{subequations}
\hspace{-0.2cm} The two alternate parameterizations shown in terms
of $\lambda$ and $ e^{\alpha}$ are sometimes
employed~\cite{c-t,Dan,Sus,inter}.

The Wightman function~(\ref{eq:wight}) for this general one complex
parameter family of squeezed states is de Sitter invariant, in the
sense that it is a function only of the $O(4,1)$ de Sitter invariant
distance between the points. However, as pointed out in Ref.~\cite{Jack}
none of these squeezed $\alpha$ states are truly invariant under the
de Sitter isometry group except for the BD state, $r=0$, since
the $r\neq 0$ states transform under $O(4,1)$ transformations by a
non-zero (in fact, infinite) phase. Since the Bogoliubov coefficient
$B_k$ in (\ref{eq:alpha}) does not approach zero at large $k$,
this class of $r$ or $\alpha$ states are {\it not} UV allowed states
unless $r \equiv 0$ identically. If Eqs.~(\ref{eq:alpha}) were taken
literally for all $k$, the integrals~(\ref{eq:I1I2})
and~(\ref{eq:finT}) would diverge {\it quartically}. Although
these states were used in various contexts, such as studying the
sign of backreaction effects of particle creation in de Sitter
space~\cite{Mot}, and have been reconsidered lately by several
authors~\cite{Dan,Sus,inter}, this severe UV divergence
is unacceptable for a physical initial state within the low energy
effective theory of gravity described by Eq.~(\ref{eq:effS}). This
is clear even at the level of field theory with no
self-interactions, provided only it is covariantly coupled to
gravity, since the stress tensor in the general $\alpha$ state has
divergences which depend on $\alpha$, and thus requires state
dependent counterterms for its renormalization~\cite{BerFol}. When
self-interactions are considered still other unphysical features
become manifest~\cite{inter}.

Computing the power spectrum for a general $(r, \theta)$ state by
using Eq.~(\ref{eq:powerphi}) with~(\ref{eq:alpha}), we obtain
\begin{subequations}
\label{eq:powalp}
\begin{eqnarray}
&& P_{\phi}(k; r, \theta) = P_{\phi}^{\rm BD} \left\{ 1 +
2\sinh^2r - \frac{\sinh 2r} { 1 + k^2\eta^2} \, {\rm Re}\Bigl( (1 +
ik\eta)^2 \, e^{-2ik\eta - i \theta}\Bigr) \right\} \,.
\end{eqnarray}
At late times, $\eta\rightarrow 0^-$ (with $k$ fixed)
\label{eq:powalpb}
\begin{eqnarray}
& & P_{\phi}(k; r, \theta) \rightarrow \frac{\hbar H^2}{4\pi^2}
\left( 1 + 2 \sinh^2 r - \sinh 2r\, \cos\theta  \right)\,.
\end{eqnarray}
\end{subequations}
Because of the non-adiabatic UV modification of the BD state by
$r$ at arbitrarily large $k$, the effects of this modification do
not redshift away, and essentially the same result is obtained if
Eq.~(\ref{eq:powalp}) is evaluated at horizon crossing time $\eta =
-1/k \rightarrow 0^-$ with $k\eta = -1$ fixed. The scale invariant
modification~(\ref{eq:powalpb}) is equivalent to that found in
Ref.~\cite{Dan} with the choice,
\begin{subequations}
\begin{eqnarray}
&&\sinh r = \frac{H}{2M}\, ,\\
&&\theta = \frac{\pi}{2} - \frac{2M}{H} -\tan^{-1}\left(\frac{H}{2M}\right)\,,
\end{eqnarray}
\label{eq:alpdan}
\end{subequations}
which gives
\begin{equation}
P_{\phi}= P_{\phi}^{\rm BD} \left\{ 1 -
\frac{H}{M} \sin \left(\frac{2M}{H}\right) + \frac{H^2}{M^2}
\sin^2\left(\frac{M}{H}\right)\right\} \simeq
P_{\phi}^{\rm BD} \left\{ 1 - \frac{H}{M} \sin \left(\frac{2M}{H}\right)\right\},
\label{eq:powerdan}
\end{equation}
where the last approximate equality holds if $H \ll M$, with $M$
the physical UV scale of new physics, denoted by $\Lambda$
in~\cite{Dan}. In the case of exact de Sitter invariance the
general de Sitter invariant squeezed state gives a simple
multiplicative correction to $P_{\phi}^{\rm BD}$ sinusoidally
varying with $H/M$, which is itself unobservable in the CMB,
since it has no $k$ dependence. Hence it could be interpreted as a
redefinition of the inflation scale $H$, with no observable consequences.

More importantly, the $k$ independence of the strictly de Sitter
invariant $r$ and $\theta$ state is untenable in the EFT framework,
since it leads to state dependent divergences in the energy-momentum
tensor. These are avoided if and only if the Bogoliubov coefficient
$B_k$ approaches zero fast enough at large $k$, for the state to be a
UV allowed state.  Thus one could consider a cutoff version of the
$(r,\theta)$ states in which $r=0$ above some large but finite
comoving cutoff,
\begin{equation}
k_M = M\, a_0\,,
\end{equation}
with $M$ the physical cutoff (in units of inverse length) at some
arbitrary initial time $t_0$. One can then assume that modes with
$k>k_M$ are in the adiabatic BD state while modes with smaller values of
$k$ are in an $(r, \theta)$ state. It is clear that such a state is
no longer de Sitter invariant and has a power spectrum identical
to~(\ref{eq:powalp}) or~(\ref{eq:powerdan}) for $k <k_M$, but
reverting back to its BD value for $k>k_M$. Thus in such a state,
\begin{equation}
P_{\phi} = P_{\phi}^{\rm BD}\left\{ 1 + \theta (k_M - k)\, \left[
2 \sinh^2r - \sinh 2r \cos \theta \right]\right\}\,,
\label{eq:powcut}
\end{equation}
instead of~(\ref{eq:powalp}). There is now a sharp break in the
power spectrum, which could be observable in principle, if we are
fortunate enough to have access to the right values of $k\sim k_M$
in the present CMB. {\it If}
we assume that this condition is satisfied
by the wavenumber of the
  present CMB, then observations would put a constraint
on the
magnitude of the deviations from scale invariance of the spectrum
of the form (\ref{eq:powcut}).
However, since this constraint is
model dependent in any given inflationary model, we do not consider
it further, and turn instead to the constraints arising from
the contributions of such a cutoff $r$ state to the energy-momentum tensor
during the de Sitter phase.

It is clear from
the divergence of the energy-momentum tensor at infinite
$k_M$
that the dominant contribution to the integrals in~(\ref{eq:I1I2})
comes from those modes close to the UV cutoff.
Substituting~(\ref{eq:alpha})
into~(\ref{eq:I1I2}) and cutting the
integrals off at $k_M=Ma_0$, gives
\begin{subequations}
\begin{eqnarray}
&&I_1=
\frac{\hbar M^4}{16 \pi^2}\,\left(\frac{a_0}{a}\right)^4\,\sinh^2 r
- \frac{\hbar M^4}{64 \pi^2}\,\left(\frac{a_0}{a}\right)^4\,\sinh 2r\,
F_{\theta}^{(3)}(x)\Big\vert_{x = \frac{Ma_0}{Ha}}\,\\
&&I_2=
\frac{\hbar M^4}{16 \pi^2}\,
\left[ \left(\frac{a_0}{a}\right)^4 + \frac{2H^2}{M^2}\,
\left(\frac{a_0}{a}\right)^2\right] \,\sinh^2 r \nonumber\\
&& \qquad - \ \frac{\hbar M^4}{64\pi^2}\left(\frac{a_0}{a}\right)^4
\,\sinh 2r \, \left[F_{\theta}^{(3)}(x) -
\frac{4}{x}\,F_{\theta}^{(2)}(x) +
\frac{4}{x^2}\, F_{\theta}^{(1)}(x)\right]_{x = \frac{Ma_0}{Ha}}
\end{eqnarray}
\label{eq:I1I2alp}
\end{subequations}
\hspace{-0.2cm} with
\begin{equation}
F^{(p)}_{\theta} (x) \equiv \frac{\partial^p}{\partial x^p}
\left(\frac{\sin x\,\sin(x-\theta)}{x}\right)\,.
\label{eq:Fpdef}
\end{equation}
\hspace{-0.2cm}
The properties of the functions $F^{(p)}_{\theta}(x)$ are discussed in
Appendix~\ref{app:int}.  All contributions to Eqs.~(\ref{eq:I1I2alp})
are finite (for finite $M$) at all times, including the initial
time. All terms redshift with the expansion at least as rapidly as
$a^{-2}$, in accordance with our general theorem in
Ref.~\cite{desitter}. The terms involving $F^{(p)}_{\theta}$ are
rapidly oscillating for early times, $Ma_0 \gg Ha$, but redshift to
zero as fast or faster than the non-oscillatory terms for late times,
$Ma_0 \ll Ha$. The transition from the oscillatory to damping behavior
occurs at a time when $Ma_0 \sim Ha$ which is of the same order
parametrically in $M/H$ as $t_{M}$, defined in Eq.~(\ref{eq:tM}). By
that time all the oscillatory terms give contributions to $I_1$ and
$I_2$ which are already of order $H^4$ and negligible.

Since the maximum of $F^{(p)}_{\theta}(x)$ is of order unity, at an
$x$ of order unity, while $\vert F^{(p)}_{\theta}(x)\vert$ is bounded
by $1/x$ as $x\rightarrow \infty$, the oscillating terms are never larger
parametrically than $HM^3$, while the non-oscillating terms make a
maximum contribution to the energy density or the pressure of order,
\begin{equation}
\frac{\hbar M^4}{16\pi^2}\,\sinh^2 r
\label{eq:Mbound}
\end{equation}
at the initial time, $a = a_0$. Comparing this with the energy density
of the inflaton field at the onset of inflation, $3\hbar H^2 M_{\rm Pl}^2
/8\pi$ and requiring that the backreaction from the additional
terms~(\ref{eq:I1I2alp}) be smaller gives the bound,
\begin{equation}
\sinh r < \sqrt{6\pi}\, \frac{H}{M} \frac{M_{\rm Pl}}{M}\,.
\label{eq:alpbound}
\end{equation}
It is possible for the right side of this inequality to be larger than
unity, even for $H \ll M$. Hence $\sinh r$ could be quite large and
the break in the power spectrum~(\ref{eq:powcut}) large enough to be
observable, without creating too large a backreaction. This kind of a
non-adiabatic initial state modification of the BD state at short
distances produces the largest effects in the power
spectrum~(\ref{eq:powcut})
of the CMB without giving an unacceptably
large backreaction during inflation.

If on the other hand $r$ is assumed small, as for example in~(\ref{eq:alpdan}),
then no term in~(\ref{eq:I1I2alp})
is larger in magnitude than
\begin{equation}
\frac{\hbar H^2M^2}{16\pi^2}\,,
\label{eq:Ibound}
\end{equation}
and we obtain only the weaker condition,
\begin{equation}
M < \sqrt{6\pi} M_{\rm Pl}\,.
\label{eq:MPbound}
\end{equation}
In this case the effects on the power spectrum would be small,
though observable in principle, and the physical momentum scale of the
cutoff $k_M/a_0=M$ at the onset of inflation need only be somewhat smaller
than the Planck scale, beyond which there would be no justification for
using the low energy effective action for gravity~(\ref{eq:effS})
in any case.

Summarizing, the $r$ states do not match the adiabatic expansion of
the mode functions or the energy-momentum tensor at any $k$ for which
$r \neq 0$. They are therefore completely non-adiabatic states. For
that very reason there is no bound on the size of their
effects on the CMB power spectrum for $k$ below the cutoff scale
$k_M$. Observations of the CMB may provide the strongest constraints
on this kind of initial state modification, but the quantitative bound
depends on the inflationary model. The only model independent constraint
for these non-adiabatic modifications of the initial state comes from the
magnitude of the backreaction produced, which is a relatively weak
  constraint,
giving Eq.~(\ref{eq:alpbound}) or~(\ref{eq:MPbound}). The largest power of the
physical cutoff $M$allowed by dimensional analysis appears
in the stress tensor, {\it i.e.,} $M^4$ in~(\ref{eq:Mbound}) for these
  states.
To illustrate how these results change if adiabatic conditions
are imposed on
the initial state, we consider next zeroth order adiabatic
states.

\subsection{Adiabatic Order Zero States}
\label{sec:dan}

A state of given adiabatic order can be obtained by first substituting
the expansion~(\ref{eq:adbtwo}) into~(\ref{eq:adbmode}) and expanding
to that adiabatic order.  The result, evaluated at some arbitrary time
$t_0$ serves as the initial condition for the exact modes $\phi_k$.
These modes will remain adiabatic to this order for all
time~\cite{birrell-davies}.

A zeroth order adiabatic vacuum state for the massless minimally
coupled scalar field can be obtained by setting $\Omega_k =
\omega_k$ in Eq.~(\ref{eq:adbmode}) and omitting terms
proportional to $\dot{a}$ in the resulting expression for
$\dot{\phi}_k$ since they are of first adiabatic order.  At the
time $t_0$ one has then
\begin{subequations}
\begin{eqnarray}
\phi_k  (0)&=& - H\eta_0\, {\sqrt\frac{\hbar}{2k}}\, e^{-ik\eta_0}\,,\\
\dot \phi_k (0) &=& \dot v_k(0) = - iH^2\eta_0^2 \, {\sqrt
\frac{\hbar k}{2}}\, e^{-ik\eta_0}\,.
\end{eqnarray}
\label{eq:zerophi}
\end{subequations}
Substituting into~(\ref{eq:abk}) with $v_k = \phi_k^{\rm BD}$
gives:
\begin{subequations}
\begin{eqnarray}
A_k &=& \left(1 + \frac{i}{2k\eta_0}\right)\, ,\\
B_k &=& \frac{i} {2 k \eta_0}e^{-2ik\eta_0}\,.
\end{eqnarray}
\label{eq:ABzero}
\end{subequations}
If the relations~(\ref{eq:ABzero}) are substituted into
Eq.~(\ref{eq:powpert}), one finds the late time power spectrum,
\begin{equation}
P_{\phi}(k;\eta\rightarrow 0^-)=
P_{\phi}^{\rm BD} \left[ 1 - \frac{\sin (2k\eta_0)}{k\eta_0}
+ \frac{\sin^2(k\eta_0)}{k^2\eta_0^2}\right] \; .
\label{eq:power-zero-general}
\end{equation}
In this case the initial state effects produce sinusoidal
modulations of the power spectrum with wave number which vanish as
$k \rightarrow \infty$. Also note that if the adiabatic order zero
initial condition is taken in the infinite past, $\eta_0 \rightarrow
-\infty$, the modifications vanish as well. This is because in
that limit the adiabatic order zero initial state becomes the BD
state with $A_k=1$, $B_k=0$, and the spectrum reverts to the
standard BD value.

Danielsson~\cite{DanBer} has considered initial data which are of the
same form as Eqs.~(\ref{eq:zerophi}) and~(\ref{eq:ABzero}), but
despite this apparently adiabatic construction, rather than viewing
$\eta_0$ as a fixed time Cauchy surface, where initial conditions are
imposed on the state for all $k$, he takes $\eta_0$ to depend on $k$
in such a way that $k\eta_0 = -M/H$, with $M$ a fixed physical
scale. The motivation seems to have been to avoid making any statement
about modes whose physical wavelength is shorter than the cutoff
$M^{-1}$, and indeed $\eta_0(k) = -M/(Hk)$ is the conformal time at
which the mode with comoving wave number $k$ first falls below the
physical cutoff $M$. However, inspection of~(\ref{eq:ABzero}) with
this substitution, shows that $|B_k|$ now behaves as a constant,
$H/2M$, as $k\rightarrow \infty$. Hence this prescription yields a
state which is {\it not} adiabatic at all, but amounts to populating
the highest frequency modes considered with a constant particle
occupation number, and choosing a cutoff $(r,\theta)$ state with
parameters given by~(\ref{eq:alpdan}). Thus the results of the
previous subsection apply. These initial conditions taken literally
for all $k$ lead to an energy-momentum tensor which is quartically
dependent on the cutoff, just as in the previous subsection. As
discussed there, this is not a physically allowed UV state if extended
to arbitrarily large $k$, {\it i.e.,} arbitrarily late times
$\eta_0(k) \rightarrow 0^-$.

This example illustrates the shortcomings of considering modifications
of the initial state of inflation and their effects on the CMB power
spectrum alone, without also considering the associated effects on the
energy-momentum tensor and backreaction. When one considers only the
power spectrum for some finite range of $k$, it may seem perfectly
reasonable to restrict attention to only those modes with a physical
wavelength larger than the short distance cutoff scale $M^{-1}$, since
no sum or integral over $k$ is required for the power
spectrum. However, the stress tensor does require such a sum over all
$k$, and some prescription for the ultra short distance modes has to
be given as these modes will redshift to wavelengths larger than the
short distance cutoff at later times. The essential question is not at
what time $\eta_0$ these UV modes have physical wavelengths larger
than $M^{-1}$, but rather what contribution do these short distance
modes make to the energy-momentum tensor, which involves a
sum/integration over {\it all} $k$, at {\it any} time. This question
requires that a choice be made about whether the stress tensor is to
be a consistent source for Einstein's equations and only if it is, can
the magnitude of the backreaction be reliably estimated. General
covariance of semi-classical gravity requires the state to be adiabatic
at the very highest trans-Planckian energies, and this adiabaticity
condition in turn constrains the possible effects of short distance
initial state modifications on the power spectrum at late times, which
might otherwise be overlooked.

Instead of taking $\eta_0$ to be a function of $k$ let us assume that
all the modes are determined at the same arbitrary but {\it fixed}
time $\eta_0$, independent of $k$, by Eq.~(\ref{eq:zerophi}).  This
defines a true adiabatic order zero state. Although $B_k$ given
by~(\ref{eq:ABzero}) now does decrease with increasing $k$, its
magnitude still does not fall off fast enough to make the state fourth
order adiabatic and UV allowed. Since $k^4 |B_k| \sim k^3$ as
$k \rightarrow \infty$, the energy-momentum tensor can depend as much
as
cubically on the comoving momentum cutoff of the mode sum. The
cubic
divergence in the state dependent mode sum means that there is
no local
(state-independent) counterterm available to absorb this
divergence. The
necessity of imposing a physical cutoff on the behavior of~(\ref{eq:ABzero})
implies that the power
spectrum~(\ref{eq:power-zero-general}) cannot be valid
for arbitrarily
large $k$ either, but instead must approach the BD spectrum
more
rapidly than~(\ref{eq:power-zero-general}) as $k\rightarrow
\infty$. If we insert a cutoff $k_M$, as in the previous subsection
such that the modes are the BD modes for $k > k_M$, then the
condition~(\ref{eq:tuning}) is necessary for these initial state
modifications to be observable in the CMB today.

If Eqs.~(\ref{eq:ABzero}) are substituted into Eqs.~(\ref{eq:I1I2})
with a cutoff $k_M$ one finds that
\begin{subequations}
\begin{eqnarray}
&&I_1 = \frac{\hbar H M^3}{32\pi^2} \left(\frac{a_0}{a}\right)^4
\left[ F^{(2)}(x) + \frac{H}{M}
-  \frac{H}{M} F^{(1)}(x)\right]_{x=k_M(\eta-\eta_0)}\, ,\\
&&I_2 = -\frac{\hbar H M^3}{32\pi^2} \left(\frac{a_0}{a}\right)^4
\left[F^{(2)}(x) - \frac{H}{M} - \frac{H}{M} F^{(1)}(x)
+ \frac{4H}{M} \left(\frac{a}{a_0}\right) F^{(1)}(x)\right.\nonumber\\
&&\qquad \left.+  \frac{4H^2}{M^2} \left(\frac{a}{a_0}\right)^2 F(x)
- \frac{4H^2}{M^2} \left(\frac{a}{a_0}\right) F(x)
- \frac{4H^3}{M^3} \left(\frac{a}{a_0}\right)^2 F^{(-1)}(x)\right]_{x=k_M(\eta-\eta_0)}\,.
\end{eqnarray}
\label{eq:I12adb0}
\end{subequations}
\hspace{-0.25cm} Here $F^{(p)} \equiv F^{(p)}_0$ with the latter
defined in Eq.~(\ref{eq:Fpdef}), $F \equiv F^{(0)}$, and
\begin{equation}
F^{(-1)}(x) \equiv \int_0^x \,\d y \, F(y) = \int_0^x\, \d y
\, \frac{\sin^2 y}{y}\,.
\label{eq:Fidef}
\end{equation}
As expected the effects of the state dependent terms
redshift away like $a^{-2}$ and $a^{-4}$, and are largest at or
near the initial time $\eta=\eta_0$ when they are of order,
\begin{equation}
\frac{\hbar H M^3}{32\pi^2}\, .
\label{eq:zeroener}
\end{equation}
Requiring this to be less than the energy density of the inflaton
field gives the bound,
\begin{equation}
M < \left(12\pi H M_{\rm Pl}^2\right)^{\frac{1}{3}}\,,
\label{eq:zerobound}
\end{equation}
for the adiabatic order zero state, in place of~(\ref{eq:alpbound})
for the non-adiabatic state.

It is clear that Eq.~(\ref{eq:zeroener}), softer by one power of $H/M$ compared to
the previous case~(\ref{eq:Mbound}) is the result of the fact that
the adiabatic order zero state has a Bogoliubov coefficient, $|B_k|$,
which approaches zero at large $k$ with one power of $1/k$ in~(\ref{eq:ABzero}).
If we had chosen a state which matches the adiabatic vacuum mode $v_k$ to first,
second, or third order, {\it i.e.,} with Bogoliubov coefficient
$|B_k|$ approaching zero at large $k$ like $k^{-2}$, $k^{-3}$ or
$k^{-4}$, respectively, then we should expect to obtain leading
contributions to the stress tensor components that behave like
$H^2M^2$, $H^3M$ or $H^4 \ln(M/H)$, respectively for large $M/H$. When
the state is a UV allowed state, the stress tensor
components are independent
of the upper limit $k_M = M a_0$ of the
  mode integrals for large $M$, so that
the integral may be extended to
infinity. In that case the stress tensor components
are of order $H^4$,
independent of the cutoff $M$, and negligible compared to the
energy
density driving the inflation for all $H \ll M_{\rm Pl}$.

\subsection{Boundary action approach}
\label{sec:boundary}

The authors of Refs.~\cite{Bound} have discussed setting
conditions of the form,
\begin{equation}
\left(\partial_t \phi_k + \kappa \phi_k\right)\big\vert_{t=t_0} = 0
\, ,
\label{eq:boundk}
\end{equation}
on the initial state mode functions, motivated by the addition
of boundary terms to the low energy EFT action functional.
Here $\kappa$ is in general a complex function of $k$.
For the BD state,
\begin{eqnarray}
\kappa_{\rm BD} &=& - \frac{\dot \phi_k^{\rm BD}(t_0)}
{\phi_k^{\rm BD}(t_0)}= \frac{Hk^2\eta_0^2}{1 + ik\eta_0}
\label{eq:kappav}
\end{eqnarray}
becomes purely imaginary in the limit $k\vert \eta_0 \vert
\rightarrow \infty$. If we make use of~(\ref{eq:abk}) we find that
the Bogoliubov coefficients are given by
\begin{subequations}
\begin{eqnarray}
A_k &=& \frac{-i(\kappa - \kappa_{\rm BD}^*)} {2 \sqrt{({\rm Im}\,
\kappa)\, ({\rm Im\,} \kappa_{\rm BD})}}
\frac {\phi_k^{\rm BD}(0)^*}{|\phi_k^{\rm BD}(0)|}\, ,\\
B_k &=& \frac{i(\kappa - \kappa_{\rm BD})} {2 \sqrt{({\rm Im}\,
\kappa)\, ({\rm Im}\, \kappa_{\rm BD})}} \frac {\phi_k^{\rm
BD}(0)}{|\phi_k^{\rm BD}(0)|}\,,
\end{eqnarray}
\label{eq:abound}
\end{subequations}
\hspace{-0.25cm}
up to an overall phase.

If attention is restricted to modifications of the BD state
corresponding to the lowest dimension local operator in the scalar EFT
on the initial time surface at $t=t_0$, namely $\beta (\nabla_i
\Phi)^2/M$ where $M$ is again the physical cutoff scale, then the
authors of Refs.~\cite{Bound} argue that this would lead to a modified
initial condition of the form~(\ref{eq:boundk}) with
\begin{equation}
\kappa = \kappa_{\rm BD} + \frac{\beta k^2}{a_0^2 M}\,.
\end{equation}
If the effective action on the boundary is real for real time it
would seem that $\beta$ must be real. The authors of
Refs.~\cite{Bound} treat $\beta$ as a real parameter, obtaining
corrections to the (real) power spectrum which are linear in
$\beta$. Treating $\beta$ as an arbitrary complex parameter, we
obtain the Bogoliubov coefficients for the case $k \vert \eta_0
\vert \gg 1$,
\begin{subequations}
\begin{eqnarray}
A_kB_k^* &\simeq & -i\beta^*\frac{k} {2Ma_0} e^{2ik\eta_0}
\left(1 - \frac{i\beta k}{2Ma_0}\right)
\left(1 + \frac{k}{Ma_0} {\rm Im}\beta\right)^{-1}\, ,\\
|B_k|^2 &\simeq& \frac{|\beta|^2 k^2}{4M^2a_0^2} \left(1 +
\frac{k}{Ma_0} {\rm Im}\beta\right)^{-1} \,.
\end{eqnarray}
\label{eq:abbeta}
\end{subequations}
\hspace{-0.25cm} As in the $\alpha$ state case discussed in
Sec.~\ref{sec:alpha}, these Bogoliubov coefficients are non-adiabatic
and would lead to a divergent stress tensor if continued to
arbitrarily large $k$. Substituting the cutoff $k_M = M a_0$ into the
power spectrum~(\ref{eq:powpert}) as before, we obtain at late times,
\begin{equation}
P_{\phi}(k; \eta \rightarrow 0^-)=
P_{\phi}^{\rm BD} \left\{ 1  +
\frac{\theta(k_M - k)\,k}{\left(k_M + k\,{\rm Im}\beta \right)}
\left({\rm Re} (i\beta^* e^{2ik\eta_0})
+ |\beta|^2\,\frac{k}{k_M}\,\cos^2(k\eta_0)
\right)\right\} \, .
\label{eq:powbou}
\end{equation}
If $|\beta|$ is of order one then both terms in Eq.~(\ref{eq:powbou})
are of the same order at $k \simeq k_M$, and the (large) deviations from
a scale invariant spectrum may be observable in present CMB data.
The same remarks about fine tuning to $k \sim k_M$ and dependence
on the specific features of the inflationary model apply to
this initial state modifications as to the cutoff $r$ states
of Sec. \ref{sec:alpha}.

If $|\beta|$ is assumed to be much less than unity, we can write
$\beta = \vert \beta \vert e^{i
\gamma}$ and obtain the modified
power spectrum to linear order in
$\vert \beta \vert$,
\begin{equation}
P_{\phi}(k) \simeq P_{\phi}^{\rm BD} \left\{ 1  - |\beta|\
\theta(k_M - k)\ \frac{k}{k_M} \, \sin(2k\eta_0 - \gamma) + {\cal
O}(|\beta|^2)\right\}  \,.
\label{eq:powlin}
\end{equation}

\begin{figure}
\includegraphics[angle=90,height=4in,width=7in]{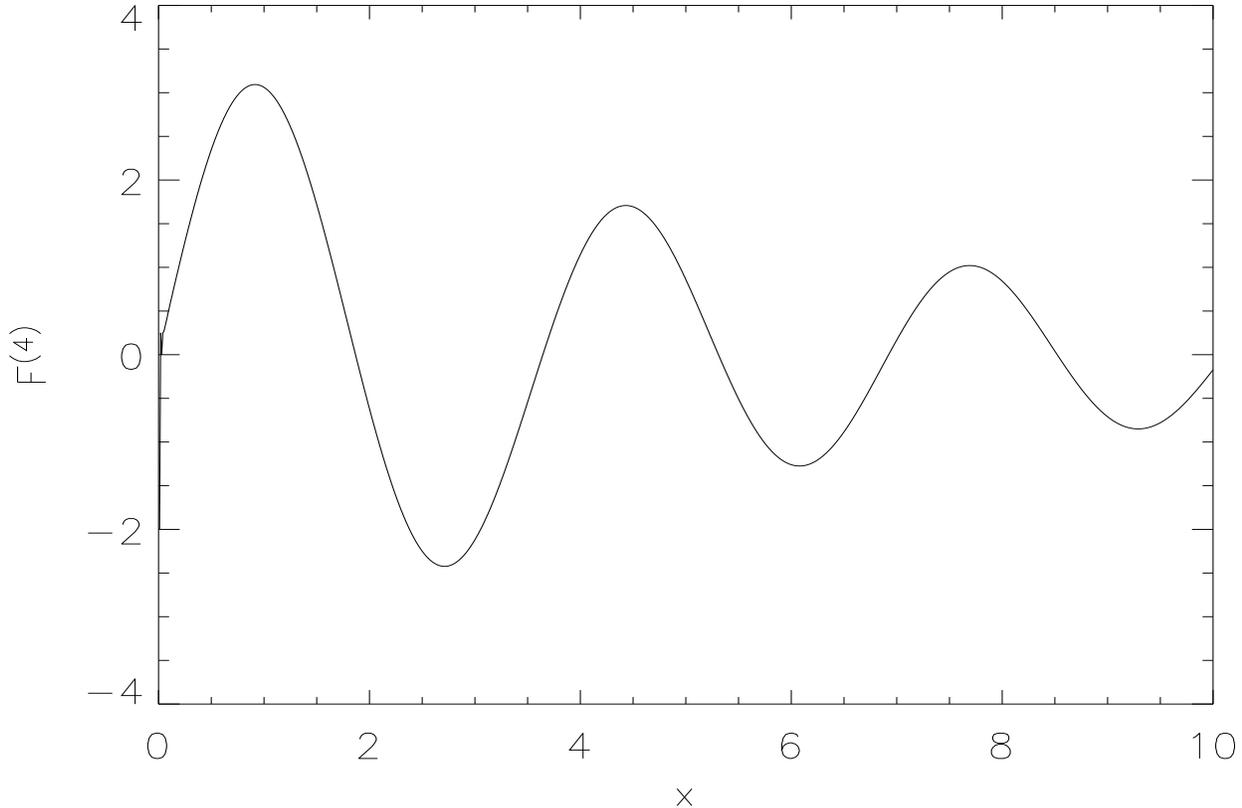}
\caption{The oscillatory function $F^{(p)}$ for $p=4$
defined by Eq.~(\ref{eq:Fp}), as a function of $x=k_M(\eta-\eta_0)$.}
\label{fig:F4}
\end{figure}

The finite state dependent contributions to the stress tensor given by
the integrals~(\ref{eq:I1I2}) are easily written down for the case of
general complex $\beta$, but because of the denominators
in~(\ref{eq:abbeta}) they are rather complicated.  In the case that
$\beta$ is purely real the integrals simplify and may be evaluated in
terms of the functions $F^{(p)} \equiv F^{(p)}_0$ with the latter
defined in Eq.~(\ref{eq:Fpdef}). The result is
\begin{subequations}
\begin{eqnarray}
&&I_1 = \frac{\hbar \beta M^4}{128\pi^2}\left(\frac{a_0}{a}\right)^4\,
F^{(4)}(x)\Big\vert_{x = k_M(\eta-\eta_0)}
+ \frac{\hbar \beta^2 M^4}{96\pi^2}\left(\frac{a_0}{a}\right)^4\,
\left[1 + \frac{3}{16} F^{(5)}(x)\right]_{x = k_M(\eta-\eta_0)} \\
&&I_2 = -\frac{\hbar \beta M^4}{128\pi^2}\left(\frac{a_0}{a}\right)^4\,
\left[F^{(4)}(x) + \frac{4H}{M}\frac{a}{a_0} F^{(3)}(x) +
\frac{4H^2}{M^2}\left(\frac{a}{a_0}\right)^2
F^{(2)}(x)\right]_{x = k_M(\eta-\eta_0)}
\nonumber\\
&& \qquad +\frac{\hbar \beta^2
M^4}{96\pi^2}\left(\frac{a_0}{a}\right)^4\, \left[1 +
\frac{3H^2}{2M^2}\left(\frac{a}{a_0}\right)^2 - \frac{3}{16}
F^{(5)}(x) - \frac{3H}{4M}\frac{a}{a_0} F^{(4)}(x) \right.
\nonumber \\
& &  \left. \qquad \qquad \qquad \qquad \qquad
-\frac{3H^2}{4M^2}\left(\frac{a}{a_0}\right)^2
F^{(3)}(x)\right]_{x = k_M(\eta-\eta_0)}\,.
\end{eqnarray}
\label{eq:stboundary}
\end{subequations}
\hspace{-0.25cm}
All terms are again finite at all times and redshift at
late times at least as rapidly as $a^{-2}$, in accordance with our
general theorem in Ref.~\cite{desitter}. The oscillatory integrals
$F^{(p)}(x)$ are discussed in Appendix B, and the illustrative
particular case of $F^{(4)}(x)$ is plotted in Fig.~\ref{fig:F4}.

Since the maximum of $F^{(p)}(x)$ is of order $3$ to $4$ at $x=k_M
(\eta - \eta_0) \sim 1$ (for $p$ even) or $x=0$ (for $p$ odd),
the maximum value of either the terms linear or quadratic in $\beta$
is of order,
\begin{equation}
{\rm max}(\beta, \beta^2)\ \frac{\hbar M^4}{32\pi^2}\,,
\label{eq:contrib}
\end{equation}
in contrast to~(\ref{eq:I1I2alp}), where the oscillatory terms
were smaller than the non-oscillatory ones.  If the maximum of the
contributions~(\ref{eq:contrib}) are required to be smaller than
the energy density driving the expansion, and $\beta \ll 1$ then
the terms linear in $\beta$ give the largest contribution and the
strongest bound, {\it viz.,}
\begin{equation}
\beta < 12\pi \left(\frac{H}{M} \right)^2
\left(\frac{M_{\rm Pl}}{M}\right)^2\,.
\label{eq:betabound}
\end{equation}
If $\beta < 1$ this bound on the term linear in $\beta$ is of the same
order of magnitude as that obtained in Ref.~\cite{Por}, but disagrees
with the bound in Ref.~\cite{Shnew}, whose authors argue that the
term linear in $\beta$ gives no bound on the bulk stress tensor away from
the initial boundary surface at $t=t_0$.

The authors of Ref.~\cite{Shnew} evaluate integrals such as $F^{(4)}$
with a Gaussian cutoff in $k$, rather than with a hard cutoff at
$k=k_M$ used here. However this by itself does not account for the
disagreement.  A Gaussian cutoff amounts to a slightly different
choice of initial state which is also perfectly UV allowed, since it
approaches the BD state faster than any power of $k$.  This different
state gives a
different finite contribution to the renormalized stress
tensor, which
is of the same order in $M$ as the contribution~(\ref{eq:contrib}),
differing only in its numerical coefficient at the initial time, and
with a smoother, less strongly oscillatory behavior in time than that
shown in Fig.~\ref{fig:F4} for the hard cutoff $k_M$. In neither case
does the contribution to the energy-momentum tensor fall off
exponentially in the conformal time difference $\eta-\eta_0$ as
claimed in Ref.~\cite{Shnew}. At late times $\eta \rightarrow 0^-$
either the Gaussian or hard cutoff of the initial state momentum
integral yields a finite stress tensor with components that fall off
as $a^{-4}$, $a^{-3}$ and $a^{-2}$, as expected by the redshift of the
RW expansion and in accordance with Ref.~\cite{desitter}.  Hence the
finite terms in the renormalized stress tensor which are first order
in $\beta$ are no more subject to renormalization ambiguities or
localized on the boundary than those which are second order in
$\beta$. Which terms give the stronger bound on $\beta$ depends
entirely on the values assumed for the parameters $k_M\eta_0$, $H$ and
$M$.

For general values of $\beta$, if the ratios $H/M$ and $M/M_{\rm Pl}$
are the same order of magnitude, then requiring that the maximum
contributions~(\ref{eq:contrib}) be smaller than the energy
density driving the expansion yields the bound,
\begin{equation}
{\rm max}(\beta, \beta^2)  < 10 \leftrightarrow 100 \;.
\end{equation}
If $\beta <1$ this bound is easily satisfied, while if $\beta >1$ the
maximum value comes from the terms quadratic in $\beta$, and the
results of Ref.~\cite{Shnew} are recovered. We conclude that although
the precise evaluation of the backreaction effects of initial state
modifications differs from the estimates given in either~\cite{Por}
or~\cite{Shnew}, the qualitative final conclusion that the
backreaction constraints are not a very severe restriction on the
parameter(s) of the boundary value action is similar to the conclusions
reached by these authors.

\section{Adiabatic Particle Creation and Dephasing}
\label{sec:adiabatic}

One of the principal physical effects
that can be described by the
semi-classical Gaussian density
matrix is particle creation by a time
varying RW scale
factor~\cite{GMM}. Although the definition of a
``particle" is
intrinsically non-unique in a time varying background,
it is possible
to use the adiabatic nature of the UV allowed RW states
to constrain
this non-uniqueness considerably. The adiabatic
particle concept was studied in some detail in Ref.~\cite{hm-pm},
where a proposal was made for the adiabatic basis. The deficiency
with that earlier proposal is that the two parts of the
stress-energy tensor corresponding to vacuum and particle
contributions are not separately conserved. Here we remedy that
defect and in the process remove almost all of the non-uniqueness
in the definition of adiabatic particle number.

Let us remark first that it is always possible to express the density
matrix $\hat \rho$ in the time {\it in}dependent number basis in which
the number operator $a^{\dagger}_{\bf k}a_{\rule{0mm}{2.3mm}\bf k}$
of Eq.~(\ref{eq:part}) is diagonal. The transformation to this particle
number basis may be derived by the methods of Refs.~\cite{BroCar,CKHM}
with the result,
\begin{equation}
\langle n\vert \hat \rho \vert n'\rangle =
\prod_{\bf k} \frac{2\, \delta_{n_{\bf k}^{\,}\,n'_{\bf k}}}{\sigma_k + 1}
\left(\frac{\sigma_k - 1}{\sigma_k + 1}\right)^{n_{\bf k}}\,,
\label{eq:tindp}
\end{equation}
where $n$ labels the set of integers $\{n_{\bf k}\}$, one for each
distinct ${\bf k}$. In this occupation number representation, the
density matrix is time independent and diagonal, since the $\sigma_k$
are constants of motion. The positive diagonal matrix elements of
$\hat \rho$ in this discrete number representation may be viewed as
the probabilities of finding exactly $n_{\bf k}$ particles in the mode
labelled by wave number $\bf k$, with the particle number basis
defined by the time independent operator, $\hat n_{\bf k} =
a^{\dagger}_{\bf k}a_{\rule{0mm}{2.3mm}\bf k}$, {\it i.e.,}
\begin{equation}
\langle a^{\dagger}_{\bf k}a_{\bf k}{\,}\rangle
= {\rm Tr}(\hat n_{\bf k}\hat\rho)
= \frac{2}{\sigma_k + 1}\sum_{n_{\bf k} = 0}^{\infty} \,n_{\bf k}\,
\left(\frac{\sigma_k - 1}{\sigma_k + 1}\right)^{n_{\bf k}}
=\frac{\sigma_k -1}{2}\,,
\end{equation}
which is equivalent to Eq.~(\ref{eq:part}), together
with~(\ref{eq:sigdef}).

Since the unitary transformation to the $n_{\bf k}$ basis
exactly undoes the action of the time evolution operator, the
preceding definition of particle number is always time independent, no
matter how rapidly the geometry changes with time.  Hence it carries
no information about particle creation in the time evolving RW
geometry, or indeed about any features of the time evolution of the
system whatsoever. Moreover, if one makes a Bogoliubov transformation
from one exact set of eigenfunctions $\phi_k (t)$ to another exact set
$v_k(t)$, then the particle number $N_k$ with respect to the new basis
remains exactly time independent, {\it cf.,}~(\ref{eq:numvac}).
Useful as these are in the description of the density matrix and
energy-momentum tensor of the field, any such time independent
parameterization of particle number is not the appropriate one to
describe particle creation in time varying RW cosmologies. For that
purpose one needs to define an appropriate time dependent number
operator with respect to an {\it approximate} adiabatic vacuum basis,
which agrees with the exact number basis only at very large $k$.

The definition of the approximate adiabatic basis is constrained by
several physical and practical requirements. First the basis should
be defined by a set of WKB-like approximate mode functions,
\begin{equation}
\tilde \phi_k = a^{-\frac{3}{2}}\tilde f_k =
{\sqrt\frac{\hbar}{2a^{3} W_k}}\,
\exp\left(-i \int^t\,\d t'\,W_k (t') - i\delta_k(t)\right)\,,
\label{eq:adbpar}
\end{equation}
similar in form to Eq.~(\ref{eq:adbmode}), for some adiabatic
frequency function $W_k$ and additional time dependent phase
$\delta_k(t)$, to be specified. If $W_k =\Omega_k$, with $\Omega_k$
a solution to Eq.~(\ref{eq:freq}), and $\delta_k$ is constant in
time, then $\tilde f_k$ would be an exact solution to the mode
equation~(\ref{eq:modefns}), and we would obtain the time independent
number basis~(\ref{eq:tindp}). However, there is no constraint on the
$\tilde \phi_k$ to satisfy the equation of motion~(\ref{eq:phikeq}),
except asymptotically for
large $k$, where they should approach the exact
mode functions. Hence
we have considerable freedom to choose the two functions,
$W_k(t)$ and
$\delta_k(t)$ which define the adiabatic basis. The most physically
meaningful
choice of basis is that which requires the vacuum energy-momentum
tensor defined
by $\tilde \phi_k$ to agree with the adiabatic expansion of the
vacuum terms in
the covariant energy-momentum tensor to a fixed adiabatic order.
It is this
tensor which couples to gravity and is covariantly conserved (as
distinguished
from the canonical Hamiltonian), so that matching the adiabatic
basis
(\ref{eq:adbpar}) to this tensor will guarantee separate conservation of
the
vacuum polarization and particle contributions to the stress-energy.

It is clear that the leading (adiabatic order zero) terms in
Eqs.~(\ref{eq:epsTtwo}) are vacuum polarization terms, and not
particle contributions, since they appear even in static or flat
spacetimes, where there can be no particle creation whatsoever.  Thus,
the frequency $W_k$ of the adiabatic modes~(\ref{eq:adbpar}) should
certainly match the zeroth order term $\omega_k$ in order for the
energy-momentum tensor of these modes to match the adiabatic order
zero terms in Eq.~(\ref{eq:epsTtwo}). However, adiabatic order zero
matching is not sufficient. If we do not require matching $\tilde \phi_k$
to at least {\it second} order in the adiabatic expansion of the vacuum
$v_k$, then we will find particle contributions to the energy-momentum
tensor which {\it diverge quadratically} in the UV cutoff $k_M$. These
divergent adiabatic order two terms are clearly part of
the state {\it in}dependent vacuum polarization contribution to
$\langle T_{ab}\rangle$, and not the particle content of the
state. Choosing the adiabatic frequency and phase in~(\ref{eq:adbpar})
to match the adiabatic order two expansion terms~(\ref{eq:epsTtwo})
{\it exactly} guarantees that they will be completely removed by
the subtractions in (\ref{eq:vrenorm}). It is this physical requirement that
the power law divergences in the conserved stress tensor components should be
associated with the state independent geometrical contributions and not the state
dependent particle content that determines the adiabatic basis functions
$\tilde \phi_k$ and renders the definition of adiabatic particle number
(almost) unique. The only remaining non-uniqueness of the particle definition
arises from the possibility of matching the vacuum contributions
to the
  stress tensor to higher than second order in the adiabatic expansion.

In order to define the adiabatic basis precisely let the exact mode functions
be expressed in terms of the adiabatic modes~(\ref{eq:adbpar}) as
\begin{subequations}
\begin{equation}
f_k =  \alpha_k(t) \tilde f_k + \beta_k(t) \tilde f_k^*\,,
\label{eq:fkalpbet}
\end{equation}
in terms of time varying Bogoliubov coefficients, $\alpha_k(t)$ and
$\beta_k(t)$. The time dependent Bogoliubov coefficients are required
to satisfy
\begin{equation}
|\alpha_k(t)|^2 - |\beta_k(t)|^2 = 1\,,
\label{eq:alpbetadb}
\end{equation}
for each $k$, in order to guarantee that the transformation of bases
is a canonical one. As in the case of the time independent change of
bases discussed in Section~\ref{sec:gen-init}, this allows a two real
parameter freedom in the choice of the $\tilde f_k$ for each $k$ (up
to an overall irrelevant phase). Rather than using the Bogoliubov
coefficients or $W_k$ and $\delta_k$ as the two independent parameters,
it is more convenient to use $W_k$ and a different independent function
of time $V_k(t)$, defined by the relation on the first time derivative of
Eq.~(\ref{eq:fkalpbet}), namely,
\begin{equation}
\dot{f}_k  = \left(- i W_k  + \frac{V_k}{2}\right)\alpha_k(t) \tilde f_k
+ \left(i W_k + \frac{V_k}{2}\right) \beta_k(t) \tilde f_k^*\,.
\end{equation}
\label{eq:timdep}
\end{subequations}
\hspace{-0.25cm}
The canonical transformation from the exact mode functions $f_k$ to
the approximate adiabatic functions $\tilde f_k$ is now completely
specified by $W_k$ and $V_k$, while $\delta_k$ of Eq.~(\ref{eq:adbpar})
is fixed (implicitly) in terms of these two. Before determining $W_k$
and $V_k$ explicitly from the stress tensor components we first compute the
particle number ${\cal N}_k$ and density matrix in the general adiabatic basis
determined by Eqs.~(\ref{eq:timdep}).

If the original field operator $\Phi(t,{\bf x})$ is expanded in terms of
the approximate adiabatic mode functions, $\tilde f_k$ and $\tilde
f_k^*$, the corresponding annihilation and creation operators,
\begin{subequations}
\begin{eqnarray}
\tilde a_{\rule{0mm}{2.3mm}\bf k} (t) &=& a_{\rule{0mm}{2.3mm}\bf k}
\alpha_{\rule{0mm}{2.3mm}k} (t)
+ a_{-\bf k}^\dag \beta_{\rule{0mm}{2mm}k}^*(t)\,,\\
\tilde a_{\bf k}^\dag (t)&=& a_{\bf k}^\dag \alpha_{\rule{0mm}{2mm}k}^*(t)
+ a_{-\rule{0mm}{2.3mm}\bf k} \beta_{\rule{0mm}{2.3mm}k}(t)\,,
\end{eqnarray}
\end{subequations}
are generally time dependent. Because of Eq.~(\ref{eq:alpbetadb}) and
the freedom to choose the phase of $\alpha_{\rule{0mm}{2.3mm}k}$ we
may set
\begin{subequations}
\begin{eqnarray}
\alpha_{\rule{0mm}{2.3mm}k} (t) &=&\cosh r_{\rule{0mm}{2.3mm}k} (t)\,,\\
\beta_{\rule{0mm}{2.3mm}k} (t) &=& e^{i\theta_k(t)}\,\sinh r_{\rule{0mm}{2.3mm}k} (t)\,,
\end{eqnarray}
\label{eq:adbsqu}
\end{subequations}
\hspace{-0.2cm}
in analogy with Eq.~(\ref{eq:squeeze}). With respect to this time dependent
adiabatic basis, the number density of particles in mode $k$ is
\begin{equation}
{\cal N}_k (t) \equiv \langle \tilde a_{\bf k}^{\dagger}
\tilde a_{\rule{0mm}{2.3mm}\bf k}\rangle = n_k + \sigma_k \,\vert \beta_k(t)\vert^2
= \sinh^2 r_k (t) + n_k \cosh 2r_k(t)\,,
\label{eq:adbnum}
\end{equation}
which also depends on time in general. However, if $W_k$ and $V_k$ are
chosen properly at large $k$, ${\cal N}_k$ will be an adiabatic
invariant with respect to the effective classical
Hamiltonian~(\ref{eq:effHam}), and therefore slowly varying in time
for a slowly varying RW scale factor. The number density of particles
spontaneously created from the vacuum with $n_k=0$ is $|\beta_k
(t)|^2$. An expression for this quantity in terms of $W_k$ and
$V_k$ is easily obtained from Eq.~(\ref{eq:timdep}) in the same way
as~(\ref{eq:abk}), with the result,
\begin{eqnarray}
\vert \beta_k (t) \vert^2 &=& \sinh^2r_k(t)=
\frac{1}{2\hbar W_k} \bigg\vert \dot f_k + \left( i W_k
-\frac{V_k}{2}\right) f_k \bigg\vert^2 \nonumber\\
&=& \frac{1}{4W_k\Omega_k}\left[ (W_k - \Omega_k)^2
+ \frac{1}{4} \left(V_k + \frac{\dot \Omega_k}{\Omega_k}\right)^2
\right]\, ,
\label{eq:betak}
\end{eqnarray}
where the last expression follows from inserting the
WKB-form~(\ref{eq:adbmode}) for the exact mode function $f_k$ in terms
of the {\it exact} frequency $\Omega_k$. From this we observe that if
(and only if) the adiabatic frequency matches the exact frequency,
$W_k = \Omega_k$ and $V_k = - \dot\Omega_k/\Omega_k$, then the
Bogoliubov coefficient $\beta_k$ vanishes identically.  In that case
there is no particle creation and ${\cal N}_k=n_k$ is strictly a
constant of the motion.

Using the value of $\vert \beta_k \vert^2$ the density matrix may
be expressed in the adiabatic particle basis, by the methods
of Refs.~\cite{BroCar,CKHM}. Unlike Eq.~(\ref{eq:tindp}) the
off-diagonal matrix elements of $\hat \rho$ do not vanish in this
basis, and are quite complicated in the general
case~\cite{BroCar,CKHM}. Although the diagonal elements are time
dependent, they depend on time only through the function $r_k(t)$.
This means that they are relatively much more slowly varying than
the corresponding phase variables $\theta_k(t)$, upon which the
off-diagonal elements of $\hat \rho$ also depend. Thus, in this
adiabatic number basis it becomes possible to argue that particle
creation is related to phase decoherence or {\it dephasing} of the
state: since macroscopic observables are generally relatively
insensitive to the process of averaging over the rapidly varying
phases, one can replace the exact $\hat \rho$ in this basis by its
more slowly varying diagonal elements only~\cite{Kandrup}. In the
pure state case, with adiabatic vacuum conditions in the infinite
past, these
diagonal elements simplify and are given explicitly
by~\cite{CKHM}
\begin{equation}
\rho_{n_k = 2l_k}(k;t) =
\frac{(2l_k-1)!!}{2^{l_k}\,l_k!}\ {\rm sech}\, r_k \tanh^{2l_k}r_k\,.
\label{eq:pair}
\end{equation}
The diagonal matrix elements are non-vanishing only for even
integers, corresponding to the fact that particles are created from
the vacuum in pairs. The positive numbers~(\ref{eq:pair}) have the
interpretation of the probabilities of finding exactly $l_k$ pairs of
adiabatic particles at time $t$ in the mode labelled by $k= |{\bf k}|$,
with vacuum initial conditions as $t \rightarrow -\infty$. If this
replacement of the exact pure state density matrix~(\ref{eq:gaussd})
by its phase averaged diagonal elements is justified, then the
von Neumann entropy suffers the replacement,
\begin{equation}
S \equiv - {\rm Tr}\, (\hat \rho\,\ln\hat\rho) \rightarrow
- \frac{1}{2\pi^2}\,\int\,\d k\,k^2\,
\sum_{l_k = 0}^{\infty}\rho_{2l_k}(k;t)\,
\ln \rho_{2l_k}(k;t)\,,
\label{eq:vN}
\end{equation}
which becomes time dependent. Although, in general the effective
von Neumann entropy does not grow strictly monotonically in time,
starting in an initial pure state with all of the $r_k =0$ leads
to a larger effective entropy at late times when some of the modes
have $r_k \neq 0$~\cite{HuPavon,Kandrup,CKHP}. This shows that
particle creation is directly related to increased squeezing of
the initial state, and the growth of entropy this entails
corresponds to the effective loss of information resulting from
averaging over the rapidly varying phases $e^{\pm i\theta_k(t)}$
in macroscopic physical observables.

The validity of a truncation of the density matrix to its diagonal
terms only in the adiabatic number basis and the associated loss
of phase information will depend on the initial state and the
details of the evolution~\cite{Kandrup}. When particle creation
takes place from initial vacuum-like states this would seem to be
quite a good approximation in the cases that it has been tested
quantitatively~\cite{CKHM,CKHP}. Conversely, if one starts from a
different type of initial state the squeezing parameters need not
increase monotonically with time, and phase averaging is not
justified. If it should happen in some special case(s) that the
squeezing coefficients $r_k$ are constant in time for all $k$, so
that the adiabatic particle number basis becomes an exact vacuum basis,
then by making the appropriate time independent Bogoliubov
transformation~(\ref{eq:bog}) to that basis one can set all the $r_k=0$.
Then it is clear that the $\theta_k$ become undefined and no phase decoherence
of the initial state can occur by particle creation or dephasing
effects in $\theta_k$.

With these general remarks on the adiabatic particle number basis
and dephasing let us fix the still undetermined functions $W_k$ and $V_k$.
Following Ref.~\cite{hm-pm} let us replace the exact mode functions
$\phi_k$ by $\tilde \phi_k$, and their time derivatives $\dot \phi_k$ by
$\d\tilde \phi_k/\d t = (-iW_k + V_k/2 - 3H/2) \tilde\phi_k$ with
$\alpha_k=1$, $\beta_k=0$, and $n_k =0$ in Eqs.~(\ref{eq:unren}) for
the stress tensor components. Using~(\ref{eq:timdep})
we obtain the (cutoff dependent) adiabatic vacuum contributions in this
basis to the energy density and trace, namely,
\begin{subequations}
\begin{eqnarray}
&& \tilde\varepsilon= \frac{\hbar}{4 \pi^2 a^3}
\int [\d k] \,k^2\ \tilde\varepsilon_k\, ,\\
&& \tilde T = \frac{\hbar}{4 \pi^2 a^3} \int  [\d k] \,k^2\ \tilde
T_k\,,
\end{eqnarray}
\end{subequations}
with $\tilde\varepsilon_k$ and $\tilde T_k$ the following functions of $W_k$ and $V_k$,
\begin{subequations}
\begin{eqnarray}
&&\tilde\varepsilon_k = \frac{1} {2W_k}
\left[\omega^2_k + W_k^2 + \frac{\left(V_k - H\right)^2}{4}
+ (6 \xi -1) \left( H V_k - 2 H^2 + \frac{\epsilon}{a^2}\right)\right]\,,\\
&& \tilde T_k = \frac{1}{W_k} \left[-m^2 +
(6 \xi -1) \left(\omega^2_k - W_k^2 - \frac{V^2_k}{4} + \frac{3 H V_k}{2}
   - \frac{H^2}{4} + \dot H\right)
+ {(6 \xi -1)}^2\,
   \frac{R}{6}\right]\,.
\nonumber\\
\end{eqnarray}
\label{eq:vacadb}
\end{subequations}
\hspace{-0.35cm}
The adiabatic vacuum basis and particle number may be
fixed now by setting these two expressions in terms of $W_k$ and $V_k$
{\it precisely equal} to the corresponding state-independent
vacuum contributions to energy-momentum tensor in a general RW spacetime
to second adiabatic order, given previously by Eq.~(\ref{eq:epsTtwo}).
That is, we require
\begin{subequations}
\begin{eqnarray}
&& \tilde\varepsilon_k = \varepsilon_k^{(2)}\,,
\label{eq:epsWV} \\
&& \tilde T_k = T_k^{(2)}\,,
\label{eq:TWV}
\end{eqnarray}
\label{eq:WV}
\end{subequations}
\hspace{-0.25cm}
thus giving two relations for the two functions, $W_k$ and $V_k$ which
determine the adiabatic particle basis.  Solving for $W_k$ we find
\begin{equation}
W_k =  \frac{-m^2 +  (6 \xi -1) \left(2\omega^2_k + H V_k + \dot H\right)
+ (6 \xi -1)^2\, \left(HV_k + \dot H + \frac{2 \epsilon\ }{a^2}\right)}
{2 (6 \xi -1)\,\varepsilon_k^{(2)} + T_k^{(2)}}\,.
\label{eq:Wsoln}
\end{equation}
If this is substituted back into either of Eqs.~(\ref{eq:WV}), we
obtain a quadratic equation for $V_k$, so that the two functions
$W_k$ and $V_k$ which determine the adiabatic vacuum modes can be determined
algebraically for any $\xi$ and $m$ for the general
RW geometry.

In this way the energy-momentum tensor components may be written
in a form analogous to~(\ref{eq:etrw}), where because of the
exact matching~(\ref{eq:WV}) the vacuum terms up to second adiabatic
order are now {\it identically zero}. Therefore the remaining
non-vacuum terms defined with respect to the time dependent adiabatic
vacuum necessarily satisfy the covariant conservation equation~(\ref{eq:cons}).
The non-vacuum terms in the adiabatic particle basis can be obtained
from~(\ref{eq:etrw}) with the replacements of $N_k \rightarrow
{\cal N}_k$, $v_k \rightarrow \tilde \phi_k$, $A_k \rightarrow
\alpha_k$ and $B_k \rightarrow \beta_k$. The non-vacuum energy-momentum
tensor components for general $W_k$ and $V_k$ are given explicitly
in~\cite{hm-pm}, and we do not to repeat them here.

Since Eq.~(\ref{eq:WV}) matches the vacuum energy-momentum tensor to
second adiabatic order, the adiabatic frequency $W_k$ obtained from
solving Eqs.~(\ref{eq:WV}) agrees with the adiabatic expansion~(\ref{eq:adbtwo})
up to and including second adiabatic order, differing
from $\Omega_k$ only at adiabatic order four. On the other hand since
$V_k^2$ and $HV_k$ appear in Eqs.~(\ref{eq:WV}), and $V_k$ contains terms
of odd adiabatic orders, $V_k$ need agree with
$-\dot\Omega_k/\Omega_k \simeq -\dot\omega_k/\omega_k$
only up to first adiabatic order, {\it i.e.,}
\begin{equation}
V_k = -\frac{\dot\omega_k}{\omega_k} + \dots = H\,
\left(1 - \frac{m^2}{\omega_k^2}\right) + \dots
\label{eq:Vk}
\end{equation}
where the ellipsis includes terms of third adiabatic order and
higher. Hence $V_k + \dot\Omega_k/\Omega_k$ is in general
non-vanishing at adiabatic order three, and the lowest order term in
the adiabatic expansion which appears in the expression for
$|\beta_k|^2$ in Eq.~(\ref{eq:betak}) is {\it sixth} order, {\it
i.e.,} the adiabatic particle number ${\cal N}_k$ defined by
Eqs.~(\ref{eq:adbnum}),~(\ref{eq:betak}) and~(\ref{eq:WV}) is a fourth
order adiabatic invariant. Time derivatives of ${\cal N}_k$ are
correspondingly highly suppressed, and particle creation is small in
slowly varying backgrounds, particularly at the highest wave numbers.
In flat spacetime, $H$, $\epsilon$ and $V_k$ vanish,
and~(\ref{eq:Wsoln}) together with~(\ref{eq:epsTtwo}) give $W_k
=\omega_k$ which is time independent. Hence the adiabatic modes become
exact modes, $\beta_k = 0$, and no particles at all are created.

In the general case, Eqs.~(\ref{eq:WV}) match the vacuum
energy-momentum tensor contributions to second adiabatic order and
therefore satisfy a weaker condition than the fourth order UV allowed
state condition. However the second order condition (\ref{eq:WV})
is sufficient to render the total
number of adiabatic particles created
finite, {\it i.e.,}
\begin{equation}
{\cal N}(t) = \frac{1}{2\pi^2}\,\int\,[\d k]\,k^2\, {\cal N}_k (t)
< \infty \;.
\label{eq:finN}
\end{equation}
Note that this would {\it not} be the case had we matched the
energy-momentum tensor components only to the lowest (zeroth) order
adiabatic expansion, or followed the Hamiltonian diagonalization
procedure of Ref.~\cite{GMM}. The requirement of matching the stress
tensor contributions~(\ref{eq:vacadb}) {\it exactly} to a fixed
adiabatic order as in~(\ref{eq:WV}) (rather than re-expanding the
algebraic expressions for $W_k$ and $V_k$) removes essentially all
of the ambiguity in the definition of the particle number, and guarantees
that the non-vacuum particle contribution is separately conserved.

Because of the weighting of the momentum integrals for the
energy-momentum tensor components~(\ref{eq:etrw}) by an extra power of
$k$ with respect to Eq.~(\ref{eq:finN}), the identification~(\ref{eq:WV})
still leaves a logarithmic cutoff dependence in the vacuum energy-momentum
tensor components. Unlike the power law divergences which are non-covariant
in form, the remaining logarithmic cutoff dependence is proportional to
the geometric tensor $^{(1)}H_{ab}$ of Eq.~(\ref{eq:Hone}). Hence it can
be viewed as a geometric vacuum polarization contribution in the low energy
EFT equipped with a short distance cutoff. Even if the short distance cutoff
is of the order of the Planck length, the higher derivatives in $^{(1)}H_{ab}$
mean that this contribution is negligible for all geometries varying more
slowly than the Planck scale. Of course one is free to define the adiabatic
particle basis  by replacing the second order terms in~(\ref{eq:WV}) with
the corresponding adiabatic order four (or higher) components $\varepsilon^{(4)}$
   and $T^{(4)}$, which would remove also the covariant logarithmic cutoff
dependence in $\langle T_{ab}\rangle$. Matching~(\ref{eq:adbpar})
to fourth order would make the resulting particle
number~(\ref{eq:adbnum}) an adiabatic invariant of even higher (in
fact, eighth) order. Matching to higher orders requires accurate
knowledge of higher and higher order time derivatives of the
metric scale factor, and hence of the entire spacetime evolution.
Because only the power law divergences are non-covariant in form
in the adiabatic procedure, matching to higher than second order
is both unnecessary from the point of view of the covariant UV
divergence structure of the stress tensor, and contrary to an EFT
approach in terms of the minimal number of spacetime derivatives
whenever EFT and the limits (\ref{eq:subPl}) apply. The tensor
$^{(1)}H_{ab}$ actually vanishes in the important special case of
de Sitter space, which renders the distinction between matching to
second or fourth adiabatic order to remove the remaining
logarithmic divergences in the stress tensor superfluous in this
case. Finally, the second order definition of particle number
through Eqs.~(\ref{eq:adbnum}), (\ref{eq:betak}) and (\ref{eq:WV})
is also the {\it minimal} one necessary to yield a finite total
number of created particles~(\ref{eq:finN}) in a general RW
spacetime. We emphasize that such a definition of particle number
is intrinsic, based on the physical requirement of identifying and
removing the vacuum contributions to the conserved stress tensor,
and does not depend critically on the existence of flat {\it in}
and {\it out} regions of the spacetime, or any extraneous notion
of particle detectors~\cite{b-d}.

We now apply this general second order definition of adiabatic
particle number to two important special cases with zero mass. When
$m=0$ the relations~(\ref{eq:WV}) simplify considerably.
For the massless conformally coupled field ($m=0$, $\xi = 1/6$), the
condition on the trace~(\ref{eq:TWV}) becomes empty but the first
condition~(\ref{eq:epsWV}) gives directly,
\begin{equation}
m=0,\ \xi=\frac{1}{6}: \qquad\left( W_k - \omega_k\right)^2 +
\frac{(V_k - H)^2}{4} \, = \, 0\,,
\end{equation}
which is solved uniquely for real $W_k$ and $V_k$ by
\begin{subequations}
\begin{eqnarray}
&& W_k = \omega_k = \frac{k}{a}\,,\\
&& V_k = H = - \frac{\dot W_k}{W_k}\,.
\end{eqnarray}
\label{eq:mccWV}
\end{subequations}
\hspace{-0.25cm} Because the fourth order adiabatic terms in
$\varepsilon^{(4)}$ and $T^{(4)}$ vanish for the massless
conformally coupled field in an arbitrary RW spacetime, this
result for $W_k$ and $V_k$ remains unchanged if the fourth order
adiabatic basis is used. Indeed as a result of the
relations~(\ref{eq:mccWV}), the adiabatic basis
functions~(\ref{eq:adbpar}) become {\it exact} solutions of
Eq.~(\ref{eq:phikeq}) for the massless, conformally coupled field
in a {\it general} RW spacetime. This means that the Bogoliubov
coefficients $\alpha_k$ and $\beta_k$ become time independent. As
in Eq.~(\ref{eq:bog}), ${\cal N}_k$ may be identified with the
time independent $N_k$ and there is no production of massless,
conformally coupled scalar particles.

In a second important special case, the massless minimally coupled field
($m=0, \xi=0$), $V_k$ drops out of Eq.~(\ref{eq:Wsoln}), and either of
Eqs.~(\ref{eq:WV}) gives a quadratic equation for $V_k$, which
is easily solved. Thus for $m=0$ and $\xi=0$ we obtain
\begin{subequations}
\begin{eqnarray}
\hspace{-2cm} && W_k = \frac{2(k^2 - \epsilon)} {2 k^2 - \epsilon
+ a^2 (\dot H + 2 H^2)}\ \frac{k}{a}\,,\\
&& \hspace{-1cm} V_k = 3 H - \frac{2\, (k^2-
\epsilon)^{\frac{1}{2}}} {2k^2 -\epsilon + a^2 (\dot H + 2 H^2)}
\left[ 4H^2k^2 - (a^2 \dot H + \epsilon) \left (\dot H + 2 H^2 -
\frac{\,\epsilon}{a^2}\right) \right]^{\frac{1}{2}}\,,
\end{eqnarray}
\label{eq:mmcWV}
\end{subequations}
\hspace{-0.3cm} in a general RW spacetime. Although this
definition of the adiabatic basis was determined by matching only
to second adiabatic order, by Eqs.~(\ref{eq:WV}), in the special
case of de Sitter space it becomes {\it exact}. Explicitly in flat
spatial sections, $\epsilon = 0$, for which $\dot H=0$, we
have
\begin{subequations}
\begin{eqnarray}
\hspace{-3cm} m=0, \xi=0, a=a_{\rm dS}:\qquad
W_k &=& \frac{k^2} {k^2 + H^2a_{\rm dS}^2}\
\frac{k}{a_{\rm dS}}\,,\\
V_k &=& H\, \frac{k^2 + 3H^2a_{\rm dS}^2}{k^2 + H^2a_{\rm dS}^2} =
- \frac{\dot W_k}{W_k}\,,
\end{eqnarray}
\label{eq:mmcdeS}
\end{subequations}
\hspace{-0.3cm}
with $a_{\rm dS}$ given by Eq.~(\ref{eq:RWds}). As a result
of these relations, the adiabatic mode function $\tilde \phi_k$ is an
{\it exact} solution of the mode equation~(\ref{eq:phikeq}) in de
Sitter space. Indeed from Eqs.~(\ref{eq:RWds}) and~(\ref{eq:mmcdeS}),
\begin{equation}
\int^t \d t' \, W_k(t')
= k^3\, \int^{\eta}\, \frac{\d \eta' \eta'^2}{1 + k^2\eta'^2} =
k\eta - \tan^{-1}(k\eta)\,.
\end{equation}
Hence,
\begin{eqnarray}
&&{\sqrt\frac{\hbar}{2a^{3} W_k}}\, \exp\left(-i \int^t\,
\d t'\,W_k (t')\right)\nonumber\\
&&= H {\sqrt\frac{\hbar\,(1 + k^2\eta^2)}{2k^3}}\ \exp
\Bigl(-ik\eta +i\tan^{-1}(k\eta)
\Bigr)\nonumber\\
&& = i\phi_k^{BD}\raisebox{-1ex}{\Big\vert}_{\stackrel{\scriptstyle m = 0}
{\scriptstyle \xi=0\ }}
\label{eq:exactmode}
\end{eqnarray}
by Eq.~(\ref{eq:mmc}). The fact that the second order adiabatic mode
functions $\tilde \phi_k$ of Eq.~(\ref{eq:adbpar}) coincide
with the exact BD mode functions up to a constant phase for the
massless, minimally coupled field in de Sitter space implies that the
Bogoliubov coefficients $\alpha_k$ and $\beta_k$ are time independent.
This may be verified explicitly by computing Eq.~(\ref{eq:betak}) with
$W_k$ and $V_k$ given by Eqs.~(\ref{eq:mmcdeS}). Thus ${\cal N}_k$
may be identified with $N_k$, and there is no production
of massless, minimally coupled scalar particles in the special case of
exact de Sitter spacetime. This implies that there is also no
phase decoherence of the free massless inflaton field in the de Sitter
epoch, with respect to the adiabatic particle basis defined by Eqs.
(\ref{eq:WV}). In the next section we will corroborate the
absence of decoherence for the massless inflaton by computing the
decoherence functional directly. A comparison of this result with the
results of earlier work, such as that of Ref.~\cite{PolSta} is given in
Appendix C.

To conclude this section we remark that the adiabatic particle
number basis should provide an efficient approximation of the low energy
semi-classical limit of the energy-momentum tensor. If we
neglect the phase correlated bilinears,
\begin{equation}
\langle \tilde a_{\rule{0mm}{2.3mm}\bf k} \tilde a_{\rule{0mm}{2.3mm}\bf k}\rangle
= \sigma_k \sinh r_k\,\cosh r_k\,e^{-i\theta_k}\,,
\label{eq:corr}
\end{equation}
which should make a relatively small contribution to $\langle T_{ab}\rangle$
in the mode sum over $k$, compared to the terms involving ${\cal N}_k$,
then the energy density becomes simply
\begin{equation}
\varepsilon \simeq \tilde\varepsilon \equiv
\varepsilon_v +
\frac{1}{2\pi^2a^3}\,\int\,[\d k]\,k^2\, \varepsilon_k^{(2)}\,{\cal N}_k\,.
\label{eq:qpareps}
\end{equation}
This should provide a useful analytic approximation to the energy density
in a general UV allowed RW state whenever the RW scale factor varies
slowly enough for the higher order adiabatic corrections to
$\varepsilon^{(2)}_k$ and ${\cal N}_k$, to be negligible.

The quantity $\varepsilon^{(2)}_k$ in Eq.~(\ref{eq:qpareps}), given
by~(\ref{eq:epstwo}) has the interpretation of the single particle
energy in the time varying RW background.  Because this second order
single particle energy satisfies
\begin{equation}
\dot\varepsilon^{(2)}_k = -H\,\varepsilon^{(2)}_k - H\, T^{(2)}_k\,,
\end{equation}
where $T^{(2)}_k$ is given by Eq.~(\ref{eq:Ttwo}), and the finite vacuum
contributions of Eqs.~(\ref{eq:vrenorm}) are separately conserved, it
follows that the conservation Eq.~(\ref{eq:cons}) is exactly
satisfied, provided that the trace in the quasi-particle approximation
corresponding to Eq.~(\ref{eq:qpareps}) is
\begin{equation}
T \simeq \tilde T \equiv T_v + \frac{1}{2\pi^2a^3}\,
\int\,[\d k]\,k^2\, T^{(2)}_k\,{\cal N}_k -
\frac{1}{2\pi^2\,Ha^3}\,\int\,[\d k]\,k^2\, \varepsilon^{(2)}_k\,
\dot{\cal N}_k\,.
\label{eq:qparT}
\end{equation}
Defining the ideal fluid pressure in this approximation to be that in
the absence of any particle creation, {\it i.e.,}
\begin{equation}
\tilde p \equiv \frac{\varepsilon_v + T_v}{3} +
\frac{1}{6\pi^2a^3}\,\int\,[\d k]\,k^2\, (\varepsilon_k^{(2)}
+ T^{(2)}_k)\,{\cal N}_k\,,
\label{eq:qparpress}
\end{equation}
the conservation Eq.~(\ref{eq:cons}) becomes then
\begin{equation}
\dot{\tilde\varepsilon} + 3H\,(\tilde\varepsilon + \tilde p) =
\frac{1}{2\pi^2a^3}\,\int\,[\d k]\,k^2\, \varepsilon^{(2)}_k\,\dot{\cal N}_k\,,
\label{eq:diss}
\end{equation}
after transposing the last term of Eq.~(\ref{eq:qparT}) to the right
hand side of Eq.~(\ref{eq:diss}). Thus in the quasi-particle limit
where it is valid to make the replacements~(\ref{eq:qpareps})
and~(\ref{eq:qparT}), the term involving $\dot{\cal N}_k$ carries the
interpretation of the rate of heat dissipation per unit volume due to
the non-conservation of adiabatic particle number ${\cal N}_k$. If the
particles are in quasi-stationary local thermodynamic equilibrium at
the slowly varying effective temperature $T_{\rm eff}(t)$, then this rate
of heat dissipation may be equated to $T_{\rm eff}$ times the rate of
effective entropy density generation $s_{\rm eff}$,
\begin{equation}
T_{\rm eff} \frac {\d s_{\rm eff}}{\d t}
=\frac{1}{2\pi^2a^3}\,\int\,[\d k]\,k^2\,
\varepsilon^{(2)}_k\,\dot{\cal N}_k\,,
\label{eq:heat}
\end{equation}
by the first law of thermodynamics. The effective entropy
generation gives rise to an effective bulk viscosity in the
energy-momentum tensor due to particle creation, even in the
absence of self-interactions of the quasi-particles.

Let us emphasize that the entropy generation and bulk viscosity are
only {\it effective}, {\it i.e.,} an approximation valid only to the
extent that the phase information contained in the off-diagonal
elements of the exact density matrix~(\ref{eq:gaussd}) in the
adiabatic particle basis cannot be recovered. Likewise the
correlations~(\ref{eq:corr}), which also depend on the rapidly varying
phases $\theta_k$ conjugate to ${\cal N}_k$ should make a negligible
contribution to macroscopic physical quantities. If the exact phase
information is retained, then the evolution remains unitary, as
required by the equivalence of the evolution to that of the effective
classical Hamiltonian~(\ref{eq:effHam}). However, the extension of the
usual description of the cosmological fluid by non-ideal terms is
suggested by the adiabatic particle creation rate and phase averaging
in the low energy EFT.  This may provide a useful phenomenological
description in some circumstances, and also shows the approximations
which are necessary in principle to pass from the fully reversible
field theory description of matter in cosmological spacetimes to an
effective, irreversible kinetic theory with a definite arrow of time.

\section{Decoherence}
\label{sec:decoherence}

The density matrix description of the evolution of arbitrary RW states
in Sec.~\ref{sec:density} allows us to describe the quantum to
classical transition, {\it i.e.,} decoherence, in a cosmological
setting. The fundamental quantity of interest is the decoherence
functional between two different histories, $\Gamma_{12}$ or
$\tilde\Gamma_{12}$~\cite{PazSin}, given by Eqs.~(\ref{eq:inner})
or~(\ref{eq:prob}), of the pure or mixed state cases,
respectively. Evaluating the Gaussian integrals needed to compute
${\rm Tr}(\hat\rho_1\hat\rho_2)$ with the measure~(\ref{eq:meas})
gives
\begin{eqnarray}
\tilde \Gamma_{12} &=& \frac{1}{4\pi^2}\,\int\,[\d
k]\,k^2\,\ln\left\{ \frac{1}{4}
\left(\sigma_{1\,k}\,\frac{\zeta_{2\,k}}{\zeta_{1\,k}} +
\sigma_{2\,k}\,\frac{\zeta_{1\,k}}{\zeta_{2\,k}}\right)^2 +
\frac{(\zeta_{1\,k}\pi_{2\,k} -
\zeta_{2\,k}\pi_{1\,k})^2}{\hbar^2} \right. \nonumber \\& & \qquad
\qquad \qquad \qquad \qquad \left.   +
\frac{(\sigma_{1\,k} -
\sigma_{2\,k})^2}{4}\right\}\,, \label{eq:decfnl}
\end{eqnarray}
where $\{\zeta_{1\,k},\pi_{1\,k};\sigma_{1\,k}\}$ and
$\{\zeta_{2\,k},\pi_{2\,k};\sigma_{2\,k}\}$ are the Gaussian state
parameters of the two different histories. This simplifies somewhat
in the case of two different pure state histories,
\begin{eqnarray}
{\rm Im}\,\Gamma_{12} &=&
\tilde\Gamma_{12}\Big\vert_{\sigma_{k\,1}=\sigma_{k\,2} =1}
\nonumber \\
   &=& \frac{1}{4\pi^2}\,\int\,[\d k]\,k^2\,\ln\left\{
\frac{1}{4} \left(\frac{\zeta_{2\,k}}{\zeta_{1\,k}} +
\frac{\zeta_{1\,k}}{\zeta_{2\,k}}\right)^2 +
\frac{(\zeta_{1\,k}\pi_{2\,k} -
\zeta_{2\,k}\pi_{1\,k})^2}{\hbar^2} \right\}\,.
\label{eq:psdfnl}
\end{eqnarray}
\hspace{-0.15cm}
In this pure state case the decoherence functional may be
expressed in terms of one complex frequency function,
\begin{subequations}
\begin{eqnarray}
\Upsilon_k &\equiv& \frac{\hbar}{2\zeta_k^2} - \frac{i\pi_k}{\zeta_k}
\label{eq:Upseqa}   \\
&=& -i a^3\, \frac{\dot\phi_k^*}{\phi_k^*} -6i\xi H a^3\,,
\label{eq:Upseqb}
\end{eqnarray}
\end{subequations}
where the last relation is derived in Appendix A. It is
straightforward to rewrite Eq.~(\ref{eq:psdfnl}) in the two alternate
forms,
\begin{subequations}
\begin{eqnarray}
{\rm Im}\,\Gamma_{12} &=&
\frac{1}{4\pi^2}\,\int\,[\d k]\,k^2\,\ln\left\{ \frac{\vert
\Upsilon_{1\,k} + \Upsilon_{2\,k}^*\vert^2}{4\,({\rm Re}\,
\Upsilon_{1\,k}) \,({\rm Re}\, \Upsilon_{2\,k})}\right\}\,
\label{eq:decups}\\
& &  \nonumber \\
   &=&\frac{1}{4\pi^2}\,\int\,[\d k]\,k^2\,\ln
\left\{\big\vert a_1^{3-6\xi}(a_1^{6\xi}\phi_{1\,k}\dot)
\phi_{2\,k}^* -
a_2^{3-6\xi}(a_2^{6\xi}\phi_{2\,k}^*\dot)\phi_{1\,k}
\big\vert^2/\hbar^2\right\}\,.
\end{eqnarray}
\label{eq:decphi}
\end{subequations}
\hspace{-0.25cm} Because of the infinite product of integrations in
the functional measure~(\ref{eq:meas}), it is clear that some
condition(s) will have to be imposed on the two states or density
matrices in these expressions, in order to insure a convergent result
for large $k$. The otherwise ill-defined divergent nature
of~(\ref{eq:inner}) has been noted by several
authors~\cite{KieLaf,PazSin,BarKam}. The divergences in the
decoherence functional are similar to those encountered in the
unrenormalized expressions for the energy-momentum tensor components,
Eqs.~(\ref{eq:unren}). In that case, the superficial degree of
divergence was quartic, whereas that of Eq.~(\ref{eq:decfnl})
or~(\ref{eq:psdfnl}) is reduced by one power of $k$, and can be no
more than cubically divergent at large $k$.  A method to handle this
cubic and subleading linear cutoff dependence of the decoherence
functional~(\ref{eq:decfnl}) or~(\ref{eq:decphi}) is needed before
meaningful results in the low energy EFT description can be obtained.

In order to study the cutoff dependence of the inner product and
decoherence functional, let us write
\begin{subequations}
\begin{eqnarray}
\Upsilon_{1\,k}&=& \Upsilon_k + \delta \Upsilon_k\,,\\
\Upsilon_{2\,k}&=& \Upsilon_k - \delta \Upsilon_k\,,
\end{eqnarray}
\label{eq:upsvar}
\end{subequations}
\hspace{-0.25cm} and expand the logarithm in Eq.~(\ref{eq:decups}) to
second order in $\delta \Upsilon_k$. We find that
\begin{equation}
{\rm Im}\,\Gamma_{12} = \frac{1}{4\pi^2}\,\int\,[\d k]\,k^2\,
\frac{|\delta \Upsilon_k|^2}{({\rm Re}\,\Upsilon_k)^2} + {\cal
O}(\delta \Upsilon_k)^4 \,.
\label{eq:secups}
\end{equation}
The leading behavior of Re $\Upsilon_k$ at large $k$ may be read from
the first order equation satisfied by this function, {\it viz.,}
\begin{equation}
\Upsilon_k^2 = ia^3 \dot\Upsilon_k + a^6\,\left[\omega_k^2
   + (6\xi-1)\left(6\xi H^2 +
\frac{\epsilon}{a^2}\right)\right] - 12i \xi Ha^3 \Upsilon_k \,,
\label{eq:upseom}
\end{equation}
which is easily derived from the definition~(\ref{eq:Upseqa}) and the
relations~(\ref{eq:hamevo3}) and~(\ref{eq:hamevo4}). From
Eq.~(\ref{eq:upseom}) we see that
\begin{equation}
{\rm Re}\, \Upsilon_k \rightarrow  ka^2
\left[1 + {\cal O}\left(\frac{1}{k^2}\right)\right]\,,
\qquad k\rightarrow \infty\,.
\label{eq:upsasym}
\end{equation}
Since $\delta\Upsilon_k$ is the same order as $\Upsilon_k$ generically
at large $k$, the decoherence functional (\ref{eq:secups}) will generally
diverge as the UV cutoff is removed. Indeed the asymptotic behavior
given by Eq.~(\ref{eq:upsasym})
implies at large $k$ that
\begin{equation}
\delta \Upsilon_k \rightarrow 2k a\, (\delta a) + \dots
\label{eq:leadel}
\end{equation}
so that (\ref{eq:secups}) will diverge cubically in the cutoff
$k_M$, unless $\delta a\equiv 0$ identically. Hence no meaningful
comparison between two different RW
geometries in the low energy EFT
can be made through the decoherence functional.

\begin{figure}
\includegraphics[angle=0,height=4.5cm,width=10cm]{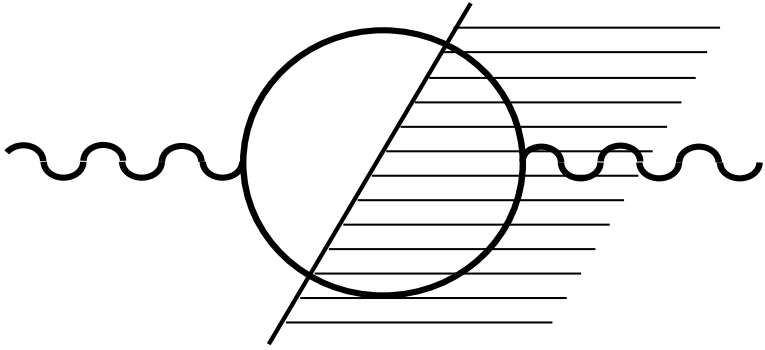}
\caption{The imaginary part of the one-loop vacuum polarization given
by Eqs.~(\ref{eq:secvar}) and~(\ref{eq:ImPi}), which enters the
decoherence functional at second order in the metric variation,
$\delta g_{ab}$ (represented by the wavy lines). The shaded part of the
diagram represents the complex conjugation of the unshaded part, and
the resulting squared amplitude is proportional to the probability for
the creation of a scalar particle/anti-particle pair, represented by the
diagonal cut through the diagram.}
\label{fig:polcut}
\end{figure}

This divergent short distance behavior of the decoherence functional
could have been anticipated from the relationship between
$\Gamma_{12}$ and the Closed Time Path (CTP)~\cite{CTP-ref} action
functional. Variation of the CTP action functional with respect to the
metric, $g_{ab}$, produces connected correlation functions of the
energy-momentum tensor with particular time orderings of their
arguments. The first variation is the same as that of $S_{\rm eff}[g]$
in Eq.~(\ref{eq:effS}), and produces the stress tensor expectation
value~(\ref{eq:Tren}). The second order variation~(\ref{eq:secups}) is
formally proportional to
\begin{equation}
{\rm Im}\,\int_0^t\,{\d}^4x\,\sqrt{-g}\,\int_0^t\,{\d}^4x'\,\sqrt{-g'}
\,\delta g_{ab}(x)\, \Pi^{abcd}(x,x')\,\delta g_{cd}(x')\,.
\label{eq:secvar}
\end{equation}
We shall be particularly interested in the specific homogeneous
variation, $\delta g_{ab} = 2 (\delta a/a) g_{ab}$ which preserves
the RW symmetries. The symmetrized expectation value of stress tensors,
\begin{equation}
{\rm Im}\,\Pi^{abcd}(x.x') = \frac{1}{2}\, \langle T^{ab}(x)T^{cd} (x') +
T^{cd} (x')T^{ab}(x)\rangle
\label{eq:ImPi}
\end{equation}
is proportional to the noise kernel of fluctuations around from the
mean $\langle T^{ab} \rangle$~\cite{noise}. In flat spacetime, where
one may transform conveniently to the momentum
representation~(\ref{eq:secvar}) is proportional to the cut in the
one-loop polarization diagram of Fig.~{\ref{fig:polcut}},
corresponding to the squared matrix element for the creation of
particle pairs by the perturbation $\delta g_{ab}$.

In coordinate space Eq.~(\ref{eq:ImPi}) is a singular distribution at
coincident points $x=x'$, involving in general up to four derivatives
of $\delta^{(4)}(x,x')$. Thus, the second variation~(\ref{eq:secvar})
exists only for metric variations, $\delta g_{ab}$, which fall off
rapidly enough in both space and time to permit integration by parts
of the derivatives of $\delta^{(4)}(x,x')$. Since the time interval
$[0,t]$ is finite in Eq.~(\ref{eq:secvar}), the surface terms
generated by these integrations by parts do not vanish at the
endpoints in general, and can generate up to two derivatives of delta
functions at equal spacetime arguments, {\it i.e.,} formal cubic (and
subleading linear) divergences in the decoherence functional, which is
just what is obtained in from
Eq.~(\ref{eq:secups}),~(\ref{eq:upsasym}), and~(\ref{eq:leadel}).

The unrenormalized decoherence functional is not only cutoff dependent
in general but also ambiguous in that it depends upon the precise
definition of the local measure~(\ref{eq:meas})~\cite{BarKam}, which does not
affect the physical content of the evolution described by the density
matrix~(\ref{eq:gaussd}). In fact, had we used the $\tilde q_{\bf
k} = a q_{\bf k}$ field to parameterize the density matrix rather than
$q_{\bf k}$, then in order to normalize the state properly the measure
would have to be replaced by a product of the $\d\tilde q_{\bf k}$.
This differs from Eq.~(\ref{eq:meas}) by an {\it infinite} number of
local factors of $a(\eta)$. Hence, we should expect the inner product and
decoherence functional defined by this measure on the field configuration
space to differ from the previous one by cutoff dependent contact terms.
Indeed, the frequency function $\Upsilon_k$ would be modified to
\begin{equation}
\tilde \Upsilon_k \equiv \frac{\Upsilon_k}{a^2}\,,
\label{eq:Upsredef}
\end{equation}
for which the leading term in the large $k$ limit~(\ref{eq:leadel})
cancels. The explicit form of $\delta \Upsilon_k$
is given below by Eq.~(\ref{eq:upschi}), with the result that
the decoherence functional~(\ref{eq:decups}) generally diverges
cubically as the comoving momentum cutoff $k_M\rightarrow\infty$.
Because of the different large $k$ behavior of $\tilde \Upsilon_k$,
if the same steps leading to Eq.~(\ref{eq:upschi}) are carried out for
$\tilde \Upsilon_k$ instead, the resulting expression lacks the last
term in Eq.~(\ref{eq:upschi}), and for that reason its contribution to
the decoherence functional is only linearly divergent~\cite{KieLaf,BarKam}.
This shows that the degree of divergence is dependent on the field
parameterization and the definition of the inner product ~(\ref{eq:inner})
on the field configuration space, which
should not have any physical
consequences for decoherence due to a slowly evolving geometry in the
low energy EFT.

Because of the appearance of divergences of odd powers, associated
with the boundaries of the region of integration, the divergences in
the decoherence functional cannot be removed by renormalization of the
bulk terms in the low energy effective action~(\ref{eq:effS}). Instead
their renormalization requires introducing counterterms in the
effective action, of dimension one and three, which are strictly
localized on the boundaries.  Adiabatic regularization may be used to
define the necessary subtractions of the decoherence functional,
corresponding to renormalization of these boundary terms, in a way
quite analogous to the adiabatic subtractions used to define the
renormalized energy-momentum tensor~\cite{parker}. Since terms in the
effective action up to dimension three are involved, we define the
renormalized decoherence functional by subtracting from its
unrenormalized value the adiabatic expansion of the functional up to
and including its {\it third} adiabatic order asymptotic expansion for
large $k$. Since the decoherence functional involves an absolute
square, this requires subtracting only up to the {\it first} adiabatic
order in the expression inside the absolute value signs in
Eq.~(\ref{eq:decphi}). That is, we define the renormalized decoherence
functional by
\begin{equation}
{\rm Im}\, \Gamma^{(R)}_{12} \equiv \frac{1}{4\pi^2} \int\,[\d k]\,
k^2\, \Bigg\vert\frac{\delta \Upsilon_k^*} {{\rm Re}\,\Upsilon_k}
- \left(\frac{\delta \Upsilon_k^*} {{\rm Re}\,\Upsilon_k}\right)_1
\Bigg\vert^2\,,
\label{eq:decren}
\end{equation}
where the subscript $1$ denotes the expansion of the quantity in
parentheses up to and including first order in its adiabatic
expansion. This leaves the leading unsubtracted behavior at large $k$
to be second adiabatic order under the absolute value signs, and
fourth adiabatic order in its square in $\tilde\Gamma_{12}^{(R)}$.
This corresponds to subtracting all surface divergences up to and
including third adiabatic order in the CTP action functional, which
is what is required.

To carry this out explicitly we begin by rewriting Eq.~(\ref{eq:Upseqb})
in conformal time in the form,
\begin{eqnarray}
\Upsilon_k^* &=& i a^2\, \frac{\phi_k^{\prime}}{\phi_k} + 6i\xi a' a\nonumber\\
&=& i a^2\, \frac{\chi_k^{\prime}}{\chi_k} + i(6\xi -1) a'a\,,
\end{eqnarray}
and computing its first variation,
\begin{equation}
\delta \Upsilon_k^* = 2i a \delta a \, \frac{\chi_k^{\prime}}{\chi_k}
+ ia^2 \frac{\delta \chi_k^{\prime}}{\chi_k} -
ia^2  \frac{\chi_k^{\prime}\delta \chi_k}{\chi_k^2} + i(6\xi -1)\delta(a'a)\,.
\label{eq:delups}
\end{equation}
The variations of $\chi_k$ and its derivative are computed by varying
Eq.~(\ref{eq:chiosc}), to obtain
\begin{equation}
\delta \chi_k'' + \left[ k^2 + m^2 a^2 + (6\xi -1)
   \left( \frac{a''}{a\,} + \epsilon\right)
\right] \delta \chi_{\rule{0mm}{2.1mm}k} = -  \left[ m^2 \delta a^2
+ (6\xi -1)\,
\delta\left( \frac{a''}{a\,}\right)\right] \chi_{\rule{0mm}{2.1mm}k}\,,
\label{eq:chivar}
\end{equation}
which is solved in terms of the retarded Green's function of the
differential operator on the left hand side,
\begin{equation}
G_R (\eta, \eta') = \frac{i}{\hbar}
\Bigl(\chi_{\rule{0mm}{2.1mm}k}(\eta)\chi_k^*(\eta') -
   \chi_k^*(\eta) \chi_{\rule{0mm}{2.1mm}k}(\eta')\Bigr)\Theta(\eta-\eta')\,
\end{equation}
in the form,
\begin{eqnarray}
\delta \chi_{\rule{0mm}{2.1mm}k}(\eta) &&= - \int\,\d\eta'\, G_R (\eta, \eta')
   \left[ m^2 \delta a^2 + (6\xi -1)\,
\delta\left( \frac{a''}{a\,}\right)\right]_{\eta'}
\chi_{\rule{0mm}{2.1mm}k}(\eta')\nonumber\\
&&= \delta\alpha_k (\eta)\, \chi_{\rule{0mm}{2.1mm}k}(\eta)
   + \delta\beta_k(\eta)\,
\chi_k^*(\eta)\,,
\label{eq:delchi}
\end{eqnarray}
where
\begin{subequations}
\begin{eqnarray}
\delta\alpha_k(\eta) &&= -\frac{i}{\hbar} \int^{\eta}\,\d\eta'\,
\left[
m^2 \delta a^2 + (6\xi -1)\,
   \delta\left( \frac{a''}{a\,}\right)\right]_{\eta'} \vert
\chi_{\rule{0mm}{2.1mm}k}(\eta')\vert^2\,,\\
\delta\beta_k(\eta) &&= \frac{i}{\hbar} \int^{\eta}\,\d\eta'\,
\left[
m^2 \delta a^2 + (6\xi -1)\,
\delta\left( \frac{a''}{a\,}\right)\right]_{\eta'}
\left[\chi_k(\eta')\right]^2\,.
\end{eqnarray}
\label{eq:delalb}
\end{subequations}
\hspace{-0.3cm}
The lower limits of the integrals in Eq.~(\ref{eq:delalb}) depend on
the initial conditions of the wave functional and give a
time independent phase in the decoherence functional below which
we shall not need to specify.
Substituting Eq.~(\ref{eq:delchi})
into~(\ref{eq:delups}) and
   using the Wronskian condition~(\ref{eq:wron}),
we find that the $\delta
\alpha_k$ term cancels and we are left with
\begin{equation}
\delta \Upsilon_k^* = 2i a \delta a \, \frac{\chi_k^{\prime}}{\chi_k}
- \frac{\hbar a^2}{\chi_k^2} \delta\beta_k  + i(6\xi -1)\delta(a'a)\,.
\end{equation}
Using Eq.~(\ref{eq:wron}) again, we have
\begin{equation}
{\rm Re}\,\Upsilon_k = \frac{\hbar a^2}{2\,\vert\chi_k\vert^2}\,,
\end{equation}
so that we secure finally,
\begin{equation}
\frac{\delta \Upsilon_k^*} {{\rm Re}\,\Upsilon_k}
= \frac{2i\chi_k^*}{\chi_{\rule{0mm}{2.1mm}k}}
\left(i\,\delta\beta_k +\frac{(6\xi-1)}{\hbar}\,
\delta\left(\frac{a'}{a}\right)\chi_k^2
+\frac{a\, \delta a}{\hbar} \left(a^{12\xi}
\phi_k^2\right)^{\prime} a^{-12\xi}\right)\,.
\label{eq:upschi}
\end{equation}
This form is valid for a scalar field of arbitrary mass and curvature
coupling in a general RW spacetime. In order to renormalize it, we
should subtract its asymptotic expansion up to adiabatic order one and
substitute the square of the subtracted quantity in
Eq.~(\ref{eq:decren}). We carry out this subtraction explicitly in two
important special cases, namely, when the mass $m=0$ and the curvature
coupling $\xi$ takes on either its conformal or minimally coupled value,
$\xi = 1/6, 0$, respectively.

Using Eq.~(\ref{eq:upschi}) together with~(\ref{eq:delalb}) and the
form of Eq.~(\ref{eq:chiosc}) in the massless, conformally coupled
case, we obtain
\begin{equation}
\frac{\delta \Upsilon_k^*} {{\rm Re}\,\Upsilon_k}\raisebox{-1ex}
{\Bigg\vert}_{\stackrel{\scriptstyle m = 0}{\scriptstyle \xi=\frac{1}{6}\ }} =
\frac{2i\chi_k^*}{\hbar\chi_{\rule{0mm}{2.1mm}k}}
\frac{\delta a}{a} \left(\chi_k^2\right)^{\prime}
= \frac{4k|\chi_k|^2 }{\hbar} \,
\frac{\delta a}{a} = 2\, \frac{\delta a}{a}\,,
\end{equation}
which involves no time derivatives and hence is clearly of adiabatic
order zero. If we were to substitute this directly into
Eq.~(\ref{eq:decups}), we would obtain a cubically divergent
decoherence functional. Since this cubic divergence can be removed by
simply redefining the canonical variables as in Eqs.~(\ref{eq:chican})
and~(\ref{eq:Upsredef}), it is clear that it can have no physical
significance. However, if we first subtract off the adiabatic order
zero part (which in this case is the entire expression), then the
renormalized decoherence functional~(\ref{eq:decren}) in an arbitrary
RW spacetime is identically {\it zero} for the massless, conformally
coupled field. This lack of decoherence corresponds to the lack of
particle creation for this field in any RW space which we have found
in the previous section.

In the massless, minimally coupled case the corresponding expression is
\begin{equation}
\frac{\delta \Upsilon_k^*} {{\rm Re}\,\Upsilon_k}\raisebox{-1ex}
{\Bigg\vert}_{\stackrel{\scriptstyle m = 0}{\scriptstyle \xi=0\ }} =
\frac{2i\chi_k^*}{\chi_{\rule{0mm}{2.1mm}k}}
\int^{\eta}\,\d\eta'\, \delta\left( \frac{a''}{a\,}\right)_{\eta'}
\left[\chi_k(\eta')\right]^2
   - \frac{2i}{\hbar}\,\delta\left(\frac{a'}{a}\right)\vert\chi_k\vert^2
+ \frac{2i\chi_k^*}{\hbar\chi_{\rule{0mm}{2.1mm}k}}\,a\,\delta a
\left(\phi_k^2\right)^{\prime} \,,
\label{eq:decmcc}
\end{equation}
which is generally non-zero. Note that the first term in this last
expression depends upon the variation of the RW scale factor over
the entire evolution from an arbitrary (unspecified) initial time
at the lower limit of the $\eta'$ integral to the final time
$\eta$, while the last two terms of Eq.~(\ref{eq:decmcc}) depend
upon the variation of the scale factor only at the final time.  It
is these two latter surface terms that generate divergences in the
unsubtracted decoherence functional~(\ref{eq:secups}), when
integrated over $k$. Such surface terms arise in the covariant
expression~(\ref{eq:secvar}) if the conservation of $T^{ab}$ is
used to express the tensorial noise correlator~(\ref{eq:ImPi}) in
terms of covariant derivatives of scalar quantities, and then an
integration by parts is performed.

Specializing to de Sitter spacetime and using the form of the BD mode
functions~(\ref{eq:BDmode}), the last term in Eq.~(\ref{eq:decmcc})
becomes
\begin{equation}
   \frac{2i\chi_k^*}{\hbar\chi_{\rule{0mm}{2.1mm}k}}\,a\,\delta a
\left(\phi_k^2\right)^{\prime}\bigg\vert_{\rm dS} = 2\,
\frac{\delta a}{a} \left(1 + \frac{i}{k\eta}\right)\,.
\end{equation}
Here the first term is of adiabatic order zero as in the previous
conformally coupled case, while the second term is of adiabatic
order one. Hence both terms are fully subtracted in the
renormalized decoherence functional. The other contact term
becomes in de Sitter space,
\begin{equation}
   - \frac{2i}{\hbar}\,\delta\left(\frac{a'}{a}\right)
\vert\chi_k\vert^2\bigg\vert_{\rm dS}
= -\frac{i}{k} \delta\left(\frac{a'}{a}\right)
\left(1 + \frac{1}{k^2\eta^2}\right)\,,
\end{equation}
which consists of an adiabatic order one and order three term. Hence
subtracting up to adiabatic order one removes the first term but
leaves the $1/k^2\eta^2$ term unaffected. Since the first term in
Eq.~(\ref{eq:decmcc}) involves two time derivatives of the scale factor or
its variation, it is adiabatic order two, and likewise unaffected by
the subtraction of up to adiabatic order one terms. Hence finally,
\begin{eqnarray}
& & \hspace{-1cm}  {\rm Im}\,
\Gamma^{(R)}_{12}\Big\vert_{\stackrel{\scriptstyle m = 0}
{\scriptstyle \xi=0\ {\rm dS}}} =\frac{1}{4\pi^2} \int\,\d k\, k^2\,
\Bigg\vert\frac{\delta \Upsilon_k^*} {{\rm Re}\,\Upsilon_k} -
\left(\frac{\delta \Upsilon_k^*} {{\rm Re}\,\Upsilon_k}\right)_1
\Bigg\vert^2_{\stackrel{\scriptstyle m = 0}
{\scriptstyle \xi=0\ {\rm dS}}}\nonumber\\
&& \hspace{-1cm} =\frac{1}{4\pi^2} \int\,\d k\,\Bigg\vert
\int^{\eta}\,\d\eta'\, e^{2ik(\eta-\eta')} \left(1 -
\frac{i}{k\eta'}\right)^2 \delta\left(
\frac{a''}{a\,}\right)_{\eta'} -
\frac{1}{k^2\eta^2}\left(\frac{k\eta - i}{k\eta +
i}\right)\delta\left(\frac{a'}{a}\right) \bigg\vert^2\,,
\label{eq:decmcds}
\end{eqnarray}
which is UV finite and non-zero for arbitrary variations of the scale
factor.

If we consider the particular variation of the scale factor in which
the de Sitter Hubble parameter $H$ is varied in Eq.~(\ref{eq:decmcds}),
while the conformal time $\eta$ is held {\it fixed}, then $a'/a = -1/\eta$
and $a''/a = 2/\eta^2$ are fixed and Eq.~(\ref{eq:decmcds})
vanishes. Hence we find that under variations of the de Sitter
curvature, which compare the wave functionals of the quantum field in
macroscopically different de Sitter universes but at the same conformal
time, the decoherence functional for the massless, minimally coupled
field {\it vanishes} identically.

The ambiguous contact terms which are field parameterization
dependent are removed by the adiabatic regularization and
renormalization procedure in Eq.~(\ref{eq:decren}), as the two
cases considered explicitly above show. Hence,
Eq.~(\ref{eq:decren}) yields both a finite decoherence functional
free of unphysical dependence on the short distance cutoff, and
one that is independent of redefinitions of the scalar field
variables and inner product. The renormalized decoherence
functional proposed here vanishes in the two special massless
cases of the conformally coupled field in a general RW background
and the minimally coupled field in a de Sitter background, the
same two cases studied in the previous section where the adiabatic
particle creation rate $\dot {\cal N}_k = 0$. Since the imaginary
part of the polarization tensor~(\ref{eq:ImPi}) is just the cut
one-loop diagram shown in Fig.~\ref{fig:polcut}, which is
proportional to the probability for the metric fluctuation to
create a particle/anti-particle pair from the vacuum, a close
correspondence between the lack of particle creation and a
vanishing decoherence functional for homogeneous metric
perturbations in the Hubble parameter is not unexpected. The fact
that the adiabatic subtraction procedure for the decoherence
functional proposed here supports this correspondence suggests
that it is the correct one to define a finite decoherence
functional for semi-classical cosmology.  In order to prove that
this is indeed the {\it unique} procedure for defining a physical
decoherence functional in the EFT approach, the adiabatic
subtractions of first and third order should be related to
definite boundary counterterms in the CTP effective action which
reside exclusively on the surfaces at the initial and final times.
These boundary terms may be related to those found recently by the
authors of Ref. \cite{h-h}. We leave the determination of these
surface terms for a future investigation.

\section{Summary and Conclusions}
\label{sec:conclusions}

The principal objective of this paper has been to place semi-classical
cosmology within a consistent EFT framework, in which possible short distance
effects can be parameterized by well-defined initial conditions at the onset of
inflation.
  Although the general principles and adiabatic methods underlying such
an EFT framework have been available for some time, we have thought it worthwhile
to make these assumptions completely explicit in this paper, and demonstrate how
they can be applied and extended in a number of different ways, which may be useful
for future cosmological models and observations. Because of the several different
applications considered in the paper, we collect here and summarize the
principal results, together with the relevant equations and sections where each
point is
discussed in detail.

\begin{itemize}
\item The general homogeneous, isotropic RW pure state is defined
by field amplitudes $\phi_k$ obeying (\ref{eq:phikeq}), which are
linear combinations of vacuum modes (\ref{eq:bog}) with Bogoliubov
coefficients satisfying (\ref{eq:bognorm}).
\item These pure RW
  states are squeezed vacuum states annihilated by
$a_{\bf k}$ in
the mode expansion (\ref{eq:phiquant}) and specified by
two real
time-independent squeezing parameters (\ref{eq:squeeze}), up to an
overall irrelevant phase.
\item The wave functionals of these pure
RW states are Gaussians
(\ref{eq:pure}) in the Schr\"odinger
  picture field coordinate basis.
\item The general RW state with a
non-zero occupation number (\ref{eq:part})
is a mixed state
described by the mixed state Gaussian density matrix
(\ref{eq:gaussd}) in the coordinate basis.
\item The general RW
mixed state requires three independent functions of $k$,
$(\zeta_k, \pi_k; \sigma_k)$, which are related to the mode
function $\phi_k$ by (\ref{eq:zetakdef}) or (\ref{eq:phibil}), and
determine the three equal time symmetrized correlators of the
field by (\ref{eq:piphibi}) in the Hamiltonian description.
\item
The first two of these functions of $k$, $(\zeta_k, \pi_k)$ are
time dependent and form a canonically conjugate pair for the
unitary evolution of the density matrix (\ref{eq:Lio}),
(\ref{eq:hamev}), described by the effective classical Hamiltonian
(\ref{eq:effHam})-(\ref{eq:Hkeff}), in which $\hbar$ appears as a
parameter.
\item The third function of $k$, $\sigma_k = 2n_k +1$
is strictly a constant of the motion.
\item The form of the
Hamiltonian of the scalar field evolution in a
fixed RW background
depends on the field parameterization, and in general
is not equal
to the covariant energy density $\varepsilon = T_{tt}$.
\item The
covariant and Hamiltonian descriptions of the evolution are
completely equivalent nonetheless, and the total Hamiltonian of
the combined matter plus geometry system (\ref{eq:Htot}) vanishes
by time reparameterization invariance for evolutions satisfying
the classical Friedman equation.
\item The power spectrum of
scalar field fluctuations in the general
homogeneous, isotropic,
mixed RW state is given by Eqs. (\ref{eq:powerv}).
\item The spectrum of the actual scalar metric fluctuations observed
in the CMB are dependent upon additional parameters which may be
different for different inflationary models. An example is the
dependence on the slow roll parameter $\epsilon$ in Eq. (\ref{eq:slowroll}).
\item The
energy density and trace of the stress tensor in the
general RW state is given by Eqs. (\ref{eq:etrw}), with
$\varepsilon_v$ and $T_v$ the values in the fiducial vacuum state.
\item In order to be a UV allowed RW state with short distance
behavior consistent with general covariance of the low energy EFT
and the Equivalence Principle, the fiducial vacuum state and all
other physical states must be fourth order adiabatic states.
\item  Any modification of the fourth
order adiabaticity condition at
short distances has the potential
to disturb the conservation of
$\langle T_{ab}\rangle$, and/or
violate the Equivalence Principle at
arbitrarily large distances
and late times, which would also violate
the decoupling hypothesis
of low energy EFT.
\item The Bunch-Davies (BD) state is a UV
allowed fourth order adiabatic
state, which is also invariant
under the full $O(4,1)$ isometry group of
global de Sitter
spacetime.
\item The general one complex parameter squeezed
$\alpha$ states of the scalar field in de Sitter space are {\it
not} UV allowed fourth order adiabatic states (even for
non-self-interacting scalar fields), except for the single value
of the parameter corresponding to the BD state.
\item Because all
UV allowed states are fourth order adiabatic, their
power spectrum
approaches that of the BD state for sufficiently large
comoving
wavenumbers $k > k_M$, and sufficiently late times $t > t_M$ after
the onset of inflation (\ref{eq:tM}), when EFT methods should
apply.
\item As a consequence of this kinematic effect of the
expansion, any modifications of the power spectrum due to initial
state effects require a coincidence of fine tuning
(\ref{eq:tuning}) in order to be observable in the CMB today.
\item Assuming such fine tuning and cutting off the squeezed
$\alpha$ state at a finite large comoving momentum scale $k_M$
produces potentially observable scale dependent modifications of
the CMB power spectrum (\ref{eq:powcut}), whose magnitude depends
in general upon additional parameters of the inflationary model.
\item Cutoff $\alpha$ states and non-adiabatic states generally
produce the
largest backreaction contributions during the inflationary epoch,
given by (\ref{eq:I1I2alp}), which are of order (\ref{eq:Mbound}).
\item States which are adiabatic order zero up to the cutoff scale
$k_M$ produce scale dependent modifications of the CMB power
spectrum (\ref{eq:power-zero-general}) which may be observable as
modulations in the CMB power spectrum.
\item Such states also
produce backreaction effects during inflation
which are somewhat
smaller in amplitude than the cutoff $\alpha$ states,
and which
can be calculated exactly from Eqs. (\ref{eq:finT}),
(\ref{eq:I1I2}) and (\ref{eq:I12adb0}).
\item The modifications of
the initial state given by the addition of a
local higher
  dimension operator with coefficient $\beta$ in the boundary
action
approach are non-adiabatic and yield in general modifications to
the CMB power spectrum at linear order, Eqs. (\ref{eq:powbou}) and
(\ref{eq:powlin}) in $\beta$.
\item The backreaction contributions
to the stress tensor during inflation
are also linear in $\beta$
in general and of order $\beta M^4$, which
may be significant,
depending on the cutoff and inflation scales $M$ and $H$,
but do not disturb inflation if (\ref{eq:betabound}) is satisfied.
\item
The adiabatic expansion of the stress tensor can be used also to
define a time dependent particle number basis for particles
created by the RW expansion, (\ref{eq:adbnum}), with parameters
$W_k$ and $V_k$ defined by Eqs. (\ref{eq:epsTtwo}),
(\ref{eq:vacadb}) and (\ref{eq:WV}) matched to the stress tensor
exactly at second adiabatic order.
\item The total particle number
defined in this way is the minimal
one that is finite,
(\ref{eq:finN}), and gives separately conserved
vacuum and
particle contributions to the covariant stress tensor.
\item In
the general massive case the particle number is not conserved
but
is a sixth order adiabatic invariant, implying that the density
matrix in the adiabatic particle representation has slowly varying
diagonal components but much more rapidly varying off-diagonal
components.
\item Although the exact evolution is unitary and
fully reversible, averaging over the rapidly varying off-diagonal
elements of the density matrix in this basis gives rise to an
approximation which is effectively dissipative, and in which the
effective von Neumann entropy of the reduced density matrix
(\ref{eq:vN}) may increase with time.
\item Neglect of these same
phase correlations in the energy-momentum
tensor via the
approximations (\ref{eq:qpareps}) and (\ref{eq:qparpress})
gives
an effective rate of heat dissipation due to particle creation
(\ref{eq:diss}) and (\ref{eq:heat}), even in the absence of matter
self-interactions.
\item Notable exceptions to this dephasing
occur in several special
massless cases, due to the absence of
particle creation for a conformally
invariant scalar field in any
RW spacetime, and for a massless, minimally
coupled scalar field
in de Sitter spacetime.
\item The latter result implies that the
quantum phase information in
the density perturbations derived in
slow roll inflationary models is
{\it not} washed out by the
expansion alone, so that the loss of phase
decoherence in such
models must be due to other effects not considered
in this paper.
\item The decoherence functional for arbitrary mixed Gaussian
states given by Eq. (\ref{eq:decfnl}) is related to the noise
kernel, or imaginary part of the second variation CTP effective
action (\ref{eq:secvar}).
\item The adiabatic method may be used
again to define the renormalized
decoherence functional
(\ref{eq:decren}) in semi-classical cosmology,
which is
independent of the short distance cutoff and field
reparameterizations.
\item This renormalized decoherence
functional vanishes in the
special cases where there is no
adiabatic particle creation,
corroborating the close connection
between particle creation,
dephasing and decoherence.
\item
Comparison of the decoherence functional defined here with a
previous result in the massless case is given in Appendix C.
\item
Verifying this adiabatic subtraction through a covariant
subtraction of the surface terms in the CTP action functional
would allow the study of decoherence effects and the quantum to
classical transition quantitatively and reliably in general RW
cosmologies for the first time.
\end{itemize}

\acknowledgments P.\ R.\ A.\ and C.\ M.-P.\ would like to thank
T-8, Los Alamos National Laboratory for its hospitality.  E.\ M.\
would like to thank the Michigan Center for Theoretical Physics
and the Aspen Center for Physics for their hospitality and
providing the venue for useful discussions with G. Shiu, J.-P. van
der Schaar and M. Porrati about their work. All authors wish to
thank the authors of~\cite{Shnew} for sharing their manuscript
with us prior to publication, to M.\ Martin and L.\ Teodoro
for careful reading of this manuscript, and to A. A. Starobinsky
for helpful comments reagrding the relation of this work to his.
This work was supported in part by grant numbers PHY-9800971 and
PHY-0070981 from the National Science Foundation, and by contract
number W-7405-ENG-36 from the Department of Energy. C.\ M.-P.\ would
like to thank the Nuffield Foundation for support by grant number
NAL/00670/G.

\appendix

\section{Equations for $\zeta_k$ and $\pi_k$ and their effective Hamiltonian}
\label{app:xi-eta}

In this appendix we compute the three independent and symmetric Gaussian
variances $\langle \Phi^2 \rangle$, $\langle \Phi \Pi_\phi + \Pi_\phi
\Phi \rangle$ and $\langle \Pi_\phi^2\rangle$, and
derive the equations of motion for the density matrix parameters
$\zeta_k$ and $\pi_k$ defined in Sec.~\ref{sec:density}.

The square of the defining relation for $\zeta_k$ in Eq.~(\ref{eq:zetakdef}) is
\begin{equation}
\sigma_k\, \vert\phi_k\vert^2 = \zeta_k^2\,.
\label{eq:zetainv}
\end{equation}
Hence using Eqs.~(\ref{eq:lmsum}) and~(\ref{eq:wight}) we find
directly for the first variance at coincident spacetime points,
\begin{equation}
\langle \Phi^2\rangle = \frac{1}{2 \pi^2} \int\, [\d k]\,  k^2\,
\sigma_k \vert \phi_k\vert^2  = \frac{1}{2 \pi^2} \int\, [\d k]\,
k^2\,  \zeta_k^2\,, \label{eq:phi2}
\end{equation}
which is explicitly real and independent of ${\bf x}$. In order to
compute the second variance we differentiate Eq.~(\ref{eq:zetainv}) to
obtain
\begin{equation}
\frac{\sigma_k}{2}(\phi_k \dot \phi_k^* + \dot \phi_k \phi_k^*)
= \sigma_k\,{\rm Re} \,(\phi_k \dot \phi_k^*) = \zeta_k \dot \zeta_k\; ,
\label{eq:phidotinv}
\end{equation}
which is the second of relations~(\ref{eq:phibil}).  Hence, using
Eq.~(\ref{eq:Piphi}) the second symmetric variance at coincident
points is
\begin{eqnarray}
&&\langle \Phi \Pi_\Phi + \Pi_\Phi \Phi \rangle =
a^3\,\langle \Phi \dot \Phi + \dot \Phi \Phi
+ 12 \xi H \Phi^2\rangle \nonumber \\
&& \qquad = \frac{a^3}{\pi^2} \int\,[\d k]\,  k^2\, \sigma_k
\left[ {\rm Re}\, (\phi_k \dot \phi_k^*)
+ 6 \xi H  \vert \phi_k\vert^2 \right]\nonumber \\
&& \qquad \qquad = \frac{1}{\pi^2} \int\,[\d k]\,  k^2\, \zeta_k
\pi_k \; . \label{eq:phipi}
\end{eqnarray}
By squaring Eq.~(\ref{eq:phidotinv}) and using the Wronskian
condition~(\ref{eq:wron}) in
\begin{eqnarray}
&&(\phi_k \dot \phi_k^* + \dot \phi_k \phi_k^*)^2 =
4 |\phi_k|^2 |\dot \phi_k|^2 + (\phi_k \dot \phi_k^* - \dot \phi_k \phi_k^*)^2\nonumber\\
&&\qquad = 4 |\phi_k|^2 |\dot \phi_k|^2 - \frac{\hbar^2}{a^6}\,,
\end{eqnarray}
we obtain
\begin{equation}
\sigma_k |\dot \phi_k|^2 = \dot\zeta_k^2 + \frac{\hbar^2\sigma_k^2}{4a^6 \zeta_k^2} \,,
\label{eq:phiddinv}
\end{equation}
which is the third member of Eq.~(\ref{eq:phibil}). Hence the third
Gaussian variance is
\begin{eqnarray}
&& \langle  \Pi_\Phi^2\rangle =
a^6 \,\langle \dot \Phi^2 + 6 \xi H (\Phi \dot\Phi + \dot\Phi \Phi)
+ 36\xi^2 H^2 \Phi^2\rangle \nonumber \\
&& \quad =\frac{a^6}{2 \pi^2} \int\,\d k\,  k^2\,
\sigma_k\,\left[\vert\dot \phi_k\vert^2 + 12 \xi H \,{\rm Re} \,(\phi_k \dot \phi_k^*)
+ 36 \xi^2 H^2 \vert\phi_k\vert^2\right] \nonumber\\
&& \qquad = \frac{a^6}{2 \pi^2} \int\,\d k\,  k^2\,
\left[\dot\zeta_k^2 + \frac{\hbar^2\sigma_k^2} {4a^6 \zeta_k^2}
+ 12 \xi H \zeta_k\dot\zeta_k + 36 \xi^2 H^2 \zeta_k^2\right] \nonumber\\
&& \qquad \qquad = \frac{1}{2\pi^2} \int\, \d k\, k^2\, \left(\pi^2_k
+ \frac{\hbar^2\sigma^2_k }{4 \zeta^2_k }\right)\, .
\label{eq:pipi}
\end{eqnarray}
This establishes Eqs.~(\ref{eq:piphibi}) of Sec.~\ref{sec:density}.

The second order differential equation for $\zeta_k$ may be
derived by differentiating~(\ref{eq:phidotinv}), and making use of
Eqs.~(\ref{eq:phiddinv}) and~(\ref{eq:phikeq}) to obtain
\begin{equation}
\ddot \zeta_k + 3 H \dot \zeta_k
+ \left(\frac{k^2 -\epsilon}{a^2} + m^2 + \xi R\right)\zeta_k
= \frac{\hbar^2\sigma_k^2}{4 a^6 \zeta_k^3}\; ,
\end{equation}
which is Eq.~(\ref{eq:zdd}) of the text.

The equations for the parameters of the Gaussian density matrix may be
compared with those arising from the purely classical
Hamiltonian~(\ref{eq:Hphi}), {\it viz.,}
\begin{subequations}
\begin{eqnarray}
\dot\Phi &=& \frac{\Pi_\Phi}{a^3} - 6\xi H \Phi \, , \\
\dot\Pi_\Phi &=& 6 \xi H \Pi_\Phi -a^3\left[\frac{-\Delta_3 +\epsilon}{a^2}
+ m^2 + (6 \xi -1)\left(\frac{\epsilon}{a^2} +6 \xi {H^2}\right) \right]
\Phi \; .
\label{eq:PiPhicl}
\end{eqnarray}
\end{subequations}
\hspace{-0.2cm} Note in particular that for $\hbar \neq 0$ the
equation of motion for $\pi_k$~(\ref{eq:hamevo4}) differs from
Eq.~(\ref{eq:PiPhicl}) of the purely classical evolution by the
last centrifugal barrier-like term in Eq.~(\ref{eq:hamevo4}) which
is a result of the uncertainty principle being enforced on the
initial data through the Wronskian condition~(\ref{eq:wron}).

The first order evolution equations for the parameters $(\bar
\phi, \bar p; \zeta_k, \pi_k; \sigma_k)$ may also be regarded as
Hamilton's equations for the effective {\it classical}
Hamiltonian,
\begin{equation}
H_{\rm eff}[\bar \phi, \bar p; \{\zeta_k, \pi_k; \sigma_k\}] =
{\rm Tr}\, ({\cal H}_{\Phi} \hat \rho) = \bar {\cal H}_{\Phi}(\bar
\phi, \bar p) + \frac{1}{2\pi^2}\int\, [\d k]\, k^2\, {\cal H}_k
(\zeta_k, \pi_k; \sigma_k)
\label{eq:effHam}
\end{equation}
where
\begin{equation}
\bar {\cal H}_{\Phi}(\bar \phi, \bar p) =
\frac{\bar p^2}{2a^3} - 6\xi H  \bar p \bar \phi +\frac{a^3}{2}
\left[ m^2 + 6 \xi\frac{\epsilon}{a^2} +6 \xi (6 \xi -1)H^2 \right] \bar \phi^2\,,
\end{equation}
is the classical Hamiltonian of the spatially independent mean values,
$(\bar \phi, \bar p)$ and
\begin{equation}
{\cal H}_k (\zeta_k, \pi_k; \sigma_k) =
   \frac{\pi_k^2}{2 a^3}
-6\xi H  \pi_k \zeta_k +\frac{a^3}{2} \left[ \omega_k^2 + (
6 \xi -1)\frac{\epsilon}{a^2} +6 \xi (6 \xi -1){H^2} \right]
   \zeta_k^2  + \frac{\hbar^2\sigma_k^2}{8 a^3 \zeta_k^2}\,,
\label{eq:Hkeff}
\end{equation}
is the effective Hamiltonian describing the Gaussian fluctuations
around the mean field for the Fourier mode $k$. It is
straightforward then to verify that Hamilton's equations for this
effective classical Hamiltonian (in which $\hbar$ appears as a
parameter), {\it viz.,}
\begin{subequations}
\begin{eqnarray}
\dot{\bar\phi}&=& \frac {\partial \bar {\cal H}_{\Phi}}{\partial \bar p}\,, \\
\dot{\bar p}&=& -\frac {\partial \bar {\cal H}_{\Phi}}{\partial \bar \phi}\,,\\
\dot\zeta_k &=&\frac {\partial {\cal H}_k}{\partial \pi_k}\,,\\
\dot\pi_k &=& -\frac {\partial {\cal H}_k}{\partial \zeta_k}\,,
\end{eqnarray}
\label{eq:zpham}
\end{subequations}
\hspace{-0.25cm}
are identical with Eqs.~(\ref{eq:hamev}) of the text. Hence
$\zeta_k$ and $\pi_k$ are conjugate variables with respect to the
effective classical Hamiltonian~(\ref{eq:effHam}).

If we define the complex frequency $\Upsilon_k$ by~(\ref{eq:Upseqa})
of the text, then by differentiating that definition and using
Eq.~(\ref{eq:hamev}) we obtain its equation of
motion~(\ref{eq:upseom}). On the other hand Eq.~(\ref{eq:phidotinv})
with $\sigma_k =1$, together with the Wronskian
condition~(\ref{eq:wron}) imply
\begin{equation}
\hbar -2ia^3\,\zeta_k\dot\zeta_k = -2ia^3\,\phi_k\dot\phi_k^*\,,
\end{equation}
so that dividing by $2 \zeta_k^2 = 2\phi_k\phi_k^*$, and using the
definition of $\pi_k$ in Eq.~(\ref{eq:zetakdef}) we obtain
\begin{equation}
\Upsilon_k = \frac{\hbar}{2\zeta_k^2} - \frac{i\pi_k}{\zeta_k} = -i a^3\,
\frac{\dot\phi_k^*}{\phi_k^*} -6i\xi H a^3\,,
\end{equation}
which establishes Eq.~(\ref{eq:Upseqb}) of the text.

Finally we remark that  ${\cal H}_{\Phi}$ depends on the choice of
variables used to represent the scalar field. Indeed, if we choose
the conformal field variable,
\begin{subequations}
\begin{equation}
\chi = a \Phi\,,
\end{equation}
instead of $\Phi$ and define the conjugate field momentum,
\begin{equation}
\Pi_{\chi}\equiv \frac{\partial S_{\rm cl}}{\partial \chi^{\prime}}
= \chi^{\prime} - \frac{a'}{a}\,\chi\, \qquad (m=0,\ \xi=0)
\end{equation}
\label{eq:chican}
\end{subequations}
\hspace{-0.25cm}
for the massless, minimally coupled field, the canonical
Hamiltonian,
\begin{equation}
{\cal H}_{\chi} = \frac{1}{2} \Pi_{\chi}^2 + \frac{k^2}{2} \chi^2 +
\frac{a'}{a}\chi\Pi_{\chi}
=a {\cal H}_{\Phi} + \dot a \Phi\Pi_{\Phi}\,,\qquad (m=0,\ \xi=0)\,,
\end{equation}
differs from ${\cal H}_{\Phi}$ defined by Eq.~(\ref{eq:Hphi}), and
neither is equal to the time component of the covariant stress tensor
which couples to gravity for general $m$ and $\xi$. This is to be
expected since the canonical transformation from ($\Phi,\Pi_{\Phi}$)
to ($\chi,\Pi_{\chi}$) is a time dependent transformation, and neither
Hamiltonian is a conserved quantity. This has the consequence
that while every representation is physically equivalent, describing
exactly the same physical time evolution, there is no spacetime or
field coordinate independent meaning to the basis which diagonalizes
the instantaneous canonical Hamiltonian in a particular set of coordinates
~\cite{GMM}, and no reason to prefer any such choice over any other as a
physical particle basis.

\section{Evaluation of Integrals}
\label{app:int}

In evaluating the integrals $I_1$ and $I_2$ in Eqs.~(\ref{eq:I1I2})
which contribute to the energy density and pressure of the massless,
minimally coupled scalar field, we encounter integrals of the form,
\begin{subequations}
\begin{eqnarray}
\int_0^{k_M}\, \d k\, k^{2n}\, \sin (2ku - \theta) = (-)^n \frac{k_M^{2n+1}}{2^{2n}}\,
F^{(2n)}_{\theta}(k_M u)\,,\\
\int_0^{k_M}\, \d k\, k^{2n+1}\, \cos (2ku - \theta)= (-)^n \frac{k_M^{2n+2}}{2^{2n+1}}\,
F^{(2n+1)}_{\theta}(k_M u)\,.
\end{eqnarray}
\label{eq:Fint}
\end{subequations}
\hspace{-0.25cm}
Integrals of this kind are easily evaluated by successive
differentiation with respect to $u$ of the elementary integral,
\begin{equation}
\int_0^{k_M}\, \d k\, \sin (2ku - \theta) = \frac {\cos \theta - \cos(2k_Mu - \theta)}{2u}
= k_M\,F_{\theta}(k_Mu)\,,
\end{equation}
where
\begin{equation}
F_{\theta}(x) \equiv F_{\theta}^{(0)}(x) = \frac {\cos \theta - \cos(2x - \theta)}{2x}
= \frac{\sin x\,\sin(x-\theta)}{x}\,.
\end{equation}
Thus, in Eq.~(\ref{eq:Fint}) we have
\begin{equation}
F^{(p)}_{\theta} (x) = \frac{\partial^p}{\partial x^p}
\left(\frac{\sin x\,\sin(x-\theta)}{x}\right)\,,
\label{eq:Fpdef1}
\end{equation}
and, in the particular case $\theta = 0$,
\begin{equation}
F^{(p)}(x) \equiv F^{(p)}_0(x) = \frac{\partial^p}{\partial x^p}
\left(\frac{\sin^2 x}{x}\right)\,.
\label{eq:Fp}
\end{equation}
We also define
\begin{equation}
F^{(-1)}(x) \equiv \int_0^x\, \d y  F(y)
= \int_0^x\, \d y \frac{\sin^2 y}{y}\,.
\end{equation}
For any $\theta$ and $p \ge 0$, $F^{(p)}_{\theta}(x)$ are damped
oscillatory functions whose maxima occur at $x=0$ for $p$ odd and
on the first oscillation for $p$ even. The values of $x$ and
$F^{(p)}_{\theta}$ at the maximum are of order unity. The leading
behavior as $x\rightarrow \infty$ is obtained by differentiating
the oscillatory numerator only, {\it i.e.,}
\begin{equation}
F^{(p)}_{\theta}(x) \rightarrow  \frac{1}{x}
\frac{\partial^p}{\partial x^p} \sin x\,\sin(x-\theta) +  {\cal
O}\left(\frac{1}{x^2}\right)\,.
\end{equation}
Thus the absolute value of $F^{(p)}_{\theta}(x)$ is bounded by
$1/x$ for $x\gg 1$. Hence the maximum value of the integrals~(\ref{eq:Fint})
are of order $k_M^{2n+1}$ and $k_M^{2n+2}$
respectively for $u\sim k_M^{-1}$, while they behave like
$k_M^{2n}$ and $k_M^{2n+1}$ respectively, multiplied by a rapidly
oscillating function of $k_Mu$, as $k_M u \rightarrow \infty$.
The form of $F^{(4)}(x)$ as a function of $x$ in the special
case of $p=4$ and $\theta=0$ is given by Fig.~\ref{fig:F4}
of the text.

\section{Comparison of Squeezing in Different Bases}
\label{app:bases}

In Sec. \ref{sec:adiabatic} we found that in the second order
adiabatic particle basis defined by Eqs. (\ref{eq:WV}) there is
no large squeezing and no particle creation of a massless, minimally
coupled scalar field in exact de Sitter spacetime. In Sec.
\ref{sec:decoherence} we corroborated the lack of true decoherence for
this field. In this Appendix we compare this result to earlier work,
in particular to Ref.~\cite{PolSta}, whose authors use a field amplitude
``pointer basis."

Let us first consider a zeroth order adiabatic basis rather than the
second order adiabatic basis defined
by Eqs.~(\ref{eq:WV}). The zeroth order basis function is obtained
by replacing $W_k$ of Eq.~(\ref{eq:mmcdeS}) by $\omega_k = k/a_{\rm
dS}$, resulting in
\begin{equation}
\tilde\phi^{(0)}_k = \sqrt\frac{\hbar}{2a^3 \omega_k}\, \exp\left(-i \int^t\,
\d t'\,\omega_k (t')\right) =
\frac{1}{a}\,\sqrt{\frac{\hbar}{2k}}\, \exp\left(-i k\eta\right)
\label{eq:basisz}
\end{equation}
instead of Eq.~(\ref{eq:exactmode}). When the exact BD mode is expressed
as a linear combination of these zeroth order adiabatic modes in the form
of the Bogoliubov transformation,
\begin{equation}
\phi_k^{BD} = \alpha^{(0)}_k \tilde\phi^{(0)}_k + \beta^{(0)}_k \tilde\phi^{(0)*}_k,
\end{equation}
we now obtain the non-trivial time dependent coefficients,
\begin{subequations}
\begin{eqnarray}
\alpha^{(0)}_k &=& \left(1 - \frac{i}{2k\eta}\right)\,,\\
\beta^{(0)}_k &=& -\frac{i}{2k\eta} \exp\left(-2i k\eta\right)\,.
\end{eqnarray}
\label{eq:bogzero}
\end{subequations}
\hspace{-0.3cm} Clearly in this adiabatic zero basis $\sinh
r_k^{(0)} = |\beta^{(0)}_k|= 1/|2k\eta|$ which does not fall off
fast enough at large $k$ to be fourth order adiabatic, and which
also approaches infinity as $\eta \rightarrow 0^-$.

The problem with the zeroth order adiabatic basis~(\ref{eq:basisz}) is
the ``particle"
number~(\ref{eq:finN}),
\begin{equation}
\frac{1}{2\pi^2} \int \d k \, k^2 \, |\beta^{(0)}_k|^2 =
\frac{1}{8\pi^2\eta^2}\int \d k \rightarrow \infty
\label{eq:Ndiv}
\end{equation}
is divergent at any finite time $\eta$. Corresponding to this linear
divergence in the total number, the energy-momentum tensor of these
``particles" is quadratically divergent at large $k$. Clearly this
quadratic divergence is a residual divergence of the {\it vacuum}
stress tensor and has nothing at all to do with physical particles,
which are unambiguously well-defined in the UV limit $k\rightarrow
\infty$.  In that limit of wavelength much smaller than the horizon
scale, the region of spacetime the modes sample may be approximated as
flat, and the effects of the time variation of the geometry should be
negligible.  The $\eta^{-2}$ factor in Eq.~(\ref{eq:Ndiv}) shows that
this mismatch only grows more severe at late times, as the zeroth
order basis~(\ref{eq:basisz}) becomes more and more different from the
second order basis~(\ref{eq:exactmode}).

The quadratic vacuum divergences can be subtracted only by matching
the energy-momentum tensor to second adiabatic order as in Eq.~(\ref{eq:WV}).
When that is done the particle number is finite and no squeezing, phase
decoherence or particle creation effect at all is obtained.
Moreover, since the second order adiabatic modes~(\ref{eq:exactmode}) are already
exact for the massless, minimally coupled scalar field in de Sitter
space, any ambiguity in the particle concept at wavelengths of
the order of the horizon scale is irrelevant here. Going to
higher orders in the adiabatic expansion will not change the result
obtained at second order.

The authors of Ref.~\cite{PolSta} define a field ``pointer basis"
for the massless, minimally coupled field in de Sitter space.
Comparing Eqs.~(15),~(19), and~(47) of Ref.~\cite{PolSta} with
relations (\ref{eq:bogzero}) above, we find that the Bogoliubov
coefficients, $\alpha_k$ and $\beta_k$ of Eq.~(15) in
Ref.~\cite{PolSta} are precisely {\it equal} to $\alpha_k^{(0)}$
and $\beta_k^{(0)}$ respectively, of the zeroth order adiabatic
basis given by (\ref{eq:bogzero}). As we have seen, the squeezing
parameter with respect to this basis, $\sinh\,r_k^{(0)} = 1/\vert
2 k \eta\vert \rightarrow \infty$ does become very large for
superhorizon modes in the late time limit. The authors of
Ref.~\cite{PolSta} argue that the large squeezing in this basis
leads to an effective decoherence of modes of the scalar field
much larger than the de Sitter horizon.

It is clear that the squeezing is very much dependent on
the basis in which it is computed. In the second order basis~(\ref{eq:exactmode}),
determined by the structure of the short distance expansion of the
covariantly conserved stress tensor through Eqs.~(\ref{eq:WV}),
there is no mixing of positive and negative frequency modes, and no
large squeezing of superhorizon modes. True decoherence of these
modes should occur through other effects, such as those considered in
Ref.~\cite{KiePolSta} at the time these modes reenter the horizon.

\end{document}